\documentclass{article}
\usepackage{bm}
\usepackage{graphicx}
\usepackage{amsmath,amsfonts,amssymb,amsthm,amsopn,amscd}
\usepackage{bbm}
\newcommand{\bd}{\begin{document}}
\newcommand{\ed}{\end{document}}
\newcommand{\beq}{\begin{equation}}
\newcommand{\eeq}{\end{equation}}
\newcommand{\nid}{\noindent}
\newcommand{\ben}{\begin{enumerate}}
\newcommand{\een}{\end{enumerate}}
\newcommand{\bit}{\begin{itemize}}
\newcommand{\eit}{\end{itemize}}
\newcommand{\baR}{\begin{array}}
\newcommand{\eaR}{\end{array}}
\newcommand{\su}[1]{\section{#1}}
\newcommand{\ssu}[1]{\subsection{#1}}
\newcommand{\sssu}[1]{\subsubsection{#1}}
\newcommand{\deq}{\stackrel {\rm def}{=}}
\newcommand{\bea}{\begin{eqnarray}}
\newcommand{\eea}{\end{eqnarray}}
\newcommand{\SOS}[1]{~\\\centerline{\bf{\large [#1]}}\\~}
\newcommand{\st}[1]{{\nid\bf #1}}
\newcommand{\emp}[1]{{\it #1}}

\newcommand\D{\displaystyle}

\newcounter{CDef}[section]
\newtheorem{definition}{Definition}
\newcommand{\bdf}{\begin{definition}}
\newcommand{\edf}{\end{definition}}
\newcommand{\bpr} {\noindent {\bf Proof }}
\newcommand{\epr}{\rule{0.25cm}{0.25cm}\\}

\newtheorem{theorem}{Theorem}
\newcommand{\bth}{\begin{theorem}}
\newcommand{\enth}{\end{theorem}}
\newcounter{Cprop}[section]
\newtheorem{proposition}{Proposition}
\newcommand{\bp}{\begin{proposition}}
\newcommand{\ep}{\end{proposition}}
\newcounter{Ccor}\newtheorem{corrollary}{Corrollary}
\newcommand{\bcor}{\begin{corrollary}}
\newcommand{\ecor}{\end{corrollary}}
\newcounter{CLem}\newtheorem{lemma}{Lemma}
\newcommand{\blem}{\begin{lemma}}
\newcommand{\elem}{\end{lemma}}
\newcounter{Cconj}\newtheorem{conjecture}{Conjecture}
\newcommand{\bconj}{\begin{conjecture}}
\newcommand{\econj}{\end{conjecture}}
\newcommand\rep{\rightharpoonup}
\newcommand\drep{\downharpoonright}

\newcommand\bbbr{{\sf I\!R}}
\newcommand\bbbrn{{\sf I\!R}^N}
\newcommand\bbbn{{\sf I\!N}}
\newcommand\Wb{\bar{W}}
\newcommand\Vm{V_{min}}
\newcommand\VM{V_{max}}
\newcommand\bt{\bar{\theta}}
\newcommand\sigt{\sigma_\theta}
\newcommand{\XI}{\mbox{\boldmath $\xi$}}

\newcommand\ei{{\bf e}_i}
\newcommand\x{{\bf x}}
\newcommand\X{{\bf X}}
\newcommand\xT{{\x^{(T)}}}
\newcommand\XTt{{\X^{(\tau)}(t)}}
\newcommand\XTtp{{\X^{(\tau)}(t+1)}}
\newcommand\VTt{{\V^{(\tau)}(t)}}
\newcommand\VTtp{{\V^{(\tau)}(t+1)}}
\newcommand\bolm{{\bf m}}
\newcommand\mT{\bolm^{(\tau)}}
\newcommand\miT{m_i^{(\tau)}}
\newcommand\mTp{\bolm^{(\tau+1)}}
\newcommand\Y{{\bf Y}}
\newcommand\F{{\bf F}}
\newcommand\bH{{\bf H}}
\newcommand\Fg{{\F_{\bg}}}
\newcommand\FgT{{\F_{\gT}}}
\newcommand\Fgt{{\F_{\bsg}^t}}
\newcommand\Fgta{{\F_{\bsg}^{t\ast}}}
\newcommand\Fgn{{\F_{\bsg}^n}}
\newcommand\Wij{W_{ij}}
\newcommand\WijT{W_{ij}^{(\tau)}}
\newcommand\WT{\cW^{(\tau)}}
\newcommand\GaT{\Gamma^{(\tau)}}
\newcommand\GijT{\Gamma_{ij}^{(T)}}
\newcommand\WTp{\cW^{(\tau+1)}}
\newcommand\dWT{\delta\cW^{(\tau)}}
\newcommand\WijTp{W_{ij}^{(\tau+1)}}
\newcommand\dWijT{\delta \WijT}
\newcommand\dWij{\delta \Wij}
\newcommand\poT{\pi_{\tom^{(\tau)}}}
\newcommand\poTp{\pi_{\tom^{(\tau+1)}}}
\newcommand\tin{\tau_i(\omega,n)}

\newcommand\ld{r_{d}}
\newcommand\lmT{\lambda_1^{(\tau)}}

\newcommand\tX{\tilde{\X}}

\newcommand\lb{\left\langle}
\newcommand\rb{\right\rangle}
\newcommand\rbT{\rb^{(\tau)}}
\newcommand\rbTp{\rb^{(\tau+1)}}

\newcommand\cA{{\cal A}}
\newcommand\cB{{\cal B}}
\newcommand\bC{{\bf C}}
\newcommand\cF{{\cal F}}
\newcommand\cL{{\cal L}}
\newcommand\cFg{{\cF_{\bg}(\phi)}}
\newcommand\cFgp{{\cF_{\bg'}(\phi)}}
\newcommand\cP{{\cal P}}
\newcommand\cR{{\cal R}}
\newcommand\cT{{\cal T}}

\newcommand\Po{{\cP_{\bom}}}
\newcommand\Pop{{\cP_{\bom'}}}
\newcommand\Pot{{\cP_{\bom(t)}}}
\newcommand\pTo{{\pi_{\tom}^{(T)}}}
\newcommand\pToT{{\pi_{\tom^{(\tau)}}^{(T)}}}
\newcommand\pTpoT{{\pi_{\tom^{(\tau+1)}}^{(T)}}}
\newcommand\po{{\pi_{\tom}}}
\newcommand\pop{{\pi_{\tom'}}}
\newcommand\piij{{\pi_{\tom;ij}}}
\newcommand\mula{{\mu^{Leb}_{\cA}}}
\newcommand\muo{{\mu_{\tom}}}
\newcommand\mg{{\mu_{\bg}}}
\newcommand\mpg{{\nu_{\bspsi}}}
\newcommand\mpgs{{\nu_{\bspsis}}}
\newcommand\Do{\cD(\bom)}
\newcommand\bDo{\bar{\cD}(\bom)}
\newcommand\Dop{\cD(\bom')}
\newcommand\bDop{\bar{\cD}(\bom')}

\newcommand\dT{{d_\Theta}}
\newcommand\dtl{\Delta^{(\tau)}[S']}

\newcommand\BA{\cB(\cA)}
\newcommand\ba{\boldsymbol\alpha}
\newcommand\Bg{\boldsymbol\Gamma}
\newcommand\bg{\boldsymbol\gamma}
\newcommand\bl{\boldsymbol\lambda}
\newcommand\bphi{\boldsymbol\phi}
\newcommand\Beta{\boldsymbol\eta}
\newcommand\bpsi{\boldsymbol\psi}
\newcommand\bPsi{\boldsymbol\Psi}
\newcommand\bom{\boldsymbol\omega}
\newcommand\bu{{\bf u}}
\newcommand\tbu{\tilde{\bu}}
\newcommand\tu{\tilde{u}}
\newcommand\sbg{\scriptstyle{\boldsymbol\gamma}}
\newcommand\gT{\bg^{(\tau)}}
\newcommand\sgT{\sbg^{(\tau)}}
\newcommand\gTp{\bg^{(\tau+1)}}
\newcommand\bsg{\sbg}
\newcommand\bsl{\scriptstyle{\boldsymbol\lambda}}
\newcommand\bspsi{\scriptstyle{\boldsymbol\psi}}
\newcommand\bspsis{\scriptstyle{\boldsymbol\psi^\ast}}
\newcommand\bls{\bsl^\ast}
\newcommand\ble{\bl^\ast}
\newcommand\bpsis{\bpsi^\ast}
\newcommand\bxi{\bm{\xi}}

\newcommand\tom{\omega}
\newcommand\sg{{\sigma_{\bg}}}
\newcommand\sTg{{\sigma_{\sgT}}}
\newcommand\sTpg{{\sigma_{\sgTp}}}
\newcommand\stg{{\sigma^t_{\bg}}}
\newcommand\sit{{\sigma^t}}
\newcommand\stgT{{\sigma^t_{\sgT}}}
\newcommand\stgo{{\sigma^t_{\bg}\tom}}
\newcommand\Gg{{G_{\bg}}}
\newcommand\Slp{\left\{0,1\right\}^\bbbn}
\newcommand\SG{\Sigma_{G}}
\newcommand\Spg{\Sigma_{\bg}}
\newcommand\Spgp{\Sigma_{\bg'}}
\newcommand\Spgm{\Sigma_{\bg_m}}
\newcommand\SpgT{\Sigma_{\bg}^{(T)}}
\newcommand\Spgn{\Sigma_{\bg}^{(n)}}
\newcommand\SpT{\Sigma^{(T)}}
\newcommand\Spn{\Sigma^{(n)}}
\newcommand\SgT{\Sigma_{\gT}}
\newcommand\Sgm{\Sigma_{\bg^{(m)}}}
\newcommand\SgTp{\Sigma_{\gTp}}
\newcommand\Minv{\cM^{inv}_{\bg}}
\newcommand\MinvT{\cM^{inv}_{\sgT}}
\newcommand\MinvTp{\cM^{inv}_{\sgTp}}
\newcommand\STpo{S^{(T)}\bpsi(\tom)}
\newcommand\Snpo{S^{(n)}\bpsi(\tom)}
\newcommand\STq{S^{(T)}_{\bg} \psi^{(4)}(.;\bl)}
\newcommand\pres{P\left[\bpsi\right]}
\newcommand\grad{\nabla_{\bg}}
\newcommand\gradPT{\nabla_{\bg}\PT}

\newcommand\phia{\phi_{i_1,j_1}}
\newcommand\phib{\phi_{i_2,j_2}}
\newcommand\psinl{\bpsi^{(l,n)}}
\newcommand\psiOl{\bpsi^{(l,0)}}
\newcommand\psilpO{\bpsi^{(l+1,0)}}
\newcommand\psilnl{\bpsi^{(l,n_l)}}
\newcommand\dpsirlpO{\delta\bpsi^{(l+1,0)}_{reg}}
\newcommand\dpsislpO{\delta\bpsi^{(l+1,0)}_{sing}}
\newcommand\psiT{\bpsi^{(\tau)}}
\newcommand\dpsiT{\delta\bpsi^{(\tau)}}
\newcommand\psiTp{\bpsi^{(\tau+1)}}
\newcommand\dPT{\delta P^{(\tau)}}
\newcommand\PT{P\left[\psiT\right]}
\newcommand\PTp{P\left[\psiTp\right]}
\newcommand\npT{\nu_{\psiT}}
\newcommand\npTp{\nu_{\psiTp}}

\newcommand\muu{\nu^{(1)}_{\bl;\bg}}
\newcommand\Pu{P^{(1)}_{\bg}(\bl)}
\newcommand\pu{\phi^{(1)}(\tom;\bl)}
\newcommand\mud{\nu^{(2)}_{\bl;\bg}}
\newcommand\Pd{P^{(2)}_{\bg(\bl)}}
\newcommand\pd{\phi^{(2)}(\tom;\bl)}
\newcommand\mut{\nu^{(3)}_{\bl;\bg}}
\newcommand\Pt{P^{(3)}_{\bg}(\bl)}
\newcommand\pt{\phi^{(3)}(\tom;\bl)}
\newcommand\muq{\nu^{(4)}_{\bl;\bg}}
\newcommand\Pq{P^{(4)}_{\bg}(\bl)}
\newcommand\pq{\phi^{(4)}(\tom;\bl)}
\newcommand\psigo{\psi(\tom;\small{\bg})}
\newcommand\phigo{\phi(\tom;\small{\bg})}
\newcommand\hchi{\hat\chi}

\newcommand\nua{\nu_{\cA}}

\newcommand\gijo{g_{ij}^{(0)}(W_{ij})}
\newcommand\gijou{g_{ij}^{(01)}(W_{ij},\omejt)}
\newcommand\gijuo{g_{ij}^{(10)}(W_{ij},\omeit)}
\newcommand\gijdu{g_{ij}^{(21)}(W_{ij},\omeit^2\times \omejt)}
\newcommand\gijud{g_{ij}^{(12)}(W_{ij},\omeit\times \omejt^2)}
\newcommand\gijuu{g_{ij}^{(11)}(W_{ij},\omeit\times \omejt)}
\newcommand\gijkl{g_{ij}^{(kl)}(W_{ij},\omeitk\times \omejtl)}
\newcommand\hijkl{h_{ij}^{(kl)}(W_{ij})}

\newcommand\aijo{a_{ij}^{(0)}}
\newcommand\aijou{a_{ij}^{(01)}(\omejt)}
\newcommand\aijuo{a_{ij}^{(10)}(\omeit)}
\newcommand\aijdu{a_{ij}^{(21)}(\omeit^2\times \omejt)}
\newcommand\aijud{a_{ij}^{(12)}(\omeit\times \omejt^2)}
\newcommand\aijuu{a_{ij}^{(11)}(\omeit\times \omejt)}
\newcommand\aijkl{a_{ij}^{(kl)}(\omeitk\times \omejtl)}

\newcommand\aijtou{a_{ij}^{(01)}(W_{ij},\pTo(\tomej))}
\newcommand\aijtuo{a_{ij}^{(10)}(W_{ij},\pTo(\tomei))}
\newcommand\aijtdu{a_{ij}^{(21)}(W_{ij},\pTo(\tomei^2\times \tomej))}
\newcommand\aijtud{a_{ij}^{(12)}(W_{ij},\pTo(\tomei\times \tomej^2))}
\newcommand\aijtuu{a_{ij}^{(11)}(W_{ij},\pTo(\tomei\times \tomej))}

\newcommand\paijou{a_{ij}^{(01)}(W_{ij},\po(\omej))}
\newcommand\paijuo{a_{ij}^{(10)}(W_{ij},\po(\omei))}
\newcommand\paijdu{a_{ij}^{(21)}(W_{ij},\po(\tomei^2\times \tomej))}
\newcommand\paijud{a_{ij}^{(12)}(W_{ij},\po(\tomei\times \tomej^2))}
\newcommand\paijuu{a_{ij}^{(11)}(W_{ij},\po(\tomei\times \tomej))}
\newcommand\paijkl{a_{ij}^{(kl)}(W_{ij},\po(\tomeik\times \tomejl))}

\newcommand\Catn{[\ba(t_1)\ba(t_2) \dots \ba(t_n)]}
\newcommand\omei{\omega_i}
\newcommand\omej{\omega_j}
\newcommand\omek{\omega_k}
\renewcommand\omit{\omega_i^{(\tau)}}
\newcommand\omjt{\omega_j^{(\tau)}}
\newcommand\tomei{\tom_i}
\newcommand\tomej{\tom_j}
\newcommand\tomt{\left[\tom\right]_0^T}
\newcommand\tomit{\left[\tom_i\right]_0^T}
\newcommand\tomjt{\left[\tom_j\right]_0^T}
\newcommand\tomeik{\tom_i^{(k)}}
\newcommand\tomejl{\tom_j^{(l)}}
\newcommand\omeit{\left[\omega_i\right]_{t-T_s,t}}
\newcommand\omejt{\left[\omega_j\right]_{t-T_s,t}}
\newcommand\omeitk{\left[\omega_i\right]^{k}_{t-T_s,t}}
\newcommand\omeitku{\left[\omega_i\right]^{k_1}_{t-T_s,t}}
\newcommand\omeitkd{\left[\omega_i\right]^{k_2}_{t-T_s,t}}
\newcommand\omejtl{\left[\omega_j\right]^{l}_{t-T_s,t}}
\newcommand\omejtlu{\left[\omega_j\right]^{l_1}_{t-T_s,t}}
\newcommand\omejtld{\left[\omega_j\right]^{l_2}_{t-T_s,t}}

\newcommand\ZTp{Z_{\bg}^{(T)}(\psi)}
\newcommand\ZT{Z^{(T)}(\psi)}
\newcommand\Zn{Z^{(n)}(\psi)}

\newcommand\Vr{V_{reset}}
\newcommand\V{{\bf V}}
\newcommand\hV{\hat{V}}

\newcommand\I{{\bf I}}
\newcommand\J{{\bf J}}
\newcommand\Iw{I^{(w)}}
\newcommand\ie{i^{(ext)}}
\newcommand\Ie{{\bf{I}}^{(ext)}}
\newcommand\Iei{I^{ext}_i}
\newcommand\Is{I^{(syn)}}
\newcommand\Isk{I^{(syn)}_k}
\newcommand\hIa{\hat{I}^{(ion)}}

\newcommand\FpT{\cF^{(\tau)}_{\small{\bphi}}}
\newcommand\FpTp{\cF^{(\tau+1)}_{\small{\bphi}}}

\newcommand\tk{\tau^{(k)}}
\newcommand\tik{t_i^{(k)}}
\newcommand\til{t_i^{(l)}}
\newcommand\tjn{t_j^{(n)}}
\newcommand\tpm{\tau^{\pm}}
\newcommand\bG{\bar{G}}
\newcommand\cC{{\cal C}}
\newcommand\cD{{\cal D}}
\newcommand\cE{{\cal E}}
\newcommand\cG{{\cal G}}
\newcommand\cH{{\cal H}}
\newcommand\cI{{\cal I}}
\newcommand\cN{{\cal N}}
\newcommand\cM{{\cal M}}
\newcommand\cQ{{\cal Q}}
\newcommand\cS{{\cal S}}
\newcommand\cU{{\cal U}}
\newcommand\cV{{\cal V}}
\newcommand\cW{{\cal W}}
\newcommand\dW{\delta \cW}
\newcommand\cWT{\cW^{(\tau)}}
\newcommand\cWTp{\cW^{(\tau+1)}}
\newcommand\cWs{\cW^{(\ast)}}
\newcommand\KT{\kappa^{(\tau)}}
\newcommand\KabT{\kappa_{i_1,j_1,i_2,j_2}^{(\tau)}}
\newcommand\GT{G_{\gT}}

\newcommand\Bev{\cB(\V,\epsilon)}
\newcommand\bmd{\cB_\cM(\delta)}

\newcommand\ket[1]{\mid #1 \rangle}
\newcommand\bra[1]{\langle  #1  \mid}
\newcommand\av[3]{\langle #1 \mid #2 \mid #3 \rangle} 
\newcommand\scal[2]{\langle #1 \mid #2  \rangle} 

\newcommand\tost{\left[\tom\right]_{s,t}}
\newcommand\ton{\left[\tom\right]_{O,n}}
\newcommand\tot{\left[\tom\right]_{0,t}}
\newcommand\CoT{\left[\tom\right]_{0,T-1}}
\newcommand\Con{\left[\tom\right]_{0,n-1}}
\newcommand\totr{\left[\tom\right]_{t,t+r-1}}
\newcommand\totm{\left[\tom\right]_{0,t-1}}
\newcommand\torm{\left[\tom\right]_{0,r-1}}
\newcommand\tos{\left[\tom\right]_{0,s}}
\newcommand\toT{\left[\tom\right]_{0,T}}
\newcommand\bomt{\bom(t)}
\newcommand\bdom{\bar{\cD}(\bom)}
\newcommand\com{\cC(\bom)}
\newcommand\comt{\cC(\bom(t))}
\newcommand\dom{\cD(\bom)}
\newcommand\mjt{M_j(t,\tom)}

\newcommand\cMo{{\cM_{\bom}}}
\newcommand\cMot{\cM_{\bom(t)}}
\newcommand\Mot{\cM_{\tot}}
\newcommand\Motm{\cM_{\totm}}
\newcommand\MoT{\cM_{\toT}}
\newcommand\Mos{\cM_{\bom(s)}}
\newcommand\oM{\Omega}

\newcommand\be{{\bf e}}
\newcommand\FT{\F^T}
\newcommand\FTp{\F^{T+1}}
\newcommand\bcF{\bar{\cF}}
\newcommand\Ft{\F\left[.;.,\tom\right]}
\newcommand\Fto{\F\left[.;t,\tom\right]}
\newcommand\Ftot{\F^{t+1}_{\tom}}
\newcommand\Ftso{\F^{t,s}_{\tom}}
\newcommand\Fkto{F_k\left(t,\tom\right)}
\newcommand\Fkso{F_k\left(s,\tom\right)}
\newcommand\FkVto{F_k\left(t,\tom\right)\V}
\newcommand\FVTo{\F^{t+1}_{\tom}(\V)}
\newcommand\FTo{\F^{t}_{\tom}}
\newcommand\FkTo{F^{t}_{k;\tom}}
\newcommand\DFFTV{\left. D\FTF\right|_{\V}}
\newcommand\DFFTpV{\left. D\FTpF\right|_{\V}}

\newcommand\dAS{d(\oM,\cS)}
\newcommand\dOS{d(\oM,\cS)}
\newcommand\dASe{d^{(exp)}(\oM,\cS)}

\newcommand\GFTe{\Gamma_{\cF,T,\epsilon}}
\newcommand\GFTpe{\Gamma_{\cF,{T+1},\epsilon}}
\newcommand\GFTep{\Gamma_{\cF,T,\epsilon'}}

\newcommand\PF{\Pi_{\cF}}
\newcommand\PbF{\Pi_{\bcF}}
\newcommand\VF{{\V_{\cF}}}
\newcommand\VFb{{\V_{\bcF}}}
\newcommand\FF{{\F_{\cF}}}
\newcommand\FTF{{\FT_{\cF}}}
\newcommand\FTpF{{\FTp_{\cF}}}
\newcommand\FFb{{\F_{\bcF}}}
\newcommand\oGF{{\Omega(\GF)}}
\newcommand\oGFTe{{\Omega(\GFTe)}}
\newcommand\oFTe{{\Omega_{\cF}(\GFTe)}}
\newcommand\cPT{\cP^{(T)}}

\newcommand\Jkto{J_k(t,\tot)}

\newcommand\gkt{\gamma_k(t,\tot)}
\newcommand\gks{\gamma_k(s,\tos)}
\newcommand\skt{\sigma_k(t,\tot)}

\newcommand\heff{h^{(eff)}}
\newcommand\PrW{\cP_\cW}
\newcommand\PrI{\cP_{\Ie}}
\newcommand\PrWi{\cP_{\cW,\Ie}}

\newcommand\ltil{\left\{\til\right\}_t}
\newcommand\ltjn{\left\{\tjn\right\}_t}

\newcommand\wsg{\cW^s(\V)}
\newcommand\wsl{\cW^s_{loc}(\V)}
\newcommand\wslf{\cW^s_{loc}(\F(\V))}
\newcommand\wse{\cW^s_{\epsilon}(\V)}
\newcommand\wset{\cW^s_{\epsilon}(\V(t))}
\newcommand\psis{\psi^\ast}
\newcommand\nus{\nu^\ast}
\newcommand\bgs{\bg^\ast}

\newcommand{\Jb}{\bar{J}}
\newcommand{\Jbab}{\bar{J}_{\alpha\beta}}
\newcommand{\Wbab}{\bar{W}_{ab}}
\newcommand{\Jbcd}{\bar{J}_{\gamma\delta}}
\newcommand{\Jab}{\sigma_{ab}}
\newcommand{\Jdab}{\sigma^2_{ab}}
\newcommand{\ma}{m_\alpha}
\newcommand{\Ma}{M_\alpha}
\newcommand{\mua}{\mu_\alpha}
\newcommand{\muax}{\mua^X}
\newcommand{\muay}{\mua^Y}
\newcommand{\va}{v_\alpha}
\newcommand{\vax}{v_\alpha^X}
\newcommand{\Ca}{C_{\alpha\alpha}}
\newcommand{\Cax}{C_{\alpha\alpha}^X}
\newcommand{\Cbx}{C_{\beta\beta}^X}
\newcommand{\Cay}{C_{\alpha\alpha}^Y}
\newcommand{\Ia}{I_\alpha}
\newcommand{\Da}{\Delta_\alpha}
\newcommand{\Deab}{\Delta_{\alpha\beta}}
\newcommand{\Deabx}{\Delta_{\alpha\beta}^X}
\newcommand{\Deabv}{\Delta_{ab}}
\newcommand{\sa}{s_\alpha}
\newcommand{\ta}{\tau_\alpha}
\newcommand{\mub}{\mu_\beta}
\newcommand{\mb}{m_{\alpha\beta}}
\newcommand{\vb}{v_\beta}
\newcommand{\vbx}{v_\beta^X}
\newcommand{\Cb}{C_{\beta\beta}}
\newcommand{\Db}{\Delta_\beta}
\newcommand{\Mb}{M_\beta}
\newcommand{\qb}{q_\beta}
\newcommand{\gb}{\gamma_\beta(\mub,\vb;\Cb)}
\newcommand{\gbt}{\gamma_\beta(\mub,\vb;\Cb(t))}
\newcommand{\tb}{\tau_\beta}
\newcommand{\Sb}{S_\beta}
\newcommand{\Ha}{H_{\alpha}}
\newcommand{\Hax}{H_{\alpha}^X}
\newcommand{\Hab}{H_{\alpha\beta}}
\newcommand{\Habx}{H_{\alpha\beta}^X}
\newcommand{\Dab}{D_{\alpha\beta}}
\newcommand{\Dabx}{D_{\alpha\beta}^X}
\newcommand{\cJ}{{\cal J}}

\newcommand{\ind}[1]{\mathbbm{1}_{#1}}          

\newcommand{\derpart}[2]{ \frac{\partial #1}{\partial #2} } 
\newcommand{\der}[2]{ \frac{\text{d} #1}{\text{d} #2} }  
\newcommand{\derparts}[2]{ \frac{\partial^2 #1}{\partial #2 ^2}}
\newcommand{\derpartsxy}[3]{ \frac{\partial^2 #1}{\partial #2 \partial #3}} 
\newcommand{\argmin}{\mathop{\mathrm{Arg\,Min}}}
\newcommand{\eps}{\varepsilon}

\DeclareMathOperator{\eqlaw}{\stackrel{\mathcal{L}}{=}}         

\newcommand{\Herm}[1]{\mathcal{H}_{#1}}            
\newcommand{\Hermnu}{\mathcal{H}_{\nu}}            
\newcommand{\erf}{\mathrm{erf}}              
\newcommand{\Cov}{\mathrm{Cov}}

\newcommand{\m}{\mathcal}

\newcommand {\image} {\includegraphics}

\newcommand{\mC}{\mathbf{C}}
\newcommand{\mD}{\mathbf{D}}
\newcommand{\mE}{\mathbf{E}}
\newcommand{\mX}{\mathbf{X}}
\newcommand{\mx}{\mathbf{x}}
\newcommand{\mY}{\mathbf{Y}}
\newcommand{\mL}{\mathbf{L}}
\newcommand{\mQ}{\mathbf{Q}}
\newcommand{\mI}{\mathbf{I}}
\newcommand{\mJ}{\mathbf{J}}
\newcommand{\mW}{\mathbf{W}}
\newcommand{\mU}{\mathbf{U}}
\newcommand{\mone}{\mathbf{1}}

\newcommand{\msig}{\boldsymbol{\sigma}}
\newcommand{\mtheta}{\boldsymbol{\theta}}
\newcommand{\mdelta}{\boldsymbol{\Delta}}
\newcommand{\mla}{\boldsymbol{\Lambda}}
\newcommand{\mga}{\boldsymbol{\Gamma}}
\newcommand{\mxi}{\boldsymbol{\xi}}

\newcommand{\proc}{\mathcal{M}_1^+(C([t_0,T], \R)}
\newcommand{\procP}[1]{\mathcal{M}_1^+(C([t_0,T], \R^{#1}))}
\newcommand{\proci}{\mathcal{M}_1^+(C((-\infty,T], \R)}
\newcommand{\prociP}[1]{\mathcal{M}_1^+(C((-\infty,T], \R^{#1})}
\newcommand{\procs}[1]{\mathcal{M}_1^+(C([0,#1], \R)}
\newcommand{\Exp}[1]{\mathbb{E}\left [ #1 \right]}
\newcommand{\ExpSimple}{\mathbb{E}}
\newcommand{\nvar}[1]{\left \| #1 \right\|_{\mathrm{var}}}
\newcommand{\nvarP}[2]{\left \| #1 \right\|_{\mathrm{var}, #2}}
\newcommand{\abs}[1]{\left \vert #1 \right \vert }
\newcommand{\pente}{\|S_{\beta}'\|_{\infty}}
\newcommand{\Fs}{F_{\text{stat}}}
\newcommand{\triplebar}[1]{\vert\vert\vert #1 \vert\vert\vert}
\newcommand{\procsP}[2]{\mathcal{M}_1^+(C([t_0,#1], \R^{#2})}
\newcommand{\dvar}[2]{d_{\rm{var}}\left(#1,\; #2\right)}
\newcommand{\dt}[2]{d_t\left(#1,\; #2\right)}
\newcommand{\norm}[1]{\left \| #1 \right \|}
\newcommand{\Si}{S}
\newcommand{\T}{\mathcal{T}}
\newcommand{\IT}{\mathcal{IT}}
\newcommand{\tV}{\widetilde{V}}

\newcommand {\VV} {\mathbf{V}}
\newcommand {\G} {\mathcal{G}}
\newcommand{\R}{\mathbbm{R}}
\newcommand{\N}{\mathbbm{N}}
\newcommand{\Lop}{\mathcal{L}}
\newcommand{\Pro}{\mathbbm{P}}
\newcommand{\f}{\varphi}
\newcommand{\Diag}{{\mathbf{\rm Diag}}}
\newcommand{\mA}{\mathbf{A}}

\begin{document}

\title{
 A  view of Neural Networks as dynamical systems.
}

\author{
B. Cessac
\thanks{Laboratoire J. A. Dieudonn\'e,
U.M.R. C.N.R.S. N° 6621,
Universit\'e de Nice Sophia-Antipolis },
\thanks{INRIA, 2004 Route des Lucioles, 06902 Sophia-Antipolis, France.}
}

\date{\today}

\maketitle

\begin{abstract}
We present some recent investigations  resulting 
from the modelling of neural networks as \textit{dynamical systems},
and dealing with the following questions,
adressed in the context of specific \textit{models}.

\renewcommand {\theenumi}{(\roman{enumi})}
\ben
\item  Characterizing the collective dynamics;
\item  Statistical analysis of spikes trains;
\item  Interplay between dynamics and network structure;
\item  Effects of synaptic plasticity.
\een
\end{abstract}

\pagebreak

\tableofcontents
\pagebreak

The study of neural networks is certainly a prominent example of
interdisciplinary research field. From biologists, neurophysiologists, pharmacologists,
to mathematicians, theoretical physicists, including engineers, computer scientists, robot designers,
a lot of people with distinct  motivations and questions are interacting. 
With maybe a common ``Graal'': to understand one day how brain is working.
At the present stage, and though significant progresses  are made regularly,
this promised day is however still in a far future. But, beyond the comprehension of brain or even of simpler neural systems
in less evolved animals, there is also the desire to exhibit general mechanisms
or principles that could be applied to such artificial systems  as computers, robots,
or ``cyborgs'' (we think of the promising research field of brain-control of artificial
prostheses, see for example the web page http://www-sop.inria.fr/demar/index\_fr.shtml). Again, there are many way of tracking these principles or mechanisms.

One possible strategy is to propose mathematical models of neural activity, at different
space and time scales, depending on the type of phenomenon under consideration.
However, beyond the mere proposal of new models, which can rapidly results in a plethora,
there is also a need to understand some fundamental keys ruling the behaviour of neural networks,
and, from this, to  extract new ideas that can be tested in real experiments. Therefore,
there is a need to make a thorough analysis of these models. This can be done by numerical
investigations, with, very often, the need of inventing clever algorithms
to fight the hard problem of simulating, in a reasonable time, and with
a reasonable accuracy, the tremendous number of degree of freedom
and the even larger number of parameters that neural networks have.
A complementary issue relies in developing a mathematical analysis,  whenever possible.

In this spirit, we present in this paper some recent investigations from 
the authors and his collaborators, resulting
from the modelling of neural networks as \textit{dynamical systems}. We warn the reader that this paper does not intend to be exhaustive
and  we shall only briefly mention many works which certainly would have deserved a longer presentation
in a more extensive review: the works by Ermentrout and Kopell on phase response theory \cite{ermentrout-koppel:84}, 
van Vreeswijk, Sompolinsky and collaborators  
\cite{vanvresswijk-sompolinsky:96,vanvresswijk-sompolinsky:98,vanvresswijk-hansel:97,vanvreeswijk:04},
 \cite{litvak-sompolinsky-etal:03},
Brunel  \cite{brunel-sergi:98,brunel-hakim:99,fourcaud-brunel:02,fourcaud-trocme-hansel-etal:03,brunel-latham:03,renart-brunel-etal:04}, 
and many others on  neural activity, 
theory of synchronization and spike patterns by Seung \cite{xie-hahnloser-etal:02}, Bressloff
and Coombes \cite{bressloff-coombes:00,bressloff-coombes:00b,bressloff-coombes:03} 
Timme \cite{timme-wolf-etal:02,ashwin-timme:05,memmesheimer-timme:06,jahnke-memmesheimer-etal:08}, Jin 
\cite{gupta-jin-etal:03}, Diesmann \cite{diesmann-gewaltig-etal:99} are only a few examples
of these omissions.

Beyond the presentation of those results 
there is also the willing of raising interesting questions emerging
from this point of view.  
After a short presentation of neural networks, and how they can be
indeed modeled as dynamical systems (section \ref{neurodyn}), we list
$4$ of these questions, and address them in \textit{specific models}.

\bit

\item \textbf{Characterizing the collective dynamics of neural networks models}. When considering neural networks as dynamical systems, a first,
natural issue is to ask about the (generic) dynamics exhibited by the system
when control parameters  vary. This is discussed in section  \ref{genedyn}.

\item \textbf{Statistical analysis of spikes trains}. Neurons respond to excitations or stimuli by finite sequences of spikes (spike trains).
Characterizing spike trains statistics is a crucial issue in neuroscience. We approach this question
considering simple models. This is discussed in section  \ref{SpikesStat}.

\item \textbf{Interplay between dynamics and synaptic network structure}. Neural network are  highly dynamical
object and their behavior is the result of a complex interplay between the neurons dynamics and the
synaptic network structure. In this context, we discuss how the mere analysis of synaptic network structure
may not be sufficient to analyse such effects as the propagation of a signal inside the network.
We also present new tools based on linear response theory \cite{ruelle:99}, useful to analysing this interwoven evolution.
This is discussed in section  \ref{network}.

\item \textbf{Effects of synaptic plasticity}. Synapses evolve according to neurons activity.
Addressing the effect of synaptic plasticity in neural networks where dynamics is \textit{emerging} from collective effects and where spikes statistics are \textit{constrained} by this dynamics
seems  to be of central importance. We present recent results in this context. 
This is discussed in section  \ref{adapt}.
\eit

Obviously, the scope of this paper is not to address these questions in a general
context. Instead, we choose to present simple examples, that one may consider as rather ``academic'',
for which one can go relatively deep, with the idea that such investigations
may reveal useful, when transposed to ``realistic'' neural networks.

\su{Neural Networks as dynamical systems} \label{neurodyn}

\ssu{From biological neurons and synapses $\dots$} \label{biolneur}

\begin{figure}[htb]
 \centerline{\includegraphics[height=8cm,width=8cm,clip=false]{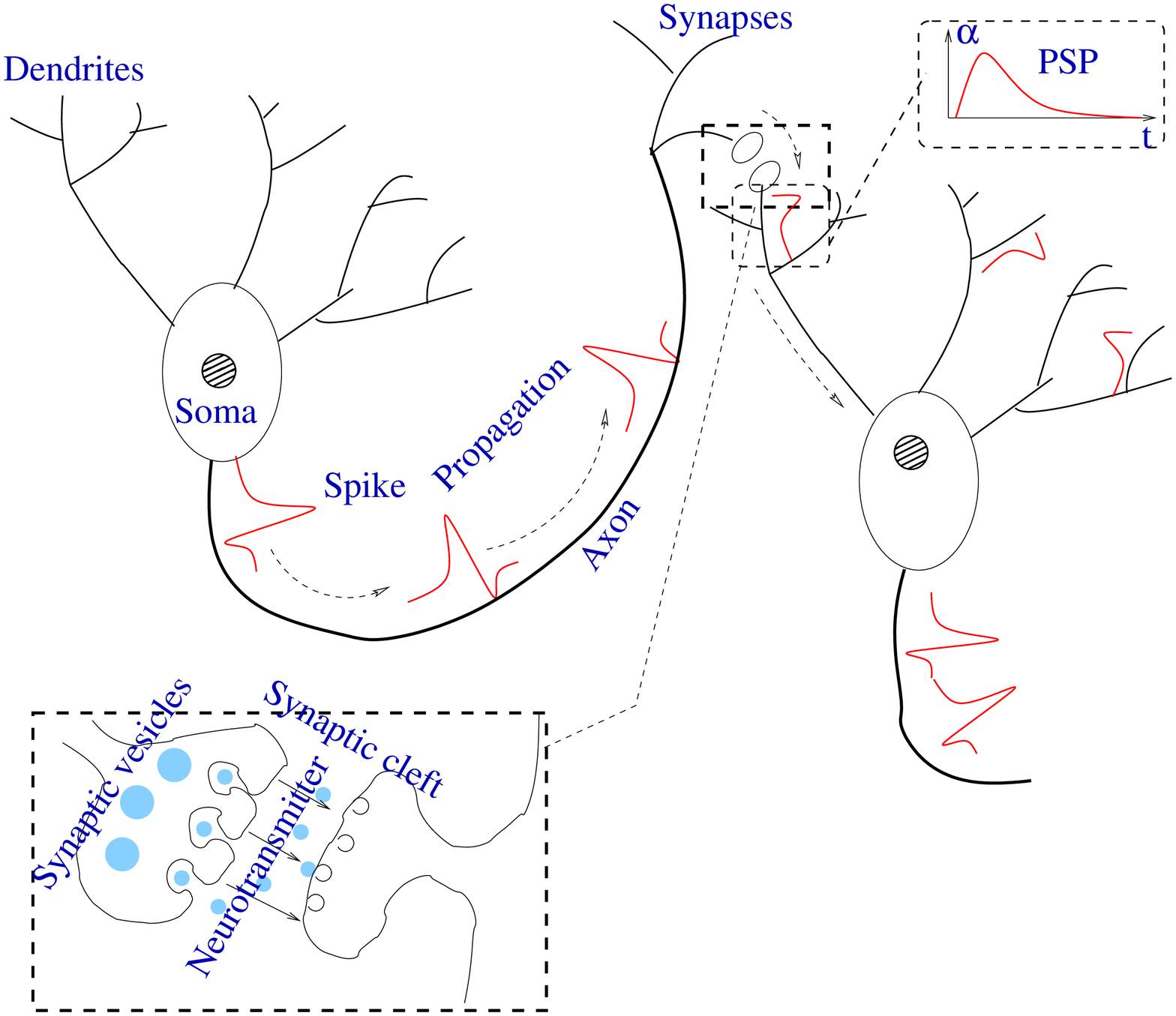}}

\caption{\footnotesize{Sketch of the neuron structure.\label{Fneur} }
}
\end{figure}

A neuron is an \textit{excitable cell}. Its activity is manifested by local variations (in space and time)  of its  \textit{membrane potential},
 called  ``action potentials'' or ``spikes''.
These variations are due to an exchange of ions species (basically $Na^+$,$K^+$,$Cl^-$) which move, through the membrane,
from the region of highest concentration (outside for $Na^+$, inside for $K^+$) to the region of lowest concentration. This 
motion does not occur spontaneously. It requires the opening/closing of specific \textit{gates} in specific ionic \textit{channels}.
The probability that a gate is open depends on the local membrane potential, whose variations can be elicited by local excitations,
induced by external currents, or coming from neighbours pieces of membrane (spike propagation). 
Neurons have a spatial structure,
depicted in fig \ref{Fneur}, and spikes propagates along this structure,  from  \textit{dendrites} to  \textit{soma},  and from soma to  \textit{synapses},
 along the  \textit{axon}.

The response of a given neuron to excitations has a wide variability. This variability is not manifested by the shape of the action potential,
which is relatively constant for a given neuron. Instead, it is revealed by the various sequences of spikes a neuron is able to emit.
Depending on the excitation, the response can be an isolated spike, a periodic spike train, a burst, etc... About twenty different spike trains
forms are classified in the literature \cite{izhikevich:04}.

Neurons are connected together. When a spike train is emitted from the soma toward the synapses, via the axon, it eventually reaches the \textit{synaptic
vesicles}. Here, a local variation of the membrane potential triggers the release of a \textit{neurotransmitter} into the synaptic cleft. This neurotransmitter
reaches by diffusion the post-synaptic receptors, located on the \textit{dendritic spines}, and generates a \textit{post-synaptic potential} (PSP). Contrarily to spikes,
PSP have an amplitude which depends on the amplitude of the excitation and on the \textit{synaptic efficacy}. Efficacy evolves according to various mechanisms
depending on the activity of pre- and post-synaptic neurons. Depending on the pre-synaptic neuron and the neurotransmitter used by this neuron, the PSP can be either positive or negative.
In the first case the pre-synaptic neuron and its synaptic connections are called \textit{excitatory}. 
Spikes coming from pre-synaptic neuron increase the membrane potential
of the post-synaptic neuron which is more keen on generating spike trains. Or PSP can be negative, corresponding to an \textit{inhibitory} pre-synaptic neuron.

Typically, a  neuron is connected to many pre-synaptic neurons and receives therefore many excitatory or inhibitory signals.
The cumulative effects of these signals eventually generate a response of this neuron's soma that propagates along the axon up to the synaptic tree,
then acting on other neurons, and so on.\\

From this short description, we can make the following summary.

\bit

\item Neurons are connected to each others in a synaptic network with \textit{causal (action/reaction)} interactions.

\item Signals exchanged by neurons are \textit{spike trains}. Spike trains coming from pre-synaptic neurons generate
a  spike train response of the post-synaptic neuron which propagates to other neurons.

\item Spike trains have a wide variability which generates an overwhelming repertoire of collective dynamical responses.

\eit

As an additional level of complexity the structure of the network constituted by
synaptic connections can also
have a wide range of forms\footnote{In mathematical models there is no a priori constraint, while
in the real world the network structure is constrained by genetics.}, with multiple layers, different species of neurons, etc.
Also, a very salient property is the capacity that synapses have to \textit{evolve and adapt}\footnote{Note
that not only synapses, but also neurons have the capacity of adaptation (intrinsic plasticity \cite{mahon-et-al:03}).
We shall not discuss this aspect in the present paper.}, according to \textit{plasticity} mechanisms.
\textit{Synaptic plasticity} occurs at many levels of organisation and time scales in the nervous system
\cite{bienenstock-etal:82}.
It is of course
involved in memory and learning mechanisms, but it also alters excitability of brain area
and regulates behavioural states (e.g. transition between sleep and wakeful activity).
On experimental grounds, synaptic changes can be induced by \textit{specific} simulations conditions
defined through  the firing frequency of pre- and post-synaptic
neurons \cite{bliss-gardner:73,dudek-bear:93}, the membrane potential of the post-synaptic
neuron \cite{artola-etal:90}, spike timing \cite{levy-stewart:83,markram-etal:97,bi-poo:01}
(see \cite{malenka-nicoll:99} for a review). Different mechanisms have been exhibited from the Hebbian's ones \cite{hebb:49} to Long Term
Potentiation (LTP) and Long Term Depression (LTD), and more recently to Spike Time Dependent
Plasticity (STDP) \cite{markram-etal:97,bi-poo:01} (see \cite{dayan-abbott:01,gerstner-kistler:02,cooper-etal:04}
for a review). 

\ssu{$\dots$ to models} \label{Models}

Regarding the overwhelming richness of behaviors  that neuronal networks are able to display, the theoretical (mathematical
or numerical) analysis of these systems is at a rather early stage. Nevertheless, some significant breakthrough have been made
within the last 50 years, as we shall see in a few examples. For this, a preliminary modeling/simplification strategy is necessary,
that we summarize as follows.

\sssu{Fix a model of neuron} \label{NeurModel}

This essentially means: fix an equation or a set of equations describing the evolution of neuron's membrane potential,
plus, possibly, additional variables (such as the probability of opening/closing ionic gates in Hodgkin-Huxley's
like models \cite{hodgkin-huxley:52}).
This choice can be guided by different and, often, mutually incompatible constraints.

\bit

\item Biological plausibility.

\item Mathematical tractability.

\item Numerical efficiency.
\eit

Regarding the first aspect one may also only focus on  a few biological features. Do we want a model that 
reproduces accurately
spike shape, or do we simply want to reproduce the variability in spike trains responses whereas spike shape is neglected
(e.g. represented by a ``Dirac'' peak) ? Do we want to focus on one specific characteristic of spike trains
(probability that a neuron fires, probability that two neurons fire within a certain time delay....) ? Clearly,
there is a large number of neuron models and, as usual, models depend on the questions that you ask.
Here are a few examples.

\paragraph{Hodgkin-Huxley model.} This model, dating back to 1952 \cite{hodgkin-huxley:52}, is still one of the best description of neuron
spike generation and propagation. Thus, it is very good from the point of view of biological plausibility.
Unfortunately, its mathematical analysis has not been completed yet and it is computational time consuming.
In this model, the dynamics of a piece of membrane  with capacity $C_m$ and potential $V$ is given by:
\bea 
C_m \frac{dV}{dt}&=& - g_{Na} m^3 h (V - E_{Na}) - g_K n^4(V - E_{K}) - g_L(V - E_L)+ I_{ext} \label{HHV}\\
\frac{1}{\gamma(T)}\frac{dn}{dt}&=&\alpha_n(V)(1-n) - \beta_n(V)n= \frac{n^\infty(V)-n}{\tau_n(V)}\label{HHn}\\
\frac{1}{\gamma(T)}\frac{dm}{dt}&=&\alpha_m(V)(1-m) - \beta_m(V)m= \frac{m^\infty(V)-m}{\tau_m(V)}\label{HHm}\\
\frac{1}{\gamma(T)}\frac{dh}{dt}&=&\alpha_h(V)(1-h) - \beta_h(V)h= \frac{h^\infty(V)-h}{\tau_h(V)}\label{HHh}
\eea
\nid where $m,h,n$ are additional variables, describing the ionic channels activity (see 
\cite{cronin:87,gerstner-kistler:02b,hille:01,keener-sneyd:98,koch:99b,nelson-rinzel:95}). $E_{Na},E_{K},E_L$ are respectively
 the Nernst potentials of $Na^+,K^+$ ions and additional ionic species (like $Cl^-$) grouped together
in a leak potential $E_L$. $g_Na$,$g_K$, $g_L$ are the corresponding conductances.
$\gamma(T)$ is a temperature dependent time scale (equal to $1$ at $6.3^o$C).
$\alpha,\beta$ are transitions rates in the masters equations (\ref{HHn},\ref{HHm},\ref{HHh})
used to model the transition open/close of ionic channels. 
Though the complete mathematical analysis of this model has not been performed yet, important results
can be found in \cite{cronin:87,keener-sneyd:98,guckenheimer-labouriau:93,guckenheimer-oliva:02}.

\paragraph{Fitzhugh-Nagumo model} One can reduce the Hodgkin-Huxley equations
in order to obtain an analytically tractable model.
In this spirit  Fitzhugh \cite{fitzhugh:66} and independently Nagumo, Arimoto \& Yoshizawa \cite{nagumo-etal:62},
 considered reductions of the
Hodgkin-Huxley model and introduced an analytically tractable
two variables model
$$
\baR{lll}
\epsilon \frac{dv}{dt}&=& f_\lambda(v,w), \label{SystExca}\\
\frac{dw}{dt}&=& g_\lambda(v,w),\label{SystExcb}
\eaR
$$
\nid where $\epsilon$ is a small parameter. 
The index $\lambda$ refers to the control parameters of the system. In the FitzHugh-Nagumo model 
$f_\lambda(v,w)=v-v^3-w+I$ is a cubic polynomial in $v$ and is linear in $w$,
while $g_\lambda(v,w)=(v-a-bw)$. The parameters $\lambda=(a,b,I)$ are deduced from 
the physiological characteristics of the neuron.

\paragraph{Integrate and Fire models} Here, one fixes a real number $\theta$, called the firing threshold of the neuron,
such that if $V_k(t) \geq \theta$ then neuron membrane potential is reset 
\textit{instantaneously} to some \textit{constant} reset value $\Vr$ and a spike is emitted 
toward post-synaptic neurons. Below the threshold, $V_k<\theta$, neuron $k$'s dynamics 
is driven by an equation of form: 

\beq \label{EqQCV}
C_k\frac{dV_k}{dt}+g_k V_k=i_k,
\eeq

\nid where $C_k$ is the membrane capacity of neuron $k$, $g_k$ its conductance
and $i_k$ a current, including various term, depending on modeling choices (external current,
ionic current, adaptation current). 

In its simplest form equation (\ref{EqQCV})
reads:

\beq\label{LIF}
\frac{dV_k}{dt}=-\frac{V_k}{\tau_k}+\frac{i_k}{C_k},
\eeq

\nid where $g_k$ is a constant, and $\tau_k=\frac{g_k}{C_k}$
is the characteristic time for membrane potential
decay, when no current is present.
This model has been introduced in \cite{lapicque:07}.
More generally, conductances and currents depend on  the
previous \textit{firing times} of the  pre-synaptic neurons \cite{rudolph-destexhe:06}
(see section \ref{GIF} for an example).

\paragraph{Discrete time models} In many papers, researchers use sooner or later numerical simulations
to guess or validate original results. Most often this corresponds to a time discretisation
with standard schemes like Euler, or Runge-Kutta. Even when seeking more elaborated schemes
such as event based integrations schemes \cite{brette-rudolph-etal:07,rochel-martinez:03}, which 
\textit{in principle} allows one to handle continuous time, there is in fact a minimal time
scale, due to numerical round-off error, below which the numerical scheme is not usable
anymore. On more fundamental grounds, in all models presented above including Hodgkin-Huxley,
there is a minimal time scale imposed by Physics.  
Thus,  although the mathematical definition of $\frac{d}{dt}$ assumes a limit $dt \to 0$,
there is a time scale below which the ordinary differential equations lose
their meaning. Actually, the mere notion of ``membrane potential'' already assumes
an average over microscopic time and space scales. Another reason
justifying time discretisation in models is the use of ``raster plot'' to characterize neurons activity.

\paragraph{Raster plots}

A raster plot is a graph where  the activity of a neuron is represented by a mere vertical bar
each ``time'' this neuron emits a spike.
When focusing on spiking neurons models, spikes are often characterized by their ``time'' of occurrence.
Except for IF models, where the notion of ``instantaneous'' firing and reset leads to nice pathologies\footnote{
Consider a loop with two neurons, one excitatory and the other inhibitory. Depending on the synaptic weights,
it is possible to have the following situation. The first neuron fires instantaneously, excites instantaneously the second one, which fires
instantaneously and inhibits instantaneously the first, which does not fire... This type of causal paradoxes, common in science-fiction
novels \cite{barjavel:44}, can also be found in IF models (eq. (\ref{LIF})) without refractory period and time delays.},
 a spike has
some duration and spike time has some uncertainty $\delta$. Therefore, the statement ``neuron $i$ fires
at time $t$'' must be understood as ``neuron  $i$ fires at time $t$ within a precision $\delta >0$''.
Moreover, a neuron cannot fire more than once
within a time period $r$ called ``refractory period''. Therefore, one can fix a positive
time scale $\delta>0$ which can be mathematically arbitrary small,
such that (i) a neuron can fire at most once between
$[t, t+\delta[$ (i.e. $\delta << r$, the refractory period);
(ii) $dt << \delta$, so that we can keep the continuous time evolution of membrane 
potentials, taking into account
time scales smaller than $\delta$, and 
 integrating membrane potential dynamics on the intervals $[t,t+\delta[$;
(iii) the spike time is known within a precision $\delta$
(see \cite{kirst-timme:09} for an interesting discussion on this approach).

 At this stage let us introduce a concept/notation used throughout this paper.
One can associate to each
neuron $k$ a variable $\omega_k(t)=1$ if neuron $k$ fires between $[t,t+\delta[$ and  $\omega_k(t)=0$
otherwise.  A ``spiking pattern'' is a vector  $\bom(t) \deq \left[\omega_k(t) \right]_{k=1}^N$
which tells us which neurons are firing at time $t$. A ``raster plot'' 
 is a sequence $\tom \deq \left\{\bom(t)\right\}_{t=0}^{+\infty}$,
of spiking patterns. 
We denote $\tot=\left\{\bom(s)\right\}_{s=0}^{t}$,  the raster plot  from time $0$  to time $t$.

\sssu{Fix a model of synapse}  \label{SynModel}
 
\paragraph{Voltage- and activity-based models} 
A single action potential from a pre-synaptic neuron  $j$ is seen as a
post-synaptic potential by a post-synaptic neuron  $i$ (see Fig.  \ref{Fneur}).  The  conductance time-course
after the arrival of a post-synaptic potential is typically given by a function $\alpha_{ij}(t-s)$
where $s$ is the time of the spike hitting the synapse and $t$ the
time after the spike. (We neglect here the delays due to the distance
travelled down the axon by the spikes). 
In  \textit{voltage-based models} one assumes that the post-synaptic potential
has the same shape no matter which pre-synaptic population caused it,
the sign and amplitude may vary though \cite{ermentrout:98}.  This leads to the relation:
$$\alpha_{ij}(t)=W_{ij} \alpha_i(t),$$
\nid where $\alpha_{i}$ represents the unweighted shape (called a $\alpha$-shape) of the post-synaptic potentials.
Known examples  of $\alpha$-shapes are $\alpha_i(t)=K_i e^{-t/{\tau_i}} H(t)$ 
or $\alpha_i(t)=K_ite^{-t/{\tau_i}} H(t)$ where $H$ is the Heaviside function.
More generally this is a polynomial in $t$ and this is  the Green function of a linear differential equation of order $k$:
\begin{equation}\label{eq:Green}
\sum_{l=0}^k a^{(l)}_i  \frac{d^l \alpha_i}{d t^l}(t)=\delta(t).
\end{equation}
$W_{ij}$ is the strength of the post-synaptic potentials elicited by neuron  $j$ on neuron $i$ (synaptic efficacy or
``synaptic weight'').

In \textit{activity-based models} the shape of a PSP depends only on the nature of the pre-synaptic cell, 
that is \cite{ermentrout:98}:
$$ \alpha_{ij}(t)=W_{ij} \alpha_j(t).$$

Assuming that the post-synaptic potentials sum linearly, the average membrane potential
of neuron\footnote{One should instead write neuron $i$'s \textit{soma}. In the sequel we shall consider
neurons as punctual, without spatial structure.} $i$ is:

\beq \label{Vsyn}
V_i(t)=\sum_{j,n} \alpha_{ij}(t-\tjn),
\eeq
where the sum is taken over the arrival times $\tjn \leq t$ of the spikes produced
by  the neurons $j$. 

\paragraph{Synaptic plasticity.}
 Most often, the mechanisms involved in synaptic plasticity have been revealed by  simulation  performed in isolated
neurons in \textit{in vitro} conditions. Extrapolating the action of these mechanisms to
in vivo neural networks requires both a bottom-up and top-down approach.
This issue is
tackled, on theoretical grounds, by inferring ``synaptic updates rules'' or ``learning rules'' from
biological observations \cite{vondermalsburg:73,bienenstock-etal:82,miller-etal:89}
 (see \cite{dayan-abbott:01,gerstner-kistler:02,cooper-etal:04}
for a review) and extrapolating, by theoretical or numerical investigations, what are the effects
of such synaptic rule on such neural network \textit{model}. This approach relies on the belief that
there are ``canonical neural models'' and ``canonical plasticity rules'' capturing the most essential
features of biology.
When considering synaptic adaptation, one proposes evolution rules for the $\alpha_{ij}$ profiles.
Most often, the mere evolution of the $W_{ij}$'s are considered. Here are a few typical examples.

\paragraph{Generic synaptic update} 
Synaptic plasticity corresponds to the 
evolution of  synaptic
efficacy  which evolve in time according to the spikes emitted 
by the pre- and post- synaptic neuron. 
In other words,
the variation of $W_{ij}$ at time $t$ is a function
of the spiking sequences of neurons $i$ and $j$ from time $t-T_s$ to
time $t$, where $T_s$ is time scale characterizing the width of the spike
trains influencing the synaptic change. In its more general form  synapse update reads:
\[
\delta Wij=g\left(\Wij(t),\ltil,\ltjn \right), \ t > T_s,
\]
where $\ltil$, ($\ltjn$) are the lists of spikes times emitted by the pre-synaptic neuron $i$,
(the post-synaptic neuron $j$), up to time $t$. Thus, synaptic adaptation results from an integration of spikes 
over the time scale $T_s$. 

With the concept of ``raster plot'' introduced at the end of section  \ref{SynModel}, we may also write:

\beq\label{DSyn}
\delta Wij=g\left(\Wij(t),\omeit,\omejt \right), \ t > T_s.
\eeq

Let us now give a few examples  of synaptic adaptation ``rules''.

\paragraph{Hebbian learning}\footnote{For further explanations of this terminology,
see section \ref{SHebb}.} In this case, synapses changes depend on the firing rate of neuron $i$,$j$.
 A typical example corresponds to 

\beq\label{HebbCorr}
g_{ij}(W_{ij},\left[\omega_i\right]_{t-T_s,t},\left[\omega_j\right]_{t-T_s,t})=\frac{1}{T_s}\sum_{s_1,s_2=t-T_s}^{t}(\omei(s_1)-r_i(s_1))(\omej(s_2)-r_j(s_2),
\eeq
 
\nid (correlation rule \cite{rao-sejnowski:01}) where $r_i(t)=\frac{1}{T_s}\sum_{s=t-T_s}^{t} \omega_i(s)$ is  the frequency
rate of neuron $i$ in the raster plot $\tom$, computed in the time window $[t-T_s,t]$. 

\paragraph{Spike-Time Dependent Plasticity} as derived from Bi and Poo \cite{bi-poo:01}
provides the average amount of synaptic variation given 
the delay between the pre- and post-synaptic spike.
Thus, ``classical'' STDP reads \cite{gerstner-kistler:02,izhikevich-desai:03}:
\beq\label{STDPFC}
g\left(W_{ij},\omeit,\omejt \right)=\sum_{s_1,s_2=t-T_s}^{t} f(s_1-s_2)\omega_i(s_1)\omega_j(s_2),
\eeq
\nid with:
\beq\label{fSTDP}
f(x)=
\left\{
\baR{lll}
A_- e^{\frac{x}{\tau_-}}, \ x <0;\\
A_+ e^{-\frac{x}{\tau_+}}, \ x >0;\\
0, \ x=0;
\eaR
\right.
\eeq 
\nid where $A_-<0$ and $A_+>0$. The shape of $f$  has been obtained from statistical extrapolations of experimental data.
 Hence STDP is based on a second order statistics (spikes correlations).
There is, in this case, an evident time scale $T_s=\max(\tau_-,\tau_+)$, beyond which
$f$ is essentially zero. 

Many other examples can be found in the literature \cite{izhikevich-desai:03} .

\sssu{Fix a synaptic graph structure}\label{Syngraph}

This point is closely related to the previous one. In particular, this structure can be fixed or evolve in time
(synaptic plasticity). In this last case, there is a complex interaction between neuron dynamics and synapses dynamics.
This structure can be guided from biological/anatomical data, or it can be random. In this last case, one is more interested in generic mathematical properties than by biological
considerations. The intermediate case can also be considered as well: deterministic synaptic architecture with random fluctuations
of the synaptic efficacy (see section \ref{meanfield} for an example). 

At this stage an interesting issue is : ``what is the effect of the synaptic graph structure on neurons dynamics ?''
This question is closely related to the actual research trend studying dynamical systems interacting on complex networks
where most studies have focused on the influence of a network structure 
on the global dynamics (for a review, see~\cite{boccaletti-etal:06}).
 In particular, much effort
has been devoted to the relationships between node synchronization and the classical statistical quantifiers of complex networks (degree 
distribution, average clustering index, mean shortest path, motifs, modularity...)
~\cite{grinstein-linsker:05,nishikawa-etal:03,lago-etal:00}. The core idea,  that the impact of network topology on global dynamics 
might be prominent, so that these structural statistics may be good indicators of global dynamics,  proved however incorrect
 and some of the related studies yielded contradictory results~\cite{nishikawa-etal:03,hong-etal:02}. Actually, synchronization 
properties cannot be systematically deduced from topology statistics but may be inferred from the spectrum of the
network~\cite{atay-biyikoglu-etal:06}. Moreover, most of these 
studies have considered diffusive coupling between the nodes~\cite{hasegawa:05}. In this case,
 the adjacency matrix has real non-negative eigenvalues, and global properties, such as stability of the 
synchronized states~\cite{barahona-pecora:02} can easily be inferred from its spectral properties.

 Unfortunately, this wisdom cannot be easily transposed to the field of neural networks where coupling between 
neurons (synaptic weights) in neural networks
 is not diffusive,  the corresponding matrix is not symmetric
 and may contain positive and negative elements.
More generally, as exemplified in sections \ref{network} and \ref{adapt}, neural networks constitute nice examples where 
the analysis of the synaptic graph  with tools coming from the field of ``complex networks''  provides poor information on dynamics. 
The main reason of this failure
is that the synaptic graph does not take into account nonlinear dynamics. In section \ref{network}
we introduce a different concept of network, based on linear response theory, which provides
much more information on the conjugated effects of topology and dynamics.

\sssu{Neural networks as dynamical systems}\label{neurasdyn}

To summarize, we shall adopt in this paper, the following point of view.
``A neural network is formally a graph where the nodes are the neurons and the edges the synapses,
each edge being weighted by the corresponding synaptic efficacy. 
Thus synapses constitute a signed and oriented graph.
Each node is characterized by an evolution equation
where the neuron state depends on its neighbours (pre-synaptic neurons).
Synaptic weights  can be fixed or evolve in time (synaptic plasticity)
according to the state/history of the two nodes it connects (pre- and post-synaptic neuron).''    

As indicated by the title of this paper we adopt here the point of view that neural networks
are dynamical systems and we analyse them in this spirit. This point of view is not necessarily
completely appropriate, but it nevertheless allows some significant insights in neuronal
dynamics. More precisely, we consider the following setting.

\paragraph{Canonical formulation of neurons dynamics} Each neuron $i$ is characterized by
its state, $X_i$, which  belongs to some compact set $\cI \in \bbbr^M$.
  $M$ is the number of variables characterizing
the state of one neuron (we assume that all neurons are described by the same number of variables).
A typical example is  $M=1$ and $X_i=V_i$ is the membrane potential
of neuron $i$ and $\cI=[\Vm,\VM]$. Other examples are provided by conductances based models of
 Hodgkin-Huxley type (\ref{HHV}) then
$X_i=(V_i,m_i,n_i,h_i)$ where $m_i,n_i$ are respectively the activation variable for Sodium
and Potassium channels and $h_i$ is the inactivation variable for the Sodium channel.

We consider the evolution of $N$ neurons,
given by a deterministic dynamical system
of type:

\beq\label{CDNN}
\frac{d\X}{dt}=\Fg(\X,t), \quad \mbox{continuous time},
\eeq

\nid or,

\beq\label{DNN}
\X(t+1)=\Fg\left[\X(t),t\right], \quad \mbox{discrete time}.
\eeq

The variable  $\X=\left\{X_i\right\}_{i=1}^N$  represents the dynamical state of a network
of $N$ neurons at time $t$. We use the notation $\V$ instead of $\X$ when neuron's state is only determined by membrane potential
whereas we use the general notation $\X$ when additional variables are involved.   

Typically $\X \in \cM=\cI^N$ where $\cM$ is the phase space of (\ref{DNN}), and  $\Fg(\cM) \subset \cM$.
 The map $\Fg: \cM \to \cM$ depends on a set of parameters $\bg \in \bbbr^P$.
The typical case considered here is $\bg = \left( \cW,\Ie \right)$ where  $\cW$ is the matrix of synaptic weights and $\Ie$ 
is some external current or stimulus.
 Thus $\bg$ is a point in a $P=N^2+N$ dimensional space of control parameters.

\paragraph{Correspondence between membrane potential trajectories and raster plots}

Typically, a neuron $i$ ``fires'' (emits a spike or
action potential), whenever its state $X_i$ belong to some connected
region $\cP_1$ of its phase space. Otherwise, it is quiescent ($X \in \cP_0=
\cI \setminus \cP_1$).
For $N$ identical neurons this leads to a ``natural partition''
$\cP$ of the product phase space $\cM$. Call $\Lambda=\left\{0,1\right\}^N$,
$\bom=\left[\omega_i\right]_{i=1}^N \in \Lambda$. Then, $\cP=\left\{\Po \right\}_{\omega \in \Lambda}$,
where $\Po = \cP_{\omega_1} \times \cP_{\omega_2} \times \dots \times \cP_{\omega_N}$.
Equivalently, if $\X \in \Po$, then all neurons such that $\omega_i=1$ are firing
while neurons such that $\omega_k=0$ are quiescent.

To each initial condition $\X \in \cM$
 we associate a ``raster plot'' $\tom=\left\{\bom(t)\right\}_{t=0}^{+\infty}$
such that $\X(t) \in \Pot, \forall t \geq 0$. We write $\X \rep \tom$.
Thus, $\tom$ is the sequence of spiking patterns
displayed by the neural network when prepared with the initial condition $\X$.
On the other way round, we say that an infinite sequence $\tom=\left\{\bom(t)\right\}_{t=0}^{+\infty}$
is \textit{an admissible raster plot} if there exists $\X \in \cM$ such that   
$\X \rep \tom$. We call $\Spg$ the set of admissible raster plots for the set
of parameters $\bg$.   The dynamics of $\X$  induces a dynamics on the set of raster plot 
given by the left shit $\sg$ such that $\sg \tom = \tom' \Leftrightarrow
\bom'(t)=\bom(t+1), \forall t \geq 0$.
Thus,  in some sense,
raster plots provide a code for the orbits of (\ref{DNN}). 
Note that the correspondence may not be one-to-one.  

\su{Collective dynamics} \label{genedyn}

When considering neural networks as dynamical systems, a first,
natural issue is to ask about the (generic) dynamics exhibited by the system
when control parameters (summarised by the symbol
 $\bg$ in the section \ref{neurasdyn}) vary.
However, at the present stage, this question is essentially unsolvable, taking into account
the very large number of degree of freedom and the even larger number of parameters.
Also, the mere notion of genericity has to be clarified. In dynamical systems theory
``generic'' has two  distinct meanings. Either one is seeking
properties holding in a residual\footnote{A set is residual if is the countable intersection
of open dense sets. In this context, ``generic'' means ``holding on a dense set of parameters''\label{fdefgen}.} 
set, in which case one deals with genericity in a topological sense.
Or one is interested in properties holding on a set of parameters having probability one, 
for a smooth and ``natural'' probability distribution defined on the space
of control parameters (e.g. Lebesgue or Gauss distribution). In this case, one speaks about
``probabilistic genericity''. These two notions of genericity usually do not coincide.
(An attempt to unifying  these two concepts has been proposed in \cite{sauer-etal:92}
under the name of ``prevalence''). 

Genericity results are relatively seldom in the field of neural networks, unless
considering some specific situations (e.g. weakly coupled  neural networks,
where some neurons of the uncoupled system, are close to the same codimension one bifurcation
point \cite{hoppenstaedt-izhikevich:97}). We present here two genericity results in this section,
and the related techniques.  For a wider review see \cite{samuelides-cessac:07,cessac-samuelides:07}.
See also \cite{soula-chow:07} for a new and recent approach.

\ssu{Mean-field methods.} \label{meanfield}

As a first example let us describe within details the so-called
dynamic mean-field theory. This method, well known in the field
of statistical physics and quantum field theory, is used in the field
of neural networks dynamics with the aim of modeling neural activity at scales integrating the effect
of thousands of neurons. This is of central importance for several reasons. First, most
imaging techniques are not able to measure individual neuron activity (``microscopic'' scale),
but are instead measuring mesoscopic effects
resulting from the activity of several hundreds to several hundreds of thousands of
neurons. Second, anatomical data recorded in the cortex reveal the existence
of structures, such as  cortical columns\footnote{Cortical columns are small cylinders, of diameter $\sim 0.1-1$ mm, that
cross transversely cortex layers. They  are involved in elementary 
sensory-motor functions such as  vision. They are composed
of several hundred to thousand neurons, belonging to a few different populations belonging to distinct cortex layers.
The electrical activity of cortical columns can be measured using different techniques. In Optical Imaging,
one uses Voltage-Sensitive Dyes (VSDs). The dye molecules act as molecular transducer that transform changes in membrane potential into optical signals
with a high temporal resolution, $< 1$ ms, and a high spatial resolution, $\sim 50$ $\mu m$. 
The measured optical signal is locally proportional to the membrane potential of all neuronal components 
and proportional to the excited membrane surface of all neuronal components \cite{grimbert:08}.
It is possible to propose phenomenological models characterising the mesoscopic electrical  activity
 of cortical columns. This is useful to  predict the behaviour of the local field potential generated by neurons activity 
and to compare this behaviour to measures.}, with a diameter of about
$50 \mu m$ to $1 mm$, containing of the order of one hundred to one  thousand neurons
belonging to a few different species. In this case,
information processing does not occur at the  scale of individual neurons
but rather corresponds to an  activity integrating the collective dynamics
of many interacting neurons and resulting in a mesoscopic signal. 

However, obtaining the equations of evolution of the effective mean-field
 from microscopic dynamics is far from being evident. In simple physical models this can
be achieved via the law of large numbers and the central limit theorem, provided
that time correlations decrease sufficiently fast. The idea of applying mean-field methods coming from
statistical physics to neural networks
dates back to Amari \cite{amari:72,amari-yoshida-etal:77}. 
Later on, Crisanti, Sompolinsky and coworkers \cite{sompolinsky-crisanti-etal:88} used a dynamic mean-field
approach to conjecture the existence of chaos in an homogeneous neural network
with random independent synaptic weights. This approach was
formerly developed by Sompolinsky and coworkers for spin-glasses
\cite{sompolinsky-zippelius:82,crisanti-sompolinsky:87,crisanti-sompolinsky:87b}.
Later on, the mean-field equations derived by Sompolinsky and Zippelius \cite{sompolinsky-zippelius:82} for spin-glasses were rigorously obtained by Ben Arous and Guionnet 
\cite{ben-arous-guionnet:95,ben-arous-guionnet:97,guionnet:97}. The application of their method to a discrete time version of the neural network considered in  \cite{sompolinsky-crisanti-etal:88}
and in \cite{molgedey-schuchardt-etal:92} was done by Moynot and Samuelides \cite{moynot-samuelides:02}. Alternative approaches have been used to get a mean-field description of a given neural network and to find its solutions. A static mean-field study of multi-population network activity was developed by Treves in \cite{treves:93}.  His analysis was completed in \cite{abbott-van-vreeswijk:93}, where the authors considered a unique population of nonlinear oscillators subject to a noisy input current. They proved, using a stationary Fokker-Planck formalism, the stability of an asynchronous state in the network. Later on, Gerstner in \cite{gerstner:95} built a new approach to characterize the mean-field dynamics for the Spike Response Model, via the introduction of suitable kernels propagating the collective activity of a neural population in time.
Brunel and Hakim considered a network composed of integrate-and-fire neurons connected with constant synaptic weights \cite{brunel-hakim:99}. In the case of sparse connectivity, stationarity, and considering a regime where individual neurons emit spikes at low rate, they were able to study analytically the dynamics of the network and to show that the network exhibits a sharp transition between a stationary regime and a regime of fast collective oscillations weakly synchronized. Their approach was based on a perturbative analysis of the Fokker-Planck equation. A similar formalism was used in \cite{mattia-del-giudice:02} which, when complemented with self-consistency equations, resulted in the dynamical description of the mean-field equations of the network, and was extended to 
a multi-population network.
Finally, Chizhov and Graham \cite{chizhov-graham:07} have recently proposed
a new method, based on a population density approach, allowing to characterize the mesoscopic
behaviour of neuron populations in conductance-based models.

The motivations of this section are twofold.
 On the one hand,  we present an example of  dynamic mean-field approach applied to plausible
models of mesoscopic neural structures in the brain  \cite{faugeras-touboul-etal:08}. Especially, we insist on
the rich phenomenology brought by this method.  On the other
hand we present some examples of bifurcations analysis of dynamical
mean-field equations and what this tells us about the generic dynamics of
the underlying neural network.

\sssu{Multi-populations dynamics}\label{multipop}

Brain structures such as cortical columns
 are made of several species of neurons (with different
physical and biological characteristics) linked together in a specific architecture \cite{thomson-lamy:07}.
We model this in the following way.
 We consider a network composed of $N$ neurons indexed by $i \in \{1,\,\ldots,\,N\}$ belonging to $P$ populations indexed by $a \in \{1,\,\ldots,\, P\}$.  Let $N_a$ be the number of neurons in population $a$. 
We have $N=\sum_{a=1}^P N_{a}$. In the following we are interested in the limit $N \to \infty$. We assume that the proportions of neurons in each population are non-trivial, i.e. :

\[\lim\limits_{N\to\infty} \frac{N_a}{N} = \rho_a \;\in \;]0,1] \quad; \forall a \in \{1,\,\ldots,\,P\}.\]
On the opposite, were $\rho_a$ to vanish, would the corresponding population  not affect the global behavior of the system and could it be neglected.
We introduce the function $p: \{1,\,\ldots,\,N\} \to \{1,\,\ldots,\, P\}$ such that $p(i)$ is the index of the population which the neuron $i$ belongs to.

\paragraph{Firing rates models.} In many examples the spiking activity is resumed by \textit{spike rates}.
Call $\nu_j(t)$ the spikes rate of neuron $j$ at time $t$ such that the number of spikes arriving
between $t$ and $t+dt$ is $\nu_j(t)dt$. Moreover,  the relation between the membrane potential of neuron $i$, $V_i$ and $\nu_i$
takes the form: 

\beq\label{eq:nuV}
\nu_i(t)=S_i(V_i(t)),
\eeq \cite{gerstner-kistler:02,dayan-abbott:01}
\nid , where $S_i$ is sigmoidal.
Therefore,
 we have, for voltage-based models,
\beq\label{eq:VBM}
V_i(t)=  
   \int_{-\infty}^t   \alpha_i(t-s) \left(\sum_j W_{ij}S_j(V_j(s))  +I_i(s)+B_i(s)\right)\,ds=\alpha_i * \left[\sum_j W_{ij}S_j(V_j)  +I_i+B_i\right](t),
\eeq
\nid where $*$ is the convolution product. Here  we have assumed that neuron $i$ receives also an external current  $I_i(t)$ and (white)
 noise $B_i(t)$.  For activity based models, defining the activity as:
\[
 A_j(t)=\int_{-\infty}^t \alpha_j(t-s) \nu_j(s)\,ds,
\]
one has
\beq\label{eq:ABM}
 A_i(t)= \alpha_i*S_i\left(\sum_j W_{ij} A_j + \alpha_i *I_i+\alpha_i *B_i\right). 
\eeq

\paragraph{Model dynamics} Applying the Green relation (\ref{eq:Green}) to the membrane potential of the voltage based model (\ref{eq:VBM})
one obtains:

$$\sum_{l=0}^k a^{(l)}_i  \frac{d^l V_i}{d t^l}(t)=\sum_{j=1}^N W_{ij} S_j(V_j(t))+ I_i(t)+B_i(t).$$

The first term of the l.h.s. is the contribution of the pre-synaptic neurons to the time variation of the membrane potential.
Under the assumption that the $\alpha$-shape, sigmoidal shape, external current and noise only depend only on the neuron's population we may write, for each neuron in the population
$a$:

\beq\label{eq:DVBM}
\sum_{l=0}^k a^{(l)}_a  \frac{d^l V_i}{d t^l}(t)=\sum_{b=1}^P\sum_{j=1}^{N_b} W_{ij} S_b(V_j(t))+ I_a(t)+B_a(t), \quad i \in a.
\eeq

In the case where $\alpha_a=e^{-\frac{t}{\tau_a}}H(t)$  (\ref{eq:DVBM})
becomes:

\beq\label{eq:SM}
\frac{d V_i}{dt}=-\frac{V_i}{\tau_a}+\sum_{b=1}^P\sum_{j=1}^{N_b} W_{ij} S_b(V_j(t))+ I_a(t)+B_a(t), \quad i \in a.
\eeq 

\nid called the ``simple model'' in the sequel.

\paragraph{Synaptic weights} When investigating the structure of mesoscopic neural assemblies such as cortical columns, experimentalists
are able to provide the average value of the synaptic efficacy from a neural population to another one \cite{thomson-lamy:07}. Obviously, these values are
 submitted
to some indeterminacy (error bars). We model this situation in the following way. 
Each synaptic weight $W_{ij}$ is modeled as a Gaussian random variable whose mean and variance
depend only on the population pair $a=p(i),b=p(j)$, and on the total number of neurons  $N_b$ of population $b$:
\[W_{ij} \sim \m N \Big(\frac{\bar{W_b}}{N_b}, \frac{\sigma^2_b} {N_b}\Big),\]
where $\m N(m,\sigma)$ denotes the Gaussian law with mean $m$ and variance $\sigma$.
We assume that the $W_{ij}$'s are uncorrelated.
 We use the convention $W_{ij}=0$ whenever there is no synaptic
connection from $j$ to $i$. 
 
\sssu{Mean-Field approach}

\paragraph{Local interaction field}
The collective behaviour of neurons in eq. (\ref{eq:DVBM}) is determined by the term:

$$\eta_i(t)=\sum_{b=1}^P\sum_{j=1}^{N_b} W_{ij} S_b(V_j(t)),$$

\nid called the ``local interaction field'' of neuron $i$. When the $W_{ij}$'s are fixed, its evolution depends
on the evolution of all neurons (i.e. the trajectory of the corresponding dynamical system).
 If the  trajectory is prescribed, and if the $W_{ij}$'s vary,
$\eta_i(t)$ becomes a random process whose law is constrained by the law of the $W_{ij}$'s.  Let us analyse
this within more details. We first make a qualitative description explaining the basic ideas without mathematical rigor.
Especially, we assume that there is a well defined ``thermodynamic limit'' ($N \to \infty$) for the quantities we consider.
Then we quote a rigorous result validating this qualitative description \cite{faugeras-touboul-etal:08}.
It uses large deviations techniques developed in \cite{ben-arous-guionnet:95,ben-arous-guionnet:97,guionnet:97,moynot-samuelides:02}
(see \cite{samuelides-cessac:07} for a review).

\paragraph{Non random synaptic weights} Assume that $\Jab=0$, namely we neglect the errors in the synaptic weights
determination. Then, we may write:

\beq\label{eq:etai}
\eta_i(t)=\sum_{b=1}^P \frac{\Wbab}{N_b}\sum_{j=1}^{N_b} S_b(V_j(t)).
\eeq

As $N_b \to \infty$,

\beq\label{philim}
\frac{1}{N_b}\sum_{j=1}^{N_b} S_b(V_j(t)) \to \phi_b(\V(t)),
\eeq
\nid assuming that the limit exists. The quantity $\phi_b(\V(t))$ is the average firing rate of population $b$ at time $t$. 
In this limit, eq.  (\ref{eq:DVBM})
becomes:

$$\sum_{l=0}^k a^{(l)}_a  \frac{d^l V_i}{d t^l}(t)=\sum_{b=1}^P \Wbab\phi_b(\V(t)) + I_a(t)+B_a(t), \quad i \in a.$$

In this equation the membrane potential evolution only depends on the neuron's $i$ population. Thus, 
setting $V_a(t) = \lim_{N_a \to \infty } \frac{1}{N_a} \sum_{i=1}^{N_a} V_i(t)$, we have:

\beq\label{FDMF}
\sum_{l=0}^k a^{(l)}_a  \frac{d^l V_a}{d t^l}(t)=\sum_{b=1}^P \Wbab\phi_b(\V(t)) + I_a(t)+B_a(t), \, a=1 \dots P,
\eeq
\nid called the ``first order mean-field'' equations in the sequel.

This equation resembles very much eq. (\ref{eq:DVBM}) if one makes the following reasoning. ``Since $\phi_b(\V(t))$ 
is the frequency rate of neurons in population $b$, averaged over this population, and since, for one neuron, the frequency
rate is $\nu_i(t)=S_i(V_i(t))$, let us write 

$$\phi_b(\V(t))=S_b(V_b(t)).$$

This leads to:

$$\sum_{l=0}^k a^{(l)}_a  \frac{d^l V_a}{d t^l}(t)=\sum_{b=1}^P \Wbab S_b(V_b(t)) + I_a(t)+B_a(t), \, a=1 \dots P,$$

\nid which has exactly the same form as eq. (\ref{eq:DVBM}) but at the level of a neurons population.
Equations of this type, called  ``naive  mean-field'' equations in the sequel, are therefore obtained
via a ``questionable'' assumption:
%
$$\frac{1}{N_b} \sum_{j=1}^{N_b} S_b(V_j(t))=S_b\left( \frac{1}{N_b} \sum_{j=1}^{N_b} V_j(t)\right).$$
%
There are many examples in physics where this assumption is wrong (such as spin-glasses). However, in the present context
where the $W_{ij}$'s are independent (and in particular non symmetric, contrarily to e.g. spin glasses
\cite{mezard:87}) it is  correct in some specific sense, as we develop.
Actually, naive mean-field equations are commonly used as phenomenological models in the neuroscience
literature. Here is an example.

\begin{figure}[htb]
\begin{center}
\includegraphics[height=4cm,width=6cm,clip=false]{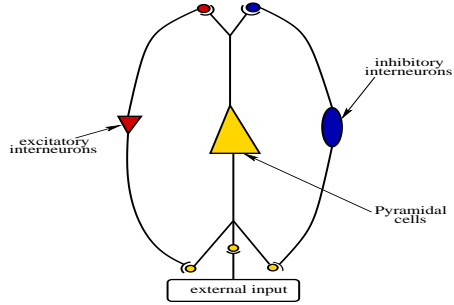}
\caption{\label{FJR} 
\footnotesize{Schematic representation of the neural
 populations and their interactions, as considered in Jansen-Rit's model \cite{jansen-rit:95}}.}
\end{center}
\end{figure}

\paragraph{The Jansen-Rit model cortical columns model} \cite{jansen-rit:95}
 features a population of pyramidal neurons that receives excitatory and inhibitory
feedback from local inter-neurons and an excitatory input from neighboring cortical units
and sub-cortical structures such as the thalamus (see Fig.  \ref{FJR}).
 The excitatory input is represented by an external stimulus with a deterministic part
$I(t)$, accounting for some specific
 activity in other cortical units, and a stochastic part $B(t)$ accounting for non specific background activity.

Denote by $\cP$ (resp $\cE$, $\cI$) 
the pyramidal (respectively excitatory, inhibitory) populations. 
Choose in population $\cP$ (respectively populations $\cE$, $\cI$) 
a particular pyramidal neuron (respectively excitatory, inhibitory inter-neuron) indexed by $i_{pyr}$ 
(respectively $i_{exc}$, $i_{inh}$). The equations of their activity variable read, in agreement with
(\ref{eq:ABM}):

\[
\begin{cases}
 A_{i_{pyr}} &= \displaystyle{\alpha_\cE \ast \Si(\sum_{j_{exc}} W_{ij} A_j+ 
\sum_{j_{inh}} W_{ij} A_j  +  \alpha_\cE \ast I(\cdot) + \alpha_\cE \ast B(\cdot) )} \\
 A_{i_{exc}} &= \displaystyle{\alpha_\cE \ast \Si(\sum_{j_{pyr}} W_{ij} A_j) }\\
 A_{i_{inh}} &= \displaystyle{\alpha_\cI \ast \Si(\sum_{j_{pyr}} W_{ij} A_j)} \\
\end{cases}
\]

 %
%
%
%
%
%
%
%
%

This is therefore an activity-based model. The transfer functions $\alpha_\cE$ and $\alpha_\cI$ correspond respectively
to excitatory and inhibitory post-synaptic potential (EPSP or IPSP). 
In the model introduced originally by Jansen and Rit,  the synaptic integration
is of first-order $\alpha(t)= K e^{-\frac{t}{\tau}}H(t)$, where
the coefficient $K,\tau$ are the same for the pyramidal and the excitatory population (denote them
by $K_\cE,\tau_\cE$), 
and different from the ones of the inhibitory population (denote them
by $K_\cI,\tau_\cI$).
The sigmoid functions are the same whatever the populations.
In Jansen-Rit's approach the connectivity weights 
are assumed to be constant, equal to their mean value. Their equations, based on a naive mean-field approach,  read therefore, with our notations \cite{jansen-rit:95,grimbert-faugeras:05}:

\begin{equation}\label{eq:JRM1}
 \begin{cases}
   \frac{dA_{\cP}}{dt}(t) &=  - \, \frac{A_{\cP}}{\tau_\cE} + K_\cE \, \Si(\Wb_{\cP\cE} \, A_{\cE}(t) 
+ \Wb_{\cP\cI} \, A_{\cI}(t)+  \alpha_\cE \ast I(\cdot)+ \alpha_\cE \ast B(\cdot)),\\
   \frac{dA_{\cE}}{dt}(t) & = - \, \frac{A_{\cE}}{\tau_\cE} + K_\cE \, \Si(\Wb_{\cE\cP} \, A_{\cP}(t)),\\
   \frac{dA_{\cI}}{dt}(t) & = - \, \frac{A_{\cI}}{\tau_\cI} + K_\cI \, \Si(\Wb_{\cI\cP} \, A_{\cP}(t)).
 \end{cases}
\end{equation}

A higher order model, where $\alpha(t)= K\frac{t}{\tau} e^{-\frac{t}{\tau}}H(t)$,
 was introduced  by van Rotterdam and colleagues \cite{rotterdam-lopes-da-silva-etal:82} to better account for the synaptic integration and to better
 reproduce the characteristics of real EPSP's and IPSP's. The bifurcation diagram of this version is quite richer than the Jansen-Rit one \cite{grimbert-faugeras:05}.  
These equations are currently used in the neuroscience community
either to provide activity models used for the analysis of signals obtained from imaging (MEG or Optical Imaging),
or to provide dynamical models of epilepsy \cite{cosandier-rimele-badier-etal:07}. 

\paragraph{Role of synaptic weights variability} 

Let us now consider the more general case where synaptic weights
have fluctuations about the mean value $\Wbab$. These variations dynamically differentiate
the neurons within a population and may induce dramatic collective effects, when
amplified by the nonlinear dynamics. Then,  the actual evolution of a population
can depart strongly from the first order mean-field approximation (not to speak of
the naive mean-field approach).

Consider the local interaction field (\ref{eq:etai}). Fix the trajectory of $\V=\left\{V_i \right\}_{i=1}^N$. Then,
the $W_{ij}$'s being Gaussian, $\eta_i(t)$ is (conditionally) Gaussian, with mean:

$$\Exp{\eta_i(t)|\V}=\sum_{b=1}^P \Wbab \frac{1}{N_b} \sum_{j=1}^{N_b} S_b(V_j(t)),$$

\nid where $\Exp{}$ is the expectation with respect to the $W_{ij}$'s distribution , and covariance:

$$Cov\left[\eta_i(t)\eta_j(s) | \V\right] =\delta_{ij} \sum_{b=1}^P \frac{\Jdab}{N_b}  \sum_{j=1}^{N_b} S_b(V_j(t)) S_b(V_j(s)),$$

\nid where we have used that the $W_{ij}$'s are independent so that $Cov(W_{ij},W_{i'j'})=\delta_{ij}\delta_{i'j'} \frac{\Jdab}{N_b}, \quad i \in a, j \in b$.
Thus, conditionally to $\V$, and still assuming that there is a well defined thermodynamic limit, $\eta_a$
 converges as $N \to \infty$ to a \textit{diagonal Gaussian process} $\eta_a$ whose law depends
only on the population, with mean:

\beq\label{eq:meanetacond}
\Exp{\eta_a(t)|\V}=\sum_{b=1}^P \Wbab \phi_b(\V(t)),
\eeq   

 \nid and covariance:

\beq\label{eq:sigetacond}
Cov\left[\eta_a(t)\eta_a(s) | \V\right] = \sum_{b=1}^P \Jdab \lim_{N_b \to \infty}\frac{1}{N_b}  \sum_{j=1}^{N_b} S_b(V_j(t)) S_b(V_j(s))
\eeq

Thus for a fixed trajectory, we find that the average value of $\eta_a$ obeys the same equation as in the first order mean-field approach,
but it has now fluctuations and correlations given by (\ref{eq:sigetacond}). \\

The main difficulty is obviously that the trajectory $\V$  is generated by dynamics including
the nonlinear and collective effects summarized in $\eta_a$. 
The following result can be shown \cite{faugeras-touboul-etal:08}.
As $N_a \to \infty$ the membrane potential of a neuron in population $V_a$ obeys the equation:
\beq\label{SDMF}
\sum_{l=0}^k a^{(l)}_a  \frac{d^l V_a}{d t^l}(t)=\sum_{b=1}^P U_{ab}(t) + I_a(t)+B_a(t)
\eeq
\nid where $U_{ab}$, called the ``mean-field interaction process'', is a \textit{Gaussian process}, (thus
entirely defined by its mean and covariance),  statistically independent of the external noise $B$
 and of the initial condition $\V(t_0)$, and defined by:
 \begin{equation}\label{eq:effectiveInteractionProcessParams}
  \begin{cases}
   \Exp{U_{ab}(t)} = \Wbab m_{b}(t) \;\; \text{where \;\;} m_{b}(t) 
\deq \mathbb{E}[S_{b}(V_{b}(t))];\\
   \Cov(U_{ab}(t), U_{ab}(s)) = \Jdab  \Deabv(t,s) \text{  where  }\\ 
   \qquad \qquad \Deabv(t,s) \deq \mathbb{E}\Big[S_{b}(V_b(t))S_b(V_b(s))\Big]; \\
   \Cov(U_{ab}(t), U_{cd}(s)) = 0 \text{ if } a\neq c \text{ or } b \neq d.
  \end{cases}
 \end{equation}

One obtains therefore a set of self-consistent equations giving the mean and covariance
of the mean-field interaction process $U_{ab}$. The interaction field of population $a$,
$\eta_a$, is given by $\eta_a=\sum_{b=1}^P U_{ab}$, so that $\eta_a$ is indeed a Gaussian process
with mean $\sum_{b=1}^P \Wbab m_b(t)$ in agreement with eq. (\ref{eq:meanetacond}), and
covariance $\sum_{b=1}^P \Jdab  \Deabv(t,s)$,  in agreement with eq. (\ref{eq:sigetacond}).
But there is a important distinction.
Eq. (\ref{SDMF}), (\ref{eq:effectiveInteractionProcessParams}) provide the law of $U_{ab}$ and $\eta_a$,
and provide a \textit{closed system of equations}
 ruling the dynamical evolution of $V_a$ averaged over the distribution of $W_{ij}$'s,
while in   equations (\ref{eq:meanetacond}),(\ref{eq:sigetacond}) we only got
the conditional law with respect to a fixed trajectory $\V$, henceforth leading
to an incomplete formulation of the problem, since, to close the equations, one needs
to know the probability distribution of the trajectories $\V$. This is
an important distinction explaining the difference of notation between
$\phi_b(\V(t))$  in eq. (\ref{eq:meanetacond}) and $m_b(t)$ in (\ref{eq:effectiveInteractionProcessParams}).

\paragraph{Example: the simple model} 
Since $U_{ab}$ is a Gaussian process it is straightforward to obtain an explicit
form for its mean and covariance as well as for the mean and covariance of $V_a$.
In the case of the simple model (eq. (\ref{eq:SM})) this leads to the following equation
for the evolution of the average value $\mu_a(t)$ of $V_a$:

\beq\label{MeanMFT}
 \frac{d\mu_a}{dt} =-\frac{\mu_a}{\tau_a}  
+ \sum_{\beta=1}^P \Wbab  \int_{-\infty}^{+\infty} S_b \left ( h\sqrt{v_b(t)} + \mu_b(t) \right) Dh
+ I_a(t),
\eeq
\nid with:
\beq\label{Dh}
Dh= \frac{e^{-\frac{h^2}{2}}}{\sqrt{2\pi}}dh,
\eeq
\nid where $v_a(t)$ is the variance of $V_a$ at time $t$.
 Let $C_{ab}(t,s)$ be the covariance
of $V_a(t),V_b(s)$. Then, $v_a(t)=C_{ab}(t,t)$.  $C_{ab}(t,s)$ is given by \cite{faugeras-touboul-etal:08}:

\bea \label{CMFT}
C_{ab}(t,s)=\delta_{ab}
e^{-(t+s)/\tau_a}\Big[v_a(0)+ \frac{\tau_a s_a^2}{2}\left(e^{\frac{2s}{\tau_a}}-1\right) + \sum_{b=1}^P \Jdab \int_{0}^t\int_{0}^s e^{(u+v)/\tau_a}\Delta_{b}(u,v)dudv\Big],
\eea
\nid where:
\bea\label{DeltaMFT}
\Delta_{b}(u,v)= 
\int_{\bbbr^2} S_{b}\left( x\frac{\sqrt{v_b(u)v_b(v) - C_{bb}(u,v)^2}}{\sqrt{v_b(v)}}
+y\frac{C_{bb}(u,v)}{\sqrt{v_b(v)}} +\mu_b (u)\right) 
S_b\left(y \sqrt{v_b(v)}+\mu_b (v)\right)\,Dx\,Dy,
\eea
\nid and where $s_a^2$ is the variance of a white noise $B_i(t)$ in (\ref{eq:VBM}) and where
$Dx,Dy$ are Gaussian integrands of type (\ref{Dh}).

These equations extend as well to more complex models, including the cortical
columns model of Jansen-Rit \cite{jansen-rit:95} and
 van Rotterdam and colleagues \cite{rotterdam-lopes-da-silva-etal:82} (see \cite{faugeras-touboul-etal:08}).
Therefore,  the introduction of fluctuations in the synaptic distribution change drastically
equations of evolution of  such neural masses models as Jansen-Rit (see \cite{faugeras-touboul-etal:08} for
further comments). 

\sssu{Bifurcations of mean-field equations: a simple but non trivial example}\label{bifmeanfield}

Let us investigate these equations within details. In the case where fluctuations 
are neglected ($\Jdab=0,\sa=0$),
equations (\ref{CMFT}),(\ref{DeltaMFT}) admit the trivial solution $C_{ab}(t,s)=0,v_b(t)=0$
and equation (\ref{MeanMFT}) reduces to the equation obtained by the naive mean-field approach. Incidentally,
this validates the naive mean-field approach in this context. However, as soon as $\Jdab >0$ dynamics
become highly non trivial since the mean-field evolution (\ref{MeanMFT}) depends on its fluctuations
via the variance $v_b(t)$. This variance is in turn given by a complex equation requiring an integration on the whole past.
Actually, unless  one assumes the stationarity of the process, 
this equation cannot be written as an ordinary differential equation and the
evolution is \textit{non-Markovian}. This result, well known in the field of spin-glasses  \cite{benarous-guionnet:95},
has only been revealed recently in the field of neural masses models \cite{faugeras-touboul-etal:08},
though mean-field approaches were formerly used \cite{sompolinsky-crisanti-etal:88,cessac-doyon-etal:94,cessac:95}. 
In these last papers, the role of mean-field fluctuations was clearly revealed and its influence on dynamics
emphasized. In particular, chaotic dynamics have been exhibited, while the mean value $\mu_a(t)$ has
a very regular and non chaotic behaviour (for example, it can be constant). 

\paragraph{The model} As an example, let us consider the following model, corresponding to a time discretisation of
(\ref{eq:SM}) with $dt=\tau$ and  only one population. Thus, synaptic weights are Gaussian with mean $\frac{\Wb}{N}$ and
variance $\frac{\sigma^2}{N}$. Dynamics is given by:

\begin{equation} \label{SDNN}
V_i(t+1) =  \sum_{j=1}^N W_{ij} S(V_j(t)) + I_i, \qquad i=1..N,
\end{equation}
\nid where $S$ is a sigmoidal function such as 
$S(x)= \tanh( gx)$ or $S(x)= \frac{1 + \tanh(gx)}{2}$. 
 The  parameter $g$ controls
the non-linearity of  $S$.  
There is a time-constant current $\I$ whose components $I_i$ are random
variables with mean $\bar{I}$ and variance $\sigma^2_I$. 

\paragraph{The mean-field equations.} They write \cite{cessac-doyon-etal:94,cessac:95}:
\beq
\mu(t+1)=\Wb \int_\bbbr S(h\sqrt{v(t)}+\mu(t))Dh + \bar{I}, \label{SKd1}
\eeq
\beq
v(t+1)=\sigma^2\int_\bbbr S^2(h\sqrt{v(t)}+\mu(t))Dh +\sigma^2_I, \label{SKd2}
\eeq
\beq
C(t+1,t'+1)= 
\sigma^2 \int_{\bbbr^2}
 S\left(
\frac{\sqrt{v(t)v(t')-C^2(t,t')}}{\sqrt{v(t')}}h+
\frac{C(t,t')}{\sqrt{v(t')}}h'+ \mu(t)\right)
S\left(h'\sqrt{v(t')}+\mu(t') \right)
DhDh' +  \sigma^2_I,\label{SKd3}
\eeq
\nid where we have made $v(t)$ explicit, though it can be obtained from (\ref{SKd3}).

Let us comment these equations. First, they contain statistical parameters determining
the  probability distribution of synaptic weights and currents, $\bar{W},\sigma,\bar{I},
\sigma_I$.
They also contain an hidden
parameter,  $g$ determining the gain of the sigmoid, which is the same for all neurons.
As we saw, deriving mean-field equations corresponds to substituting the analysis of 
the dynamical system (\ref{DNN}), with a huge number of random microscopic parameters, 
by an ``averaged'' dynamical system depending on
these few  deterministic macroscopic parameters. In this spirit, we expect
these equations to give indications about 
 the generic behavior (in a probabilistic sense)
 when the synaptic weights and couplings are drawn according these values of macroscopic parameters,
and when the number of neurons is large.

The variables $m,C$ essentially play the role of order parameters in statistical physics. 
 They
characterize the emergent behavior
of a system  with a large  number of degree of freedom and they exhibit drastic changes
corresponding, in statistical physics, to phase transitions, and in our context
to a macroscopic bifurcations.

\begin{figure}[htb]
\begin{center}
\includegraphics[height=8cm,width=10cm,clip=false]{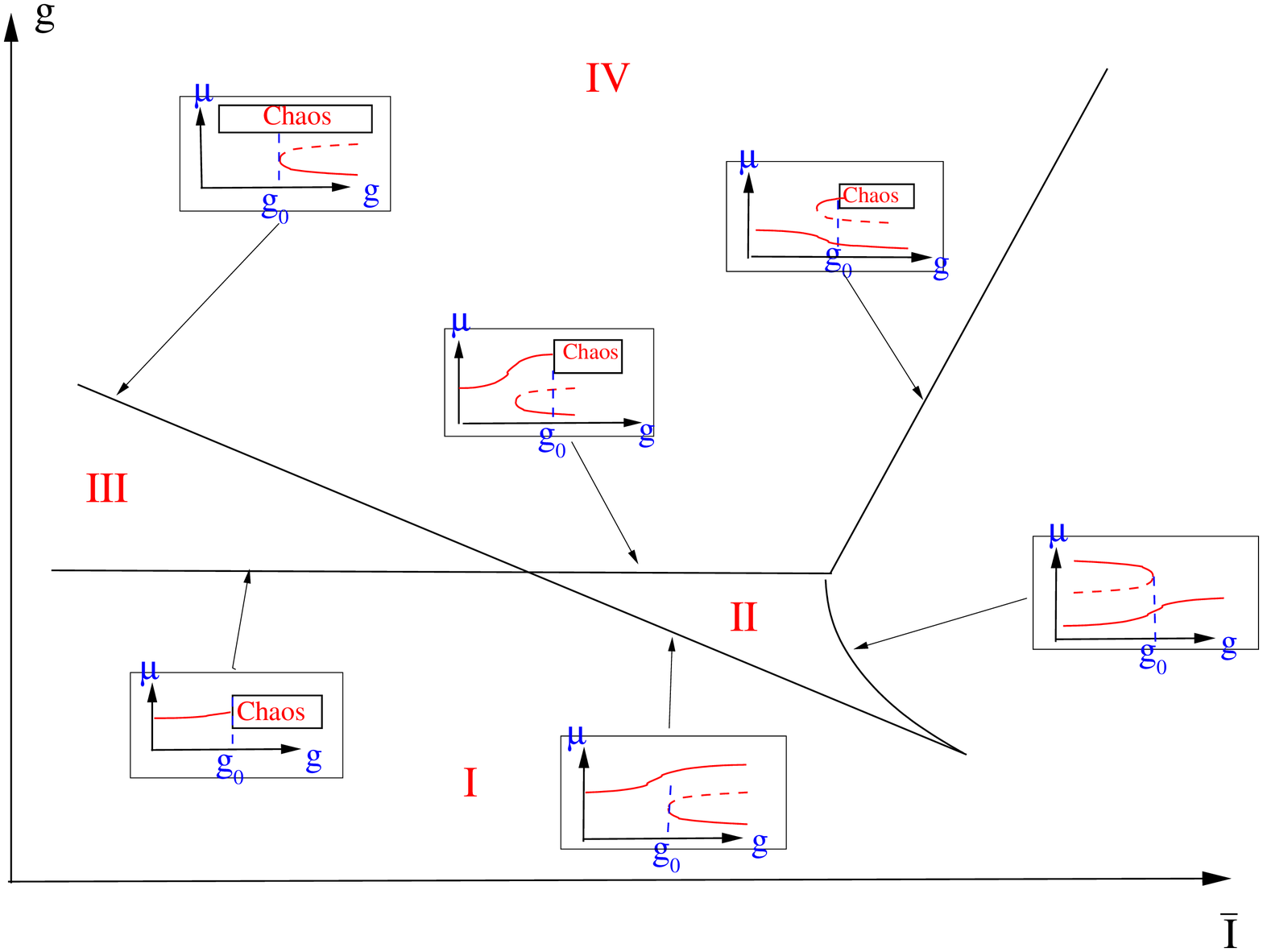}
\caption{\label{Fbif} 
\footnotesize{Schematic bifurcation map of model (\ref{SDNN}) in the plane $\bar{I},g$
(drawn from \cite{cessac-doyon-etal:94}). In the insets are represented the type of bifurcation
occuring, for the mean-field equations, when crossing the line indicated by an arrow while increasing $g$ and keeping
$\bar{I}$ constant. The bifurcation occurs for a value denoted $g_0$ in the inset.
 In these insets, the coordinates are $(g,\mu)$ where $\mu$ is given by eq. (\ref{SK1}).
The transition from a stable fixed point to chaos corresponds, for finite size systems
to a Ruelle-Takens transition \cite{ruelle-takens:71}, while in infinite dimension ($N \to \infty$)
this is a sharp transition (infinitely many eigenvalues of the Jacobian matrix at the fixed point
crossing simultaneously the frontier of stability). Region I corresponds to a regime with one stable fixed point;
region II to the coexistence of two stable fixed points; region III to a chaotic attractor and
region IV to the coexistence of a stable fixed point and of a chaotic attractor.}}
\end{center}
\end{figure}

\paragraph{Bifurcations in mean-field equations.} Having these equations in hand, the idea is now to study the reduced dynamical system  (\ref{SKd1}),(\ref{SKd2}),(\ref{SKd3})
and to infer information about the typical dynamics of (\ref{SDNN}).
In the present example there exists a  stationary regime of (\ref{SKd1}),(\ref{SKd2}),(\ref{SKd3}) and  the stationary equations are given by:
\beq\label{SK1}
\mu=\Wb \int_\bbbr S(h\sqrt{v}+\mu)Dh+\bar{I},\\
\eeq
\beq\label{SK2}
v=\sigma^2 \int_\bbbr 
S^2(h\sqrt{v}+\mu)Dh+\sigma^2_I,
\eeq
\beq\label{SK3}
C(t-t')=\sigma^2\int_{\bbbr^2}
S\left(
\frac{\sqrt{v^2-C^2(t-t')}}{\sqrt{v}}h+
\frac{C(t-t')}{\sqrt{v}}h'+ \mu\right)
S\left(h'\sqrt{v}+\mu \right)
DhDh'+\sigma^2_I.
\eeq
These equations give important information about the statistical
behavior of the model (\ref{SDNN}) with an increasing accuracy
when the size increases. For example saddle-node bifurcations
can be exhibited giving rise to bi-stability 
(see fig. \ref{Fbif} and \cite{cessac-doyon-etal:94} for more details). 
But, the most salient feature, as revealed by
  a detailed analysis of the complete set of equations (\ref{SK1}), (\ref{SK2}),(\ref{SK3}),
 and especially of the equation for the time covariance (\ref{SK3})), is the existence of a 
\textit{chaotic
regime}, occurring for a sufficiently large non-linearity $g$. This chaotic region
is delimited, in the space of  parameters  $\left(g,\Wb,\sigma,\bar{I},\sigma_I\right)$, 
by a manifold whose equation is:

\beq \label{AT}
\sigma^2
\int_\bbbr S'^2(h\sqrt{v}
+\mu)Dh=1.
\eeq
Note that, in the case where $S(x)=\tanh(gx)$ this equation gives precisely the so-called De Almeida Thouless line 
\cite{dealmeida-thouless:78}, delimiting, in the Sherrington-Kirckpatrick model of spin-glasses 
\cite{sherrington-kirkpatrick:78}, a frontier in the plane temperature-local external field, below
which dynamics becomes highly non trivial. Here the parameter corresponding to
the inverse temperature is $g$, while the external local field corresponds to $\I$. The ``low temperature regime''
of the SK model corresponds therefore to the chaotic regime of (\ref{SDNN}).  This analogy
is further discussed in \cite{cessac:94,cessac:95}.

 %
%
%
%
%
%
%
%
%

\paragraph{Interpretation.}
It can be shown that the crossing of this manifold corresponds, in the infinite system, to a sharp transition from fixed point to
infinite dimensional chaos \cite{cessac:94,cessac:95}. Considering the finite size system, one can show that
(\ref{SDNN}) exhibits generically  a Ruelle Takens \cite{ruelle-takens:71} transition to chaos as $g$ increases.
 As $N$ increases the transition to chaos occurs on a $g$ range becoming
more and more narrow, giving this sharp transition in the thermodynamic limit.
This  is related to the fact that the eigenvalues of the Jacobian
matrix accumulate on the stability circle as $N \to \infty$ \cite{girko:84} (see \cite{cessac:95,cessac-samuelides:07} for details).

The interesting remark is that, considering only the naive mean-field equation (equation
for the mean $\mu(t)$ with variance $v(t)=0$), one can easily exhibit examples (e.g. $S(x)=\tanh(gx)$ with
no current) where $\mu(t)$ is a constant ($0$), while fluctuations are chaotic. This clearly shows
the limits of the naive mean-field approach and the interest of analysing the role of fluctuations,
not only in simple models such as (\ref{SDNN}) but also for more realistic models
with several populations, like Jansen-Rit's (\ref{eq:JRM1}). Field fluctuations
could reveal effects that do not appear in the naive mean-field approach and that could be measured
in experiments. This question is under investigations (see the web page http://www-sop.inria.fr/odyssee/contracts/MACACC/macacc.html for more details).

\ssu{Dynamics of conductance based Integrate and Fire Models} \label{GIF}

Let us now investigate a second type of collective dynamics, in the
context of Integrate and Fire models introduced in section \ref{NeurModel}.
These models have known a great success due to their (apparent) conceptual simplicity and  analytical 
tractability 
\cite{mirollo-strogatz:90,ernst-pawelzik-etal:95,senn-urbanczik:01,timme-wolf-etal:02,memmesheimer-timme:06,gong-vanleuwen:07,jahnke-memmesheimer-etal:08} that can be used to explore some general principles of neurodynamics 
and coding. Surprisingly, the analysis of only one IF neuron submitted to a periodic current
reveals already an astonishing complexity and the mathematical analysis requires elaborated methods
from dynamical systems theory \cite{keener-hoppensteadt-etal:81,coombes:99,coombes-bressloff:99}. 
In the same way, the computation of the spike train probability
distribution resulting from the action of a Brownian noise on an IF neuron is not a completely straightforward exercise
\cite{knight:72,gerstner-kistler:02,brunel-sergi:98,brunel-latham:03,touboul-faugeras:07} and may require rather elaborated mathematics.
 At the level of networks the situation is even more complex, and the techniques
used for the analysis of a single neuron are not easily extensible
 to the network case. For example, Bressloff and Coombes \cite{bressloff-coombes:00} have 
extended the analysis in \cite{keener-hoppensteadt-etal:81,coombes:99,coombes-bressloff:99} 
to the dynamics of strongly coupled spiking neurons,
but restricted to networks with specific architectures and under restrictive assumptions on the firing times. Chow and Kopell
 \cite{chow-kopell:00}
studied IF neurons coupled with gap junctions but the analysis for large networks assumes constant synaptic weights.
  Brunel and Hakim \cite{brunel-hakim:99} extended the Fokker-Planck analysis combined to a mean-field
approach to the case of a network with inhibitory synaptic couplings but 
under the assumptions that all synaptic weights are equal.
However, synaptic weights variability plays a crucial role in the dynamics, as we saw in the previous section
(see also \cite{vanvresswijk-hansel:97,vanvresswijk-sompolinsky:98,vanvreeswijk:04}). 
Note that the rigorous derivation of the mean-field equations,
that requires large-deviations techniques \cite{benarous-guionnet:95},
 has not been yet done for the case of IF networks with continuous time dynamics
(for the discrete time case see \cite{soula-beslon-etal:06,samuelides-cessac:07}).   

In this section, we present a rigorous result characterising the generic dynamics of 
a Generalised Integrate and Fire model, where time has been discretized according to
the discussion of paragraph ``raster plots'' in section \ref{NeurModel}. We then
give an example where we consider  random synaptic weights. 

\sssu{Model} As we saw, the occurrence of a post-synaptic potential on synapse $j$, at time $\tjn$,
results in a change of membrane potential (eq. (\ref{Vsyn})). In conductance based models \cite{rudolph-destexhe:06} 
this change
is incorporated in the adaptation of conductances. The  evolution
of $V_k$, the membrane potential of neuron $k$, reads, setting the membrane capacity $C_k=1$
for simplicity:

\beq\label{yvettenet}
\frac{dV_k}{dt}=-\frac{1}{\tau_L} \, (V_k -E_L)-i_k^{(syn)}(V_k,t,\tot)+i_k^{(ext)}(t),
\eeq

\nid where $\tot$ is the raster plot up to time $t$. Recall that knowing $\tot$
is equivalent to knowing the list $\ltjn$ of firing times of all neurons up to time
$t$. The first term in the r.h.s. is a leak term, $i_k^{(ext)}(t)$ is an external current,
while:

$$i^{(syn)}_k(V_k,t,\tot)=(V_k-E^+)\, \sum_{j \in \cE} g_{kj}(t,\tot) + (V_k-E^-)\, \sum_{j \in \cI} g_{kj}(t,\tot),$$

\nid where $E^\pm$ are reversal potential (typically $E^+ \simeq 0 mV$ and $E^- \simeq -75 mV$).
As in the previous section, $\cE$ and $\cI$ refers respectively to excitatory and inhibitory neurons, and
the $+$ ($-$) sign is relative to excitatory (inhibitory) synapses. 
Note that conductances are always positive thus
the sign of the post-synaptic potential is determined by the reversal potentials $E^\pm$.
At rest ($V_k \sim -70mV$) the $+$ term leads to a positive PSP while $-$ leads to a negative
$PSP$.

Conductances depend on  past spikes via the relation:

$$g_{kj}(t,\tot)=  G_{kj} \sum_{n=1}^{M_j(t,\V)} \alpha_j(t-\tjn).$$ 

\nid In this equation, $M_j(t,\V)=\sum_{s=0}^t \omega_j(s)$ is the number
of times neuron $j$ has fired at time $t$.  $G_{kj}$ is a positive constant proportional to
 the synaptic efficacy
$$\left\{
\baR{ccc}
 W_{kj}=E^+G_{kj} \quad &\mbox{if}&  \quad j \in \cE, \\
 W_{kj}=E^-G_{kj} \quad &\mbox{if}&  \quad j \in \cI. 
\eaR
\right. 
$$
Recall that we use the convention $W_{kj}=0$
if there is no synapse from $j$ to $k$

Then, we may write (\ref{yvettenet}) in the form :

$$\frac{dV_k}{dt}+g_k V_k=i_k,$$

\nid (eq. (\ref{EqQCV}) introduced in section \ref{NeurModel}) with:

$$g_k(t,\tot)=\frac{1}{\tau_L}+\sum_{j=1}^N g_{kj}(t,\tot) ,$$

\nid and:

$$i_k(t,\tot)=\frac{E_L}{\tau_L} 
+  E^{+} \, \sum_{j \in \cE} g_{kj}(t,\tot)
+  E^{-} \, \sum_{j \in \cI} g_{kj}(t,\tot)
+i^{(ext)}_k(t).
$$

This equation characterises the membrane potential evolution below the threshold $\theta$.
Recall that, in Integrate and Fire models, if $V_k(t) \geq \theta$ then neuron membrane potential is reset 
\textit{instantaneously} to some \textit{constant} reset value $\Vr$ and a spike is emitted 
toward post-synaptic neurons.

\sssu{Time discretisation}
Using a time discretisation with a time step $\delta=1$,
 with the hypothesis discussed in section  \ref{NeurModel} leads
to the following  discrete-time model \cite{cessac-vieville:08}:

\beq \label{DgIF}
V_k(t+1)= \gamma_k(t,\tot)\left[1-Z(V_k(t))\right]V_k(t)+\Jkto,
\eeq

\nid where:

\beq \label{gamma}
\gamma_k(t,\tot) \deq e^{-\int_{t}^{t+1} \, g_k(s,\tot) \, ds} < 1,
\eeq

\nid is the integrated conductance over the time interval $[t,t+1[$,

$$\Jkto=\int_{t}^{t+1} i_k(s,\tot) \, \nu_k(s,t+1,\tot) \, ds,$$

\nid is the corresponding integrated current with:

$$\nu_k(s,t+1,\tot)=e^{-\int_{s}^{t+1} \, g_k(s',\tot) \, ds'},$$

\nid and where $Z$ is defined by :

\beq\label{Z}
Z(x)=\chi\left[x \geq \theta \right],
\eeq

\nid where $\chi$ is the indicator function that  will later on allows us to include the firing condition in the evolution equation
of the membrane potential (see (\ref{DNN})).

\sssu{Generic dynamics} \label{genedynIF}

It can be shown that this systems has the following properties.

\paragraph{Singularity set. }The  dynamics (\ref{DNN}) (and the
dynamics of continuous time IF models as well)
 is not smooth, but has singularities, due to the sharp
threshold definition in neurons firing.
The singularity set is:
$$\cS=\left\{\V \in \cM | \exists i=1 \dots N, \mbox{\ such \ that} \
V_i=\theta \right\}.$$
This is the set of membrane potential vectors such that at least
one of the neurons has a membrane potential exactly equal to the threshold
\footnote{A sufficient condition for a neuron $i$ to fire at time $t$
is $V_i(t)=\theta$ hence $\V(t) \in \cS$. But this is not a necessary
condition. Indeed,
there may exist discontinuous jumps in the dynamics, even if time is continuous,
 either due to noise, or  $\alpha$ profiles with jumps (e.g. $\alpha(t) = Ke^{-\frac{t}{\tau}}, \ t \geq 0$).
Thus neuron $i$ can fire with $V_i(t)>\theta$ and $\V(t) \notin \cS$.
 In the present case, this situation arises because time is discrete and
 one can have $V(t-\delta) < \theta$
and $V(t) >\theta$. 
This holds as well even if one uses  numerical
schemes using interpolations to locate more precisely the spike time \cite{hansel-etal:98}.
}.
This set has a simple structure:
it is a finite union of $N-1$ dimensional hyperplanes. 
Although $\cS$ is a ``small'' set
both from the topological (non residual set)
 and probabilistic  (zero Lebesgue measure) point of view,
it has an important effect on the dynamics.

\paragraph{Local contraction.} The other important aspect is that the dynamics is locally \textit{contracting},
because $\gamma_k(t,\tot)<1$ (see eq. (\ref{gamma})). 
This has the following  consequence. Let us consider the trajectory of a point $\V \in \cM$ and perturbations
with an amplitude $< \epsilon$ about $\V$ (this can be some fluctuation in the current,
or some additional noise, but it can also be some error due to a numerical implementation).
Equivalently, consider the 
evolution of the $\epsilon$-ball $\Bev$.
If $\Bev \cap \cS = \emptyset$ 
then the image of $\Bev$ is a ball with a smaller
diameter. This means, that, under the condition $\Bev \cap \cS = \emptyset$, a perturbation
is \textit{damped}. Now, if the images of the ball under the dynamics never intersect
$\cS$,  any $\epsilon$-perturbation around $\V$ is exponentially damped
and the perturbed trajectories about $\V$ become asymptotically indistinguishable
from the trajectory of $\V$. This means that, if the membrane potential of neurons do
not approach the threshold within a distance smaller\footnote{Since  time is discrete
a neuron can fire and nevertheless satisfy this condition.} than $\epsilon$ then perturbations
of size smaller than $\epsilon$ are damped.
 Actually, there is a more dramatic effect. If all neurons have fired after a finite time $t$ then
all perturbed trajectories collapse onto the trajectory of $\V$ after $t+1$ iterations. This loss
of initial condition in a finite time is typical for IF models and is due to the reset of the membrane
potential to a fixed value. For a discussion on IF model dynamics when this condition is relaxed
see \cite{kirst-geisel-etal:09}. See also \cite{gong-vanleuwen:07,jahnke-memmesheimer-etal:08}.

\paragraph{Initial conditions sensitivity.} On the opposite, assume that there is a time, $t_0$, such that
the image of the ball $\Bev$ intersects  $\cS$.
By definition, this means that there exists a 
subset of neurons $\left\{i_1, \dots, i_k\right\}$ and
  $\V'  \in \Bev$, such that $Z(V_i(t_0))\neq Z(V'_i(t_0))$, 
$i \in \left\{i_1, \dots, i_k\right\}$. For example, some neuron does not fire
when not perturbed but the application of an $\epsilon$-perturbation induces
it to fire (possibly with a membrane
potential strictly above the threshold).  This requires obviously this neuron to be close enough to
the threshold. Clearly, the evolution of the unperturbed and perturbed trajectory
may then become drastically different. Indeed, even if only one neuron
is lead to fire when perturbed, it may induce other neurons to fire
at the next time step, etc \dots, inducing an avalanche phenomenon
leading to unpredictability  and initial condition sensitivity\footnote{This effect
 is well known in the context of synfire chains \cite{abeles:82,abeles:91,abeles-etal:93,hertz:97}
or self-organized criticality \cite{blanchard-cessac-etal:00}.
}.

It is tempting to call this behaviour ``chaos'', but there is an important difference
with the usual notion of chaos in differentiable systems. In the present case,
due to the sharp condition defining the threshold,
initial condition only occurs at sporadic instants, whenever some neuron
is close enough to the threshold. 
Indeed, in certain periods of time the membrane potential typically is quite far below threshold, 
so that the neuron can fire only if it receives strong excitatory input over 
a short period of time. It shows then a behaviour that is robust against fluctuations.
On the other hand, when membrane potential is close to the threshold
a small perturbation may induce drastic change in the evolution.

\paragraph{Stability with respect to small perturbations.} Therefore,  depending on  parameters
such as the synaptic efficacy,
the external current, it may happen that, in the stationary regime, the 
typical trajectories stay away from the singularity set, say within a distance larger 
than $\epsilon >0$. 
 Thus, a small perturbation (smaller than $\epsilon$) does not produce any
 change in the evolution.
At a computational level, this robustness leads to stable input-output transformations.

On the other hand, if the distance between the set where the asymptotic dynamics
lives\footnote{Namely, the $\omega$-limit set, $\Omega$, which is  the set of accumulation points of $\Fgt(\cM)$,
where $\Fgt(\cM)$ is the mapping defining the dynamics (eq. (\ref{DNN})).
Since $\cM$ is closed and invariant, we have
$\oM=\bigcap_{t=0}^\infty \Fgt(\cM)$. 
In dissipative systems (i.e. a volume element in the phase space  is dynamically contracted),
   the $\omega$-limit
set typically contains the attractors of the system. }  and the singularity set
is zero (or practically, very small) then the dynamics exhibit initial conditions sensitivity,
and chaos. Typically a measure of this ``distance'' is given by \cite{cessac:08}:

\beq
\dOS=\inf_{\V \in \Omega} \inf_{t \geq 0} \min_{i=1 \dots N} |V_i(t)-\theta|,
\eeq

\nid where $\Omega$ is the $\omega$-limit set. 

\paragraph{Generic dynamics.} 

Now, the following theorem holds \cite{cessac-vieville:08}.

\bth\label{ThdAS}

If $\dOS>0$ then

\ben 

\item  $\Omega$ is composed of finitely many periodic orbits with a finite period,

\item There is a one-to-one correspondence between a trajectory on $\Omega$ and its raster plot,

\item There is a finite Markov partition.
\een

\enth

Note however that $ \dAS>0$ is a sufficient but not a necessary condition to have a periodic
dynamics. The main role of the condition $\dAS>0$ is  to avoid situations
where the membrane potential of some neuron accumulates on $\theta$ \textit{from below} (ghost orbits).
This corresponds to a situation where the membrane potential of some ``vicious'' neuron fluctuates below
the threshold, and approaches it arbitrary close, with no
possible anticipation of its first firing time. This leads to
an effective unpredictability in the network evolution,
since when this neuron eventually fire, it may drastically
change the dynamics of the other neurons, and therefore the observation
of the past evolution does not allow one to anticipate what will be
the future. In some sense, the system is in sort of a metastable state
but it is not in  a stationary state.

Now, assuming that conductances depend on past time only via a finite time horizon, one can show
that,

\bth\label{gener}
 Generically, in a probabilistic and topological sense, $\dOS >0$. 
\enth

(see \cite{cessac:08} for the proof).

\paragraph{Discussion} 

Though the previous results suggests that dynamics is rather trivial since the
first item tells us that dynamics is periodic, periods can however be quite long, depending on parameters.
 Indeed,
 following  \cite{cessac:08} an estimate for an upper bound on the orbits
period is given by:

\beq \label{nM}
T \simeq 2^{N\frac{\log(\dAS)}{\log(<\gamma>)}}
\eeq

\nid where $<\gamma>$ denotes the value of $\gamma$ averaged over time and initial
conditions.
Though this is only an upper bound this suggests that periods diverge when $\dAS \to 0$.
This is consistent with the fact that when $\dAS$ is close to 0 dynamics ``looks chaotic''.
Therefore, $\dAS$ is  what a physicist could call an ``order parameter'',
quantifying somehow the dynamics complexity.
The distance $ \dAS$ can be numerically estimated as done in \cite{cessac:08,cessac-vieville:08}.\\
 
 %
%
%
%
%
%
%
%
%
\begin{figure}[htb]
\begin{center}
\includegraphics[height=4cm,width=6cm,clip=false]{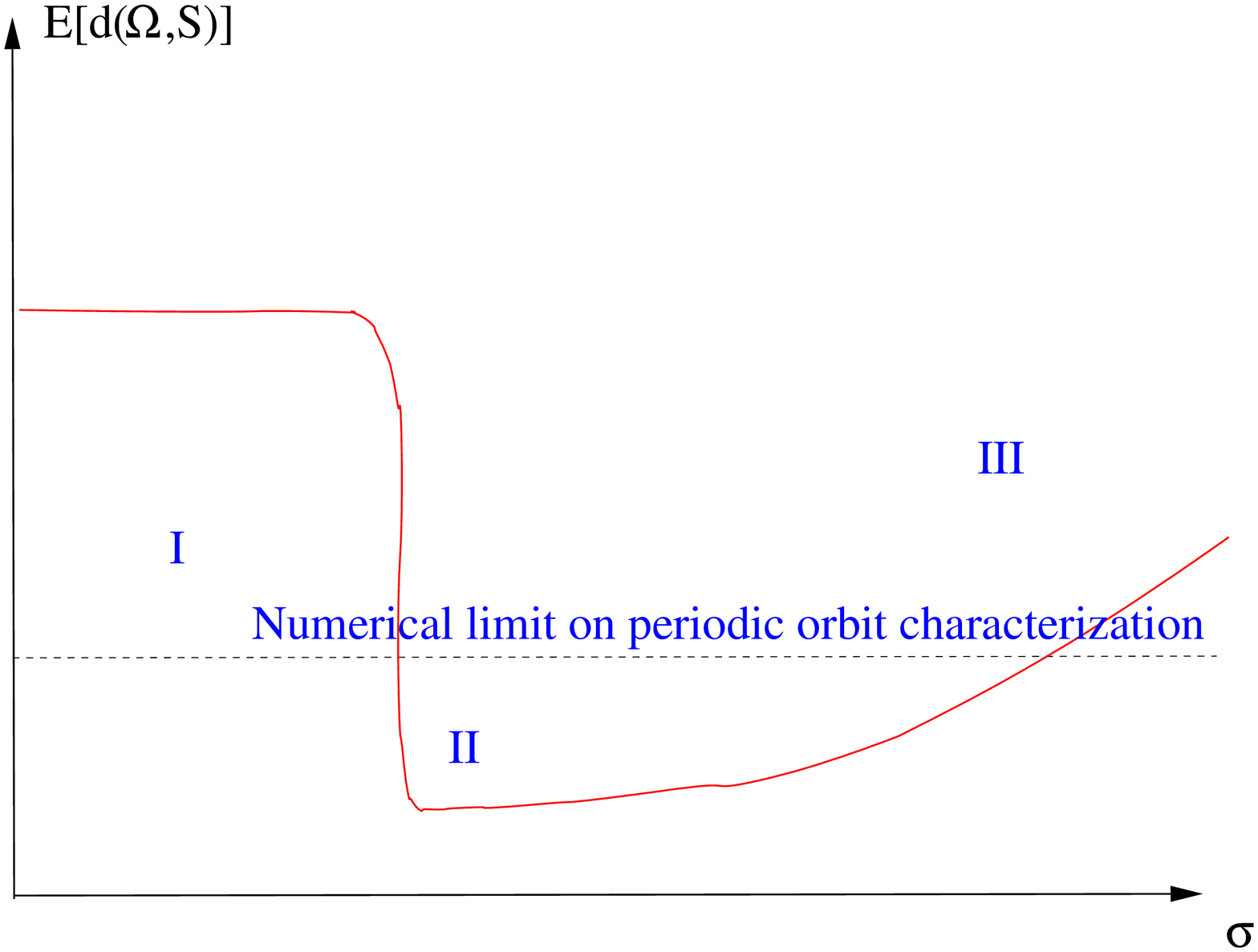}
\caption{\label{FbifIF} 
\footnotesize{Sketch of the dynamical regimes exhibited by model (\ref{DgIF}) when synaptic weights are
drawn at random with a Gaussian distribution $\cN(0,\frac{\sigma^2}{N})$
(drawn from \cite{cessac:08}). The expectation of $\dAS$ under the $W_{ij}$'s distribution, $\Exp{\dAS}$ , is drawn.
It defines three regions.  Region I corresponds to ``neural death'' (all neurons stop firing after a finite time);
region II to a regime indistinguishable from chaos where the period of orbits are quite larger than
what can be measured numerically; region III is a region where periodic orbits can be numerically detected.}}
\end{center}
\end{figure}

Let us give an example of application of this result. Consider model 
(\ref{DgIF}) where the synaptic weights are drawn at random with a Gaussian distribution 
$\cN(\frac{\Wb}{N},\frac{\sigma^2}{N})$, in the same spirit as in section \ref{meanfield}.
We have sketched the average value $\dAS$, averaged over the distribution
of the $W_{ij}$'s, as a function of $\sigma$, when $\Jb=0$ and $\gamma$ is fixed.
The curve of $\dAS$, as a function of $\sigma$, delimits 3 regions.
 Region I corresponds to ``neural death'' (all neurons stop firing after a finite time);
region II to a regime indistinguishable from chaos where the period of orbits are quite larger than
what can be measured numerically; region III is a region where periodic orbits can be numerically detected. 
This transition is reminiscent
 of the one exhibited in  \cite{keener-hoppensteadt-etal:81} for an isolated
neuron submitted to a periodic excitation, but the  present analysis hold  at the network level.
 
Let us now discuss  the second item of theorem \ref{ThdAS}. It expresses that the raster plot is a \textit{symbolic
coding} for the membrane potential trajectory. In other words there is no loss
of information about the dynamics when switching from the membrane potential
description to the raster plot description. This is not true anymore if $\dAS=0$.
This issue, as well as the existence of a Markov partition, is used in section \ref{SpikesStat}.

\ssu{Conclusion}
In this section we have shown two examples of classical neural networks models, where
the use of combined techniques from dynamical systems theory, statistical physics and probability 
theory allows the characterization of the dynamical regimes generically occurring. Moreover, considering
random and \textit{independent} synaptic weights $W_{ij}$'s we have been able to 
obtain a phenomenological ``bifurcation diagram'' where one replaces the overwhelming
number of control parameters ($N^2$ synaptic weights plus additional parameters defining the external current)
by a small set of statistical parameters controlling the probability distribution of
the $W_{ij}$'s (mean and variance).
 This diagram characterizes the average behaviour of many different copies
of the neural network when the $W_{ij}$'s are drawn at random with a
specific value of their mean and variance. It does not tell us what will be the typical behaviour
of a given network (i.e. a given realization of the $W_{ij}$'s). Moreover, 
for the mean-field approach reported in section \ref{meanfield}
the bifurcation map corresponds to taking the limit $N \to \infty$
where, e.g. the transition to chaos is easy to represent since it is sharp.
The situation is radically different for finite $N$ where the ``edge of chaos''
associated with the transition by quasi-periodicity is rather complex and results
from the overlapping of Arnold tongues \cite{mackay-tresser:86,gambaudo-tresser:88}.
For the gIF model, theorem \ref{ThdAS} and \ref{gener} hold for generic values
of the synaptic weights $W_{ij}$'s hence they apply to the huge space of parameters
$\bg$. Moreover, they characterize generic behaviours both in a topological and probabilistic sense.
However, to figure out how $\dAS$ looks like we focused actually on the same situation
as  in section \ref{meanfield} where the $W_{ij}$'s are drawn at random, independently, where
we study the effect of their variance of the average value of $\dAS$. 
It seems possible to have
an analytic expression of $\dAS$, but this requires to take the ``thermodynamic limit''
$N \to \infty$ (Cessac and Touboul, in preparation).

Thus, it appears clearly that these approaches are limited 

\ben

\item By the assumption of independence of the $W_{ij}$'s.
\item By the necessity of taking the limit $N \to \infty$ to obtain analytic expression.
\een

These limitations are further discussed in the conclusion section \ref{conclusion}.

\su{Spikes trains statistics}  \label{SpikesStat}

As we have seen in section \ref{neurodyn}, neuronal activity
 is manifested by emission of spike trains having  a wide variety
of forms (isolated spikes, periodic spiking, bursting, tonic spiking, tonic bursting, etc)
 \cite{izhikevich:04,brette-gerstner:05,touboul:08},
depending on physiological parameters, but also on excitation coming either from other
neurons or from external sources. From these evidences, it seems natural to consider
spikes as ``information  quanta'' or ``bits'' and to seek the information exchanged by
neurons in the structure of spike trains. Doing this, one switches from the description
of neurons in terms of membrane potential dynamics, to a description in terms of
spikes trains and raster plots. Though this change of description raises many questions
it is commonly admitted in the computational  neuroscience community that spike trains
contain  the ``neural code''.

Admitting this raises however other questions. 
How is ``information'' encoded in a spike train:
 rate coding \cite{adrian-zotterman:1926}, temporal coding \cite{theunissen-miller:95},
rank coding \cite{perrinet-et-al:01,delorme-et-al:01}, correlation coding \cite{johnson:80} ?
How to measure the information content of a spike train ? 
There is a wide literature dealing with these questions 
\cite{nirenberg-latham:03,johnson:04,barbieri-et-al:04,nemenman-et-al:06,arabzadeh-et-al:06,sinanovic-johnson:06,gao-et-al:08,osbone-et-al:08},
which are inherently related to the notion of  \textit{statistical characterisations} of spike trains, 
see \cite{rieke-etal:96,dayan-abbott:01,gerstner-kistler:02b}
and references therein for a a review.
As a matter of fact, a prior to handle ``information'' in a spike train is
the definition of a suitable probability distribution that matches the empirical
averages obtained from measures. 

\ssu{Spike responses of neurons}

Neurons respond to excitations or stimuli by finite sequences of spikes. 
Thus, the dynamical response $R$ of a neuronal network to a stimuli $S$ (which can be applied
to several neurons in the network), is a sequence $\bom(t) \dots \bom(t+\tau)$ of spiking
patterns. ``Reading the neural code'' means that one seeks a correspondence between
responses and stimuli. However, the spike response does not only depend on the stimulus, but also
on the network dynamics and therefore fluctuates randomly.
Thus, the spike response  is sought as
a conditional probability $P(R|S)$ \cite{rieke-etal:96} and ``reading the code'' consists of
inferring $P(S|R)$ e.g. via Bayesian approaches, providing a loose dictionary where
the observation of a fixed spikes sequences $R$ does not provide a unique possible
stimulus, but a set of stimuli, with different probabilities. 
Having models
for conditional probabilities $P(R|S)$ is therefore of central importance.
For this, one needs a good notion of statistics.

These  statistics
 can be obtained in two different ways. Either one repeats
a large number of experiments, submitting the system to the same stimulus $S$,
and performs a sample averaging. This approach relies on the assumption
that the system has the same statistical properties during the whole set of experiments 
(i.e. the system has not evolved, adapted or undergone bifurcations meanwhile). 
Or, one performs a time average. For example, to compute $P(R|S)$, one counts the number of times $n(R,T,\tom)$ when the finite sequence of
spiking patterns $R$,
appears in a spike train $\tom$ of length $T$, when the network is submitted to a stimulus $S$.
 Then, the probability $P(R|S)$ is estimated
by:
$$P(R|S) = \lim_{T \to \infty} \frac{n(R,T,\tom)}{T}.$$
This approach implicitly assumes that the system is in a stationary state. 

The empirical approach is often ``in-between''. One fixes a time window of length
$T$ to compute the time average and then performs an average over a finite number $\cN$
of experiments corresponding to selecting different initial conditions. 
In any case the implicit assumptions are  essentially impossible to control in real (biological)
experiments, and difficult to prove in models. 
So,  they are basically used as ``working'' assumptions.  
To summarise, one observes, from $\cN$ repetitions of the same experiment,
 $\cN$ raster plots $\tom_m, m=1 \dots \cN$ on a finite time horizon 
of length $T$.  From this, one computes experimental averages allowing to estimate
$P(R|S)$ or, more generally, to estimate the average value, $\langle \phi \rangle$, of some prescribed
observable $\phi(\tom)$.  These averages are estimated by : 
\beq\label{empav}
\bar{\phi}^{(\cN,T)}=\frac{1}{\cN T}\sum_{m=1}^{\cN} \sum_{t=1}^T \phi(\stg \tom_m).
\eeq
Typical examples of such observables are $\phi(\tom)=\omega_i(0)$ in which case 
$\langle \phi \rangle$ is the firing rate of neuron $i$;  
$\phi(\tom)=\omega_i(0)\omega_j(0)$ then $\langle \phi \rangle$ measures the probability of spike coincidence
for neuron $j$ and $i$; $\phi(\tom)=\omega_i(\tau)\omega_j(0)$ then $\langle \phi \rangle$ measures the probability of 
the event ``neuron $j$ fires and neuron $i$ fires $\tau$ time step later'' (or sooner according to the sign
of $\tau$). In the same way  $P(R|S)$ is the average 
of the indicatrix function $\chi_R(\tom)=1$ if $\omega \in R$ and $0$ otherwise, the statistics being
performed when
the neuronal network is submitted to $S$.
Note that in (\ref{empav}) we have used the shift $\stg$ for the time evolution of the raster plot. This notation is
more compact and more adapted to the next developments than the
classical formula, reading, e.g., for firing rates 
$\frac{1}{\cN T}\sum_{m=1}^{\cN} \sum_{t=1}^T \phi(\bom_m(t))$.

This estimation depends on $T$ and $\cN$. However, one expects that,
as $\cN,T \to \infty$, the empirical average $\bar{\phi}^{(\cN,T)} $ 
converges to the theoretical average $ \langle \phi \rangle$, as stated e.g. from the law of large numbers.
Unfortunately, one usually does not have access to these limits, and one is lead to extrapolate
theoretical averages from empirical estimations. The main difficulty is that 
these observed raster plots are produced by an underlying dynamics
which is usually not explicitly known (as it is the case in experiments) or impossible to fully characterise 
(as it is the case in most large dimensional neural networks models). 
Thus, one is constrained to propose ad hoc statistical models.
As a matter of fact, the choice of a statistical model always relies on assumptions. Here we make an attempt to formulate
these assumptions in a compact way with the widest range of application.
These assumptions are compatible
with the statistical models commonly used in the literature like Poisson models or  Ising like models \`a la Schneidman
and collaborators \cite{schneidman-etal:06}, but 
lead also us to propose more general forms of statistics. Moreover, our approach
incorporates additional elements such as the consideration of neurons dynamics,
and the fact that this dynamics severely constrain the set
of admissible raster plots, $\Spg$. This last issue is, according to us, fundamental, and, to the best
of our knowledge, has never been considered before in this field. \\ 
 
On this basis we propose the following definition.
 Fix a set $\phi_l$, $l=1 \dots K$,
of observables, i.e. functions $\Spg \to \bbbr$ which associate real numbers to sequences of spiking
patterns. Assume that the empirical average (\ref{empav}) of these functions has been computed, 
for a finite $T$ and $\cN$, and
that $\bar{\phi_l}^{(T,\cN)}=C_l$.

A \textit{statistical model} is a probability distribution $\nu$ on the set of raster plots such
that:

\ben
\item $\nu(\Spg)=1$, i.e. the set of non admissible raster plots has a  zero $\nu$-probability.
\item $\nu$ is ergodic for the left-shift $\sg$.
\item For all $l=1 \dots K$, $\nu(\phi_l)=C_l$, i.e., $\nu$ is compatible with the empirical averages.
\een

Note that item 2 amounts to assuming that statistics are invariant under time translation.
On practical grounds, this hypothesis can be relaxed  using sliding time windows.
This issue is discussed in more details in  \cite{cessac-rostro-etal:09}.
Note also that $\nu$ depends on the parameters $\bg$.
Assuming that $\nu$ is ergodic has the advantage that one does not have to average \textit{both}
over experiments \textit{and} time. It is sufficient to focus on time average for
a single raster plot, via the time-empirical average:

\beq\label{pTo}
\pi_{\omega}^{(T)}(\phi) = \frac{1}{T} \sum_{t=1}^T \phi(\stg \omega).
\eeq

\ssu{Raster plots statistics.} \label{Statmod}

A canonical way to construct statistical models comes from statistical physics \cite{jaynes:57}.
This approach has been introduced for spike train analysis by \cite{schneidman-etal:06} and generalised 
in \cite{cessac-rostro-etal:09}.
According to item (1)-(3) we are seeking a probability distribution $\nu$ which matches
the constraints $\nu(\phi_l)=C_l, \ l=1 \dots K$, where $\nu(\phi_l)$ is the average of $\phi_l$
under $\nu$. We want to stick on these
constraints, imposed by experimental results,
 without adding any other hypothesis. In the realm of statistical physics this amounts
to maximising the statistical entropy under the constraints  $\nu(\phi_l)=C_l, \ l=1 \dots K$.
In the context of the so-called thermodynamic formalism of ergodic theory,
which is a quite powerful tool to handle such statistical problems, this amounts 
to solving the following variational principle:
\beq\label{pres}
\pres=\sup_{\nu \in m^{(inv)}} (h\left[\nu\right]+\nu\left[\bpsi\right]),
\eeq
where $m^{(inv)}$ is the set of invariant (stationary) measures for the dynamics 
and $h$ is  the entropy rate. 
We have introduced a ``potential'',  

\beq\label{psi}
\bpsi=\sum_{l=1}^K \lambda_l \phi_l,
\eeq
 where
the $\lambda_l$'s are adjustable Lagrange multipliers.
 A probability measure $\mpg$ which realises the supremum, i.e.
$$\pres=h\left[\mpg\right]+\mpg\left[\bpsi\right],$$
is called an ``equilibrium state''. 
The function $\pres$ is called the ``topological pressure'' in the realm of
ergodic theory, and ``thermodynamic potential'' (free energy, free enthalpy, pressure) in statistical
physics. Note that ergodic theory imposes less constraints on dynamics
than statistical physics (the microscopic dynamics does not need to be Hamiltonian).
 From the topological pressure one computes the moments of the distribution
$\mpg$. In particular\footnote{This relations assumes that $\pres$ is differentiable, i.e.
that the system is away from a phase transition.},
\beq\label{Gener}
\frac{\partial \pres}{\partial  \lambda_l}=\mpg(\phi_l).
\eeq
This relation fixes the value of the Lagrange multipliers $\lambda_l$
in order to have $\mpg(\phi_l)=C_l$.

Moreover, in ``good cases'' (e.g. uniformly hyperbolic dynamical systems),
equilibrium states are also Gibbs states 
\cite{bowen:75,bowen:98,keller:98,parry-pollicott:90,chazottes-keller:09}.
 A Gibbs state, or Gibbs measure, is a probability measure such that,
one can find some constants $c_1,c_2$ with $0 < c_1 \leq 1 \leq c_2$ such
that for all $n \geq 1$ and for all $\tom$:
\beq\label{dGibbs}
c_1 \leq \frac{\mpg\left(\tom  \in \Con\right)}{ \exp(-n\pres+\Snpo)} \leq c_2,
\eeq
where $\Snpo = \sum_{t=0}^{n-1} \bpsi(\stg \tom)$ and where 
we denote by $\Con$ a cylinder set of length $n$, namely the set
of raster plots $\tom'$ such that  $\bom'(t)=\bom(t), t=0 \dots n-1$. Basically, this means that the probability
that a raster plot  starts with the bloc $\Con$ behaves like $\frac{\exp(\Snpo)}{Z_n}$.
One recognises the classical Gibbs form where 
space translation in lattice system is replaced by  time translation
(shift $\stg$) and where the normalisation factor $Z_n$ is the partition function.
Note that $\pres=\limsup_{n \to \infty}\frac{1}{n}\log Z_n$, so that $\pres$ is indeed the formal analog
of a thermodynamic potential (like free energy).

In this context, the probability of a spiking pattern block 
$R=\left[\bom \right]_{0,n-1}$ of length $n$ corresponding
to the response $R$ to a stimuli $S$ ``behaves like'' (in the sense of eq. (\ref{dGibbs})):

\beq\label{PRS}
P\left[R|S \right]=\nu\left[\omega \in R | S \right] \sim \frac{1}{Z_n\left[\lambda_1(S),\dots,\lambda_l(S)\right]}
\exp\left[
\sum_{l=1}^K \lambda_l(S) \sum_{t=0}^{n-1} \phi_l(\stg \tom)
\right],
\eeq
where  the $\lambda_l$'s depend on the stimulus $S$.  Obviously, for two different stimuli the probability $P(R|S)$ may drastically change.

\ssu{Examples.}

\st{Firing rates.} If $\phi_l(\tom)=\omega_l(0)$, then $\pTo(\phi_l)=r_l$ is the average firing rate of neuron $l$ within
the time period $T$. Then, the corresponding statistical model is a Bernoulli distribution where neuron $l$ has a probability
$r_l$ to fire at a given time. The probability that neuron $l$ fires $k$ times within a time delay $n$ is a binomial distribution 
and the inter-spike interval is Poisson distributed \cite{gerstner-kistler:02}.\\

\st{Spikes coincidence.} If $\phi_l(\tom)\equiv \phi_{(i,j)}(\tom)=\omega_i(0) \, \omega_j(0)$ where, here,
 the index $l$ is an enumeration for all (non-ordered) pairs
$(i,j)$, then the corresponding statistical models has the form of an Ising model, as discussed by Schneidman and collaborators in 
\cite{schneidman-etal:06,tkacik-etal:06}. As shown by these authors in experiments on the salamander retina, the probability of spike
blocs estimated from the ``Ising'' statistical model fits quite better to empirical date than the classical Poisson model.  \\

\st{Enlarged spikes coincidence.} As a generalisation one may consider the probability of co-occurrence of spikes from neuron $i$ and $j$ within
some time interval $\tau$. The corresponding functions are $\phi_l(\tom)=\omega_i(0)\omega_j(\tau)$ and the probability
of a spike bloc $R$ reads:

$$
P\left[R|S \right] = 
 \frac{1}{Z_n\left[\lambda_{1,1}(S),\dots,\lambda_{N,N}(S)\right]}
\exp\left[
\sum_{i \leq j} \lambda_{ij}(S) \sum_{t=0}^{n-1} \omega_i(t) \, \omega_j(t+\tau)
\right].
$$
\nid
An example is provided in section  \ref{plaststat}.

\nid Further generalisations can be considered as well.\\

\st{Generalised Integrate and Fire models} Due to their particular structure and especially
the fact that generically a Markov partition exists,  gIF models of type (\ref{DgIF})
are explicit examples where this theory gives striking results (see \cite{cessac-rostro-etal:09} for details
and section \ref{adapt} for an application to the effect of synaptic plasticity to spike trains
statistics.)

\ssu{Validating a statistical model}

There are currently huge debates on the way how brain encodes information.
Are frequency rates sufficient to characterise the neural code \cite{vanvreeswijk:04} ?
Are pair correlations significant ? Do higher order statistics matter ?
Actually, it might be that the answer depend on the brain process under consideration
and some peoples actually believe that ``brain speaks several languages and speak all of them
at the same time'' (Franck Grammont, private communication. For a nice illustration of this
see \cite{grammont-riehle:99}). 
These questions are inherently linked to the notion of (i) finding  statistical
models; (ii) discriminate several statistical models and select the ``best one''.

Let us consider an illustrative example, i.e. the question: {\em are correlations significant}~? 
Answering this question is a crucial issue 
for biologists/experimentalists \cite{segev-etal:04,pillow-etal:05,pillow-etal:08}. 
Note that it has absolutely no meaning to try and answer this question from empirical data
when considering ``the brain'' as a whole. But, as emphasised by \cite{roudy-nirenberg-etal:09},
 there is maybe some hope 
to  make one step forward when considering  \textit{small} neural
assemblies (e.g. small pieces of retina). 

Moreover this question has no ``absolute'' answer but a relative answer in the following sense. Let us consider the 1st order potential:

$$\bpsi_1(\tom)= \sum_{i=1}^N \lambda_i \, \omega_i(0), $$

\nid thus only taking firing-rates into account, ``against'' the 2nd order potential:

$$\bpsi_2(\tom)=\sum_{i=1}^N \lambda_i \,\omega_i(0)  + \sum_{i,j=1}^N \sum_{\tau=-T_s}^{T_s} \lambda_{ij\tau} \, \omega_i(0)\omega_j(\tau),$$

\nid where $T_s$ is a characteristic time scale. This potential form  takes both firing-rate and correlations into account. 

The realm of thermodynamic formalism offers a numerically tractable way to compare the statistical models related
to these two potentials. The relative entropy or Kullback-Leibler divergence\footnote{Let $\mu$,$\nu$ be two invariant 
measures both defined on the same set of admissible raster plot  $\Spg$.
The   relative entropy (or Kullack-Leibler divergence) between $\mu$ and $\nu$ is:
\beq\label{HKL}
h(\mu|\nu)=\limsup_{n \to \infty} \frac{1}{n}\sum_{\Con} 
\mu\left(\Con\right)
\log\left[\frac{\mu\left(\Con\right)}{\nu\left(\Con\right)} \right].
\eeq
} between a
Gibbs measure $\mpg$ and a stationary measure $\mu$ is given by \cite{keller:98,chazottes-etal:98,chazottes-keller:09}:
$$h\left(\mu|\mpg \right) = \pres - \int \bpsi d\mu - h(\mu).$$

In the present case, we are given an empirical measure, $\pTo$, (see eq. (\ref{pTo})), obtained from experiments.
To discriminate between the two potentials $\bpsi_1,\bpsi_2$ a possible criterion 
 consists of choosing the potential which minimises the relative entropy of the corresponding
Gibbs measure with respect to the empirical measure. Namely, if there is $T_0>0$ such that, for all $T \geq T_0$ :

\beq
h(\pTo \, |\, \nu_{\bpsi_1}) < h(\pTo \, |\, \nu_{\bpsi_2})
\eeq 

\nid then $\bpsi_1$ is considered as a better statistical model than $\bpsi_2$. The nice thing
is that, using the thermodynamic formalism, 
one can develop algorithms allowing such a comparison (Cessac, Vasquez, Vi\'eville, in preparation).

\ssu{Conclusion}

When analysing spike train statistics, one is lead to propose several statistical models
corresponding to distinct hypothesis. For example, characterising inter-spike interval distribution
by a homogeneous Poisson process ultimately corresponds to assuming  that correlations between
neurons and time correlations are irrelevant and that frequency rates are sufficient to characterise
statistics. In our presentation this amounts to considering a Gibbs potential of form $\sum_{i=1}^N \lambda_i \omega_i(0)$.
More general forms can be proposed as well. But this leads to two fundamental questions:

\ben
\item How to discriminate statistical models from empirical data ? 
This is a crucial issue, whose tractability was deeply raised by and Roudy and his collaborators in
a very recent paper \cite{roudy-nirenberg-etal:09}. Many criteria used in the literature rely on the computation
of the Kullback-Leibler divergence. We have shown how the use of the thermodynamic formalism,
relying on a safe recipe from statistical mechanics, could allow to compute this quantity from data.

\item Instead of defining the statistical model from \textit{ad hoc} observables,
is it possible to propose a canonical form relying on some generic principle ?
This issue is addressed in section \ref{plaststat} where we consider the effect of synaptic
plasticity on spike trains statistics. 

\een

\su{Interplay between synaptic graph structure and neurons dynamics.}\label{network}

\ssu{Causal actions} 

Since
synapses are used to transmit neural fluxes (spikes) from a neuron
to another one, the existence of synapses between a neuron (A) and another
one (B) is implicitly  attached to a notion of  ``influence'' or
causal and directed action.  However, as we saw, a neural network is a highly dynamical
object and its behavior is the result of a complex
interplay between the neurons dynamics and the
synaptic network structure. Moreover, the neuron
$B$ receives usually synapses from many other
neurons, each them being ``influenced'' by
many other neurons, possibly acting on $A$, etc...
Thus the actual ``influence''
or action of A on B has to be considered dynamically
and in a global sense, by considering $A$ and $B$
not as isolated objects, but, instead, as entities
embedded in a system with a complex interwoven dynamical
evolution. In this context it is
easy to imagine examples where  there is a synapse from $A$ to $B$
but no clear cut influence, or, in the opposite, no synapse and nevertheless
an effective action.

Thus, one has to consider topological
aspects (such as the feedback circuits) \textit{and} dynamical aspects.
One way of doing this is to compute correlations between neurons (cross-correllogramms). 
However, correlations functions do not 
provide causal information. A strong
correlation between $A$ and $B$ at time
$t$ does not tell us if $A$
acts on $B$ or if $B$ acts on $A$ (note in particular
that $C_{AB}(t)=C_{BA}(-t)$).

Another way consists of
exciting neuron $A$, say with a weak signal, and observe
the effects on $B$, e.g. by comparing its evolution
with and without the signal applied on $A$.
A natural choice for an excitatory signal
is a periodic signal, with a tunable frequency.
Thus, the response function, drawn versus frequency,
provides similar information as the complex 
susceptibility in Physics. In particular, peaks
in the susceptibility corresponds to resonances,
that is, a response of maximal amplitude. These resonances can be used to provide
an effective, frequency dependent notion of network
structure, as we now show.

\ssu{A simple but non trivial example}

Consider the model (\ref{SDNN})
where neurons are represented by frequency rates.
As we saw in section  \ref{bifmeanfield} this model exhibit, in finite dimension
$N$ a generic transition to chaos by quasi-periodicity, when increasing the non-linearity
of the sigmoidal transfer function $S$.

\begin{figure}[htb]
\begin{center}
\includegraphics[height=4cm,width=4cm,clip=false]{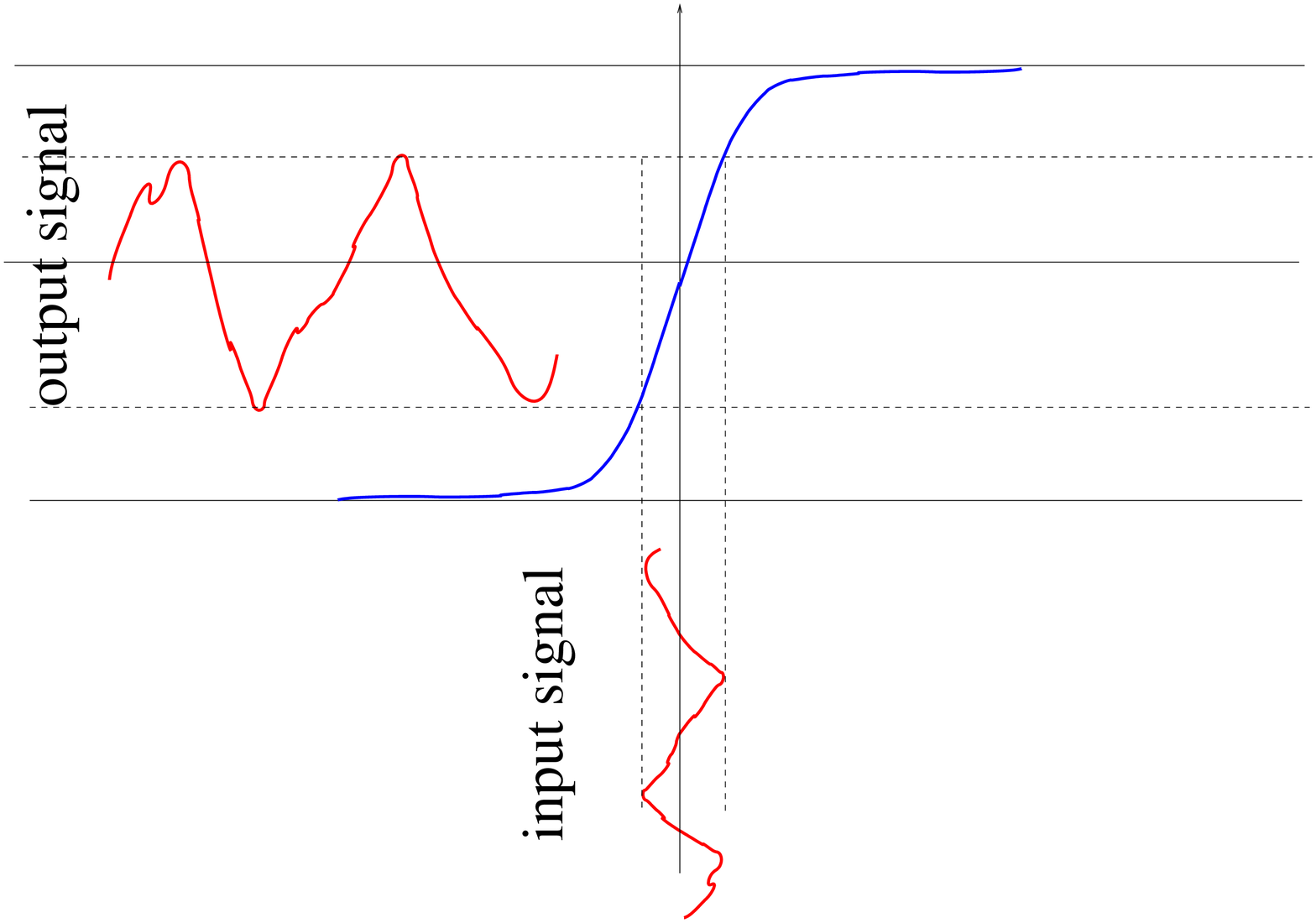}
\hspace{1cm}
\includegraphics[height=4cm,width=4cm,clip=false]{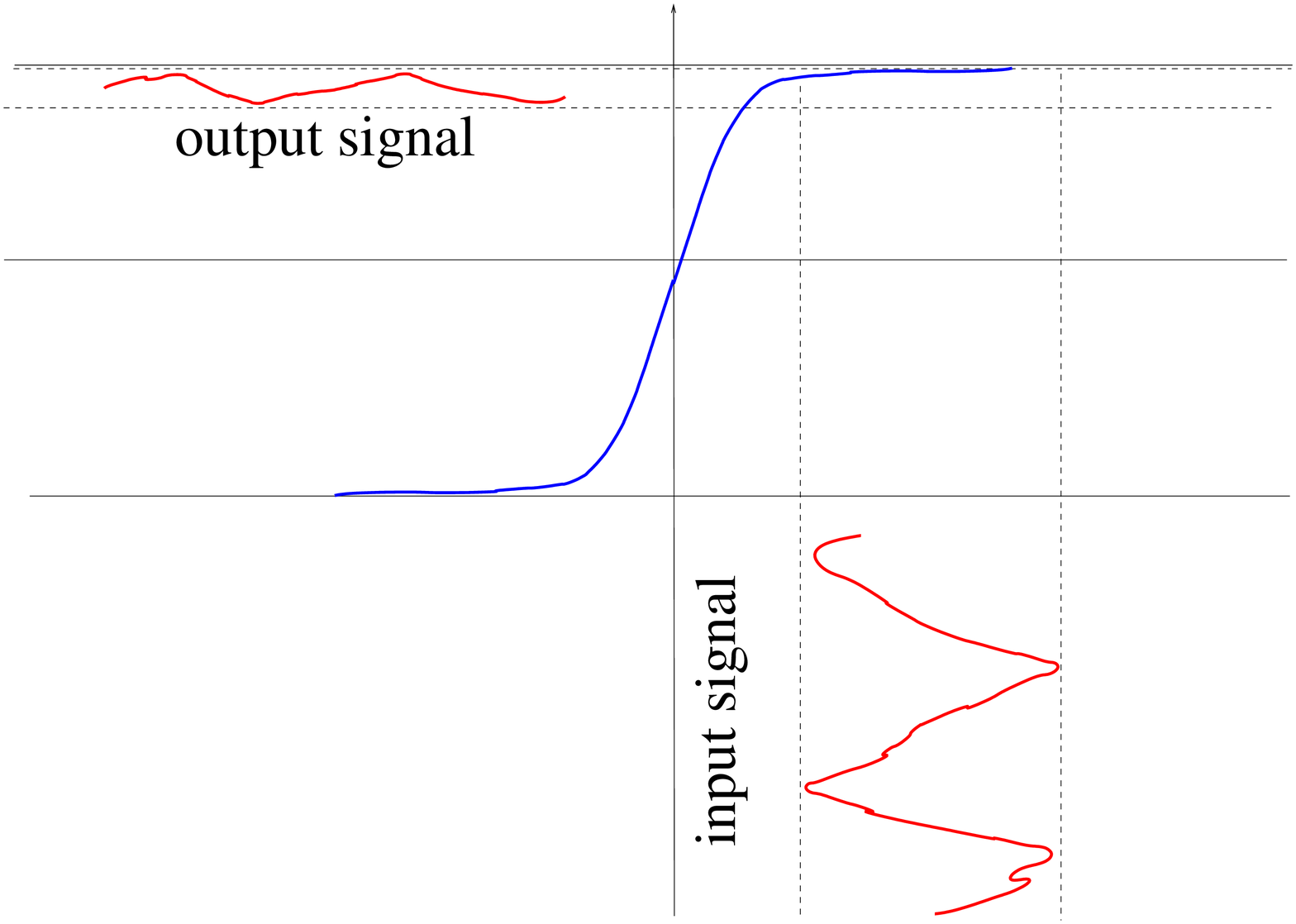}
\hspace{1cm}
\includegraphics[height=4cm,width=4cm,clip=false]{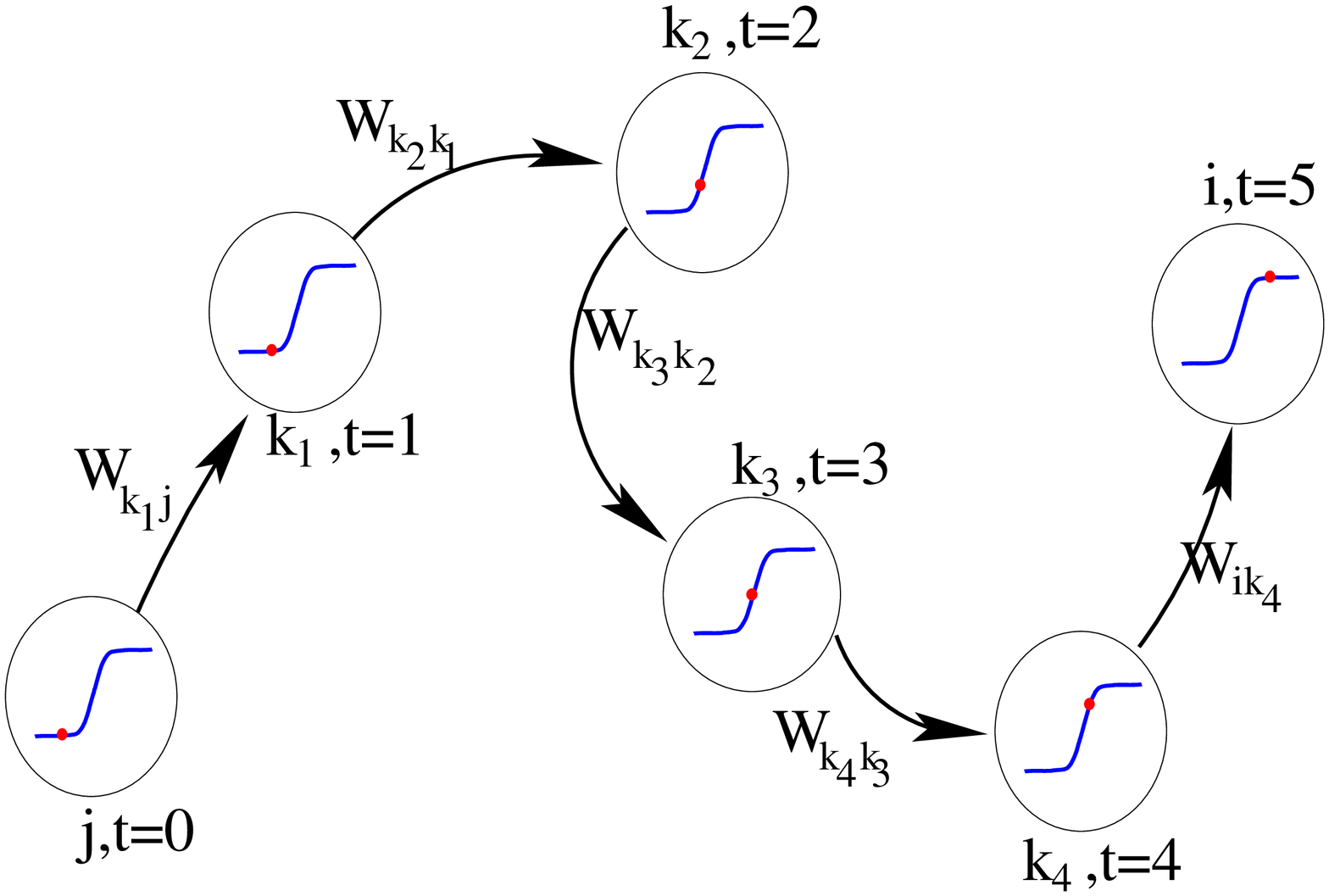}
\caption{\label{FSat} \footnotesize{
Nonlinear effects induced by a transfer function with a sigmoidal shape on signal
transmission. Fig. \ref{FSat}a (left). Amplification. Fig. \ref{FSat}b (middle). Saturation.
 Fig. \ref{FSat}c (right). The propagation of a signal along a path in the network
depends not only on the weights of the links but also on the level
of saturation of the nodes that the signal meets. The level of saturation
depends on the current state of the node (schematically represented as a
red point in the figure). This state evolves with time.}}
\end{center}
\end{figure}

\paragraph{Signal propagation and effects of non-linearity.} 
Assume  that this system is in the chaotic regime.
 Note that the corresponding Fourier spectrum
is not flat but contains peaks (resonances) reminiscent of the transition by quasi-periodicity
\cite{cessac-sepulchre:06}.
Assume now that we superimpose upon the membrane potential  
$V_j(t)$ of the neuron $j$ a small external signal $\xi_j(t)$.
Does this signal have an effect which propagates  inside the network ? and how ?
 Because of the sigmoidal shape of the transfer functions the answer 
depends crucially, not only on the connectivity of the network, but also on the value of the $V_k$'s.
 Assume, for the moment and for simplicity, that the time-dependent signal $\xi_j(t)$ has variations
substantially faster than the variations of $V_j$. Consider then the cases depicted in
Fig. \ref{FSat}.
 In the first case (a) the signal $\xi_j(t)$ is amplified by $S$, without distortion if  $\xi_j(t)$ is weak enough.
In the second case (Fig. \ref{FSat}b), it is damped and distorted by  the saturation of the sigmoid.
More generally, when considering the propagation of this signal from the node $j$ to some node $i$
one has to take into account the level of saturation of the nodes encountered in the path, but
the analysis is complicated by the fact that the nodes have their own dynamical evolution
 (Fig. \ref{FSat}c).

 %
%
%
%
%
%
%
%
%

\begin{figure}[htb]
\begin{center}
\includegraphics[height=3cm,width=6cm,clip=false]{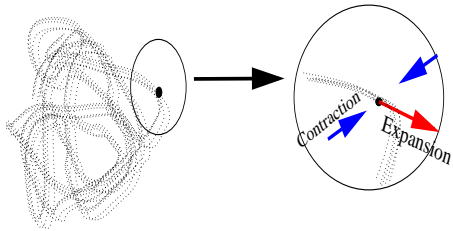}
\caption{\label{FAtt} 
\footnotesize{Sketch of an attractor and of the decomposition of the tangent space into
contracting and expanding directions. A perturbation in the contracting 
space leads to a trajectory which converges back to the attractor.
A perturbation in the transverse direction leads to a trajectory staying
onto the attractor but which separates from the mother trajectory.
 This representation is made for a \textit{discrete time}
system. Note that the regular spacing between points is an artefact of the representation.  }}
\end{center}
\end{figure}

\paragraph{Tangent space splitting.}
In this context we would like to measure the \textit{average} ``influence''
of neuron $A$ on neuron $B$ (namely how a weak
signal applied on $A$ perturbs on average the state of $B$), including
the effects of the nonlinear dynamics.
Typically, in  dissipative systems, such as neural networks models, 
 where volume in the phase space is dynamically contracted, dynamics
asymptotically settle ``onto''  attractors\footnote{Let $\cM$ be the (compact) phase
space of the dynamical system. A set $\cA \in \cM$ is called an \textit{attractor}
if it is invariant ($\F^t(\cA)=\cA$) and if
if there exists an open set $\cU \in \cM$ such that $\cA = \cap_{t \geq 0} \F^t(\cU)$.
This definition affords several non equivalent extensions and variants 
\cite{williams:74,milnor:85,eckmann-ruelle:85,cosnard-demongeot:85,cosnard-etal:93,katok-hasselblatt:98}}.
Examples of attractors are stable fixed points, or stable periodic orbits.
Chaotic attractors have moreover the following property. While
dynamics transverse to the attractor is contracting (corresponding precisely
to the attractivity property), dynamics ``parallel'' to the attractor is 
expanding, corresponding to initial condition sensitivity. In other words,
the tangent space of attractor points can be split into a contracting
and an expanding part (see fig. \ref{FAtt}). 

Structural properties of chaotic attractors are usually characterized by statistical
quantities such as Lyapunov exponents.
There is indeed a natural notion of average in chaotic systems 
related to the  Sinai-Ruelle-Bowen measure $\rho$
(SRB) \cite{sinai:72,bowen:75,ruelle:78}
 which is obtained as the (weak) limit of the Lebesgue measure $\mu$ under the dynamical evolution\footnote{
A crucial property is that a SRB measure
 has a density along the unstable manifolds,
 but it is singular in the directions transverse to the attractor. }:
$$\rho=\lim_{n \to +\infty} \F_{\bg}^{n} \mu.$$
\nid where $ \F_{\bg}^{n}$ is the $n$-th iterate of  $\F_{\bg}$.
Following an orbit upon the attractor it is possible
to characterize the average expansion and contraction rates for this orbit
via Lyapunov exponents. A positive Lyapunov exponent indicates
local expansion while a negative one indicates contraction.

 %
%
%
%
%
%
%
%
%

In the following we will assume that all Lyapunov exponents are bounded away from zero\footnote{This 
formalism requires, on rigorous grounds, that the system is uniformly hyperbolic \cite{ruelle:99},
and examples of diverging susceptibility can be exhibited for the logistic map \cite{ruelle:05}
or the H\'enon map \cite{cessac:07}. Also, it does not hold at a bifurcation point where susceptibility can diverge.
On practical grounds we require that all Lyapunov exponents are bounded away from zero. 
}.
  Then for each $\V \in supp\, \rho$, 
where $supp\, \rho$ is the support of $\rho$, there exists a splitting 
$E^{(s)}_{\V} \oplus E^{(u)}_\V$ such that $E^{(u)}_\V$,
 the unstable space, is locally tangent to the attractor (the local unstable manifold)
and $E^{(s)}_\V$, the stable space, is transverse to the attractor (locally tangent to the local stable manifold).
 Let us emphasize that the stable and
unstable  spaces depend on $\V$ (while the Lyapunov exponents are $\mu$ almost surely constant).

Let us consider a point $\V$ upon the attractor
and make a small perturbation $\delta_\V$. This perturbation can be decomposed as
$\delta_\V = \delta_\V^{(u)} + \delta_\V^{(s)}$ where $\delta_\V^{(u)} \in E^{(u)}_{\V}$ 
 and $\delta_\V^{(s)} \in E^{(s)}_\V$. 
$\delta_\V^{(u)}$ is locally amplified with an exponential rate (given by the largest
positive Lyapunov exponent). On the other hand 
 $\delta_\V^{(s)} $  is damped with an exponential speed (given by the smallest
negative Lyapunov exponent)

\paragraph{Linear response.}
 Assume now that we superimpose a  signal of weak amplitude, considered as an external current, upon some neuron ($k$) 
in such a way that the dynamics is still chaotic  (with only a tiny variation of the Lyapunov exponents).
For simplicity,
we suppose  that the signal does not depend on the state of the system, but we can consider this generalization without difficulty
(linear response still applies in this case, but  equations (\ref{dro}), (\ref{Chi}) do not hold anymore).
Denote by $\bxi$ the vector\footnote{\label{fxi} Though this term acts in equation (\ref{SDNN}) as an external current, we use the notation
$\bxi$ throughout the paper, to distinguish between an arbitrary external current ($\Ie$) and some specific stimulus
intended to carry some ``information'' in the network.} $\left\{\xi_i \right\}_{i=1}^N$. The new dynamical
system is described by the equation:
$$\tV(t+1)=\Fg\left[\tV(t)\right]+\bxi(t).$$
The weak signal $\bxi(t)$ may be viewed as a small perturbation of the trajectories of the unperturbed system (\ref{SDNN}).
  At each time this perturbation has a decomposition $\bxi(t) = \bxi^{(s)}(t)+\bxi^{(u)}(t) $ on the local stable and unstable spaces.
The stable component $\bxi^{(s)}(t)$ is exponentially damped. The unstable one $\bxi^{(u)}(t)$  is 
 amplified by the dynamics and then  scrambled by the nonlinear terms.  
Consequently, it is impossible to predict the long term effect of signal $\bxi(t)$ on the global dynamics.

This is true for  \textit{individual trajectories}.    However, the situation
is substantially different if one considers the \textit{average} effect of the signal, the average being performed with respect
to the SRB measure $\rho$ of the unperturbed system.  Indeed, as an application of the general theory~\cite{ruelle:99},  
 it has been established in \cite{cessac-sepulchre:04,cessac-sepulchre:06,cessac-sepulchre:07} that
the \textit{average} variation $\delta_{V_i}(t)$ of the membrane potential $V_i$ under the influence
of the signal is given, to the linear order, by:
\beq\label{dro}
\left< \tV_i(t) - V_i(t) \right>=\sum_{\sigma=0}^{\infty}\sum_{j}\chi_{ij}(\sigma)\xi_j(t-\sigma-1).
\eeq
\nid We used the shortened notation $< \ >$ for the average with respect
to $\rho$. In this expression  $\chi_{ij}(\sigma)$ are  the matrix elements of :
\beq\label{Chi}
\chi(\sigma) = \int \rho(d\V)  D\F^{\sigma}_{\V}.
\eeq
Thus $\chi(\sigma) $ is a matrix representing the average value of the iterate $\sigma$ of the \textit{Jacobian matrix}.  
 Let us note that the 
 fact that $\chi(\sigma) $   remains bounded for $\sigma \to \infty$ is not a trivial result  because $D\F^{\sigma}_{\V} $ diverges
 exponentially with $\sigma$.    The convergence of  $\chi(\sigma) $ has been rigorously shown by Ruelle under the hypothesis of 
uniform hyperbolicity.
It results from the exponential correlation decay (mixing) in the unstable directions and on the exponential contraction.

This means that, provided that $\bxi(t)$ is sufficiently small,
and for any smooth
 observable $A$, the  variation $< A >_t-< A >$ is \textit{proportional} to $\bxi(t)$ up to small 
nonlinear corrections.
In other words,  $\rho_t $ is \textit{differentiable}  with respect to the perturbation.
The derivative is called the  \textit{linear} response.

\paragraph{Causal circuits.}
In the case of the dynamical system~(\ref{SDNN}) one can decompose $\chi_{ij}(\tau)$ as :
\beq\label{chiij}
\chi_{ij}(\tau)=
\sum_{\gamma_{ij}(\tau)}
        \prod_{l=1}^{\tau}W_{k_l k_{l-1}}
\left< \prod_{l=1}^{\tau}S'(V_{k_{l-1}}(l-1))\right>,
\eeq
\nid  The sum holds on each possible paths
$\gamma_{ij}(\tau)$, of length $\tau$, connecting the
neuron $k_0=j$ to the neuron $k_\tau=i$, in $\tau$ steps.
One remarks that each
path is weighted by the product of a \textit{topological} contribution
depending only on the weight $W_{ij}$ and
a \textit{dynamical} contribution. Since, in the kind of systems we consider, functions $S$ are sigmoid, 
the weight of a path $\gamma_{ij}(\tau)$ depends crucially on
the state of saturation of the neurons $k_0, \dots, k_{\tau-1}$ at times $0, \dots, \tau-1$.
Especially, if $S'(V_{k_{l-1}}(l-1))>1$ a signal is amplified while it is damped if
$S'(V_{k_{l-1}}(l-1)) < 1$. Thus, though a signal has many possibilities for going from $j$ to $i$ in $\tau$ time steps,
some paths may be ``better'' than some others, in the sense
that their contribution to $\chi_{ij}(\tau)$ is higher.  Therefore eq.~(\ref{chiij}), which quantifies
the intuition raised in fig. \ref{FSat},  underlines a key point.
  The analysis of signal transmission in a coupled network of 
dynamical neurons with nonlinear transfer functions
 requires to consider both the topology of the interaction graph {\em and}  the nonlinear dynamical regime of the system.

As a remark note that since the derivatives $S'$ in (\ref{chiij}) are bounded by some constant
(proportional to $g$), one can bound the Jacobian matrix component $|DF_{ij}|$ by some $\lambda_{ij}$.
This provides a bound on  (\ref{chiij}) which resembles very much to an expression obtained
by Afraimovich and Bunimovich in \cite{afraimovich-bunimovich:07} (lemma 3) from which
they derive, using the thermodynamic formalism, a topological pressure characterizing the stability
of the dynamical system. Actually, their analysis fully applies here when
the attractor is a fixed point and their theorem 1 typically provides 
a parametric condition for the stability of the  fixed point.
 Note that  then eq. (\ref{chiij}) reduces
to $\chi_{ij}(\tau)=\sum_{\gamma_{ij}(\tau)}\prod_{l=1}^{\tau}W_{k_l k_{l-1}}$ and expresses
 the Jacobian matrix $D\F^\tau$ at the fixed point,
in terms of graph loops. This can be related, to the so-called cyclic expansions used
in dynamical system and ergodic theory (see http://chaosbook.org/ for a very nice presentation).
 Actually we believe that  Afraimovich and Bunimovich approach
can be extended to our case also in the case of chaotic dynamics. But
one has to use a double cyclic expansion: on the loops of the graph, and on the unstable
periodic orbits which can be used to approximate the SRB measure (Cessac, in preparation).

\paragraph{Complex susceptibility.} The existence of this linear response theory  opens up the way to  applications involving  chaotic neural networks 
 {\it used as a linear filter}.  Indeed  eq.~(\ref{dro})  describes a linear system which   transforms an input signal $\bxi(t)$ 
of small amplitude into  an output signal 
$ \left< \tV_i(t) - V_i(t) \right> $ according to a standard convolution product.    
In particular, if  the external signal is chosen as: 
$$\bxi(t) = \epsilon e^{-i\omega t} \, \hat{\be}_j$$
 (where
$\hat{\be}_j$ is the unit vector in direction $j$), then the response of the system is also harmonic with :
$$  \left< \tV_i(t) - V_i(t) \right>  =   \epsilon \hat{\chi}_{ij}(\omega)  e^{-i\omega( t-1)}, $$
where the frequency-dependent amplitude:
\beq\label{suscep}
 \hat{\chi}_{ij}(\omega) =  \sum_{\sigma=0}^{\infty}  \chi_{ij}(\sigma)  e^{i\omega \sigma} 
\eeq
is called the {\it complex susceptibility}.  
In ref.\cite{cessac-sepulchre:04} a method have been conceived and implemented allowing to compute 
 $ \hat{\chi}_{ij}(\omega) $ numerically. 
   The knowledge of  the susceptibility matrix  is very useful as it enables 
one to detect resonances, i.e. frequencies for which  the amplitude response of the system to a periodic input signal  
is maximum.   In fact the existence of a linear response implies that $ \hat{\chi}_{ij}(\omega) $ is bounded for all
 $\omega \in [0,2\pi]$.  Moreover, in view  of eq.~(\ref{suscep}), it is analytic in the complex upper plane.   On the other 
hand, $ \hat{\chi}_{ij}(\omega) $ can have poles within a strip 
in the lower half plane, e.g. in $\omega_0 - i\lambda$, $\lambda >0$.  In this case,
 and if $\lambda$ is small, the amplitude $| \hat{\chi}_{ij}(\omega) | $ exhibits a peak of width $\lambda$  and  
height $| \hat{\chi}_{ij}(\omega_0) |$  which can be interpreted in the present context as follows:   when unit  $j$ 
 (whose state varies chaotically due to the global dynamics)   is subjected to a small periodic excitation  at 
frequency $\omega_0$ and amplitude $\epsilon$  then the \textit{average} response of  unit   $i$  behaves
 periodically with same frequency and amplitude $\epsilon | \hat{\chi}_{ij}(\omega_0) | $ which  is maximal 
in a frequency interval centered about $\omega_0$.

\paragraph{An example of resonances}

\begin{figure}
\begin{center}
\includegraphics[height=6cm,width=6cm,clip=false]{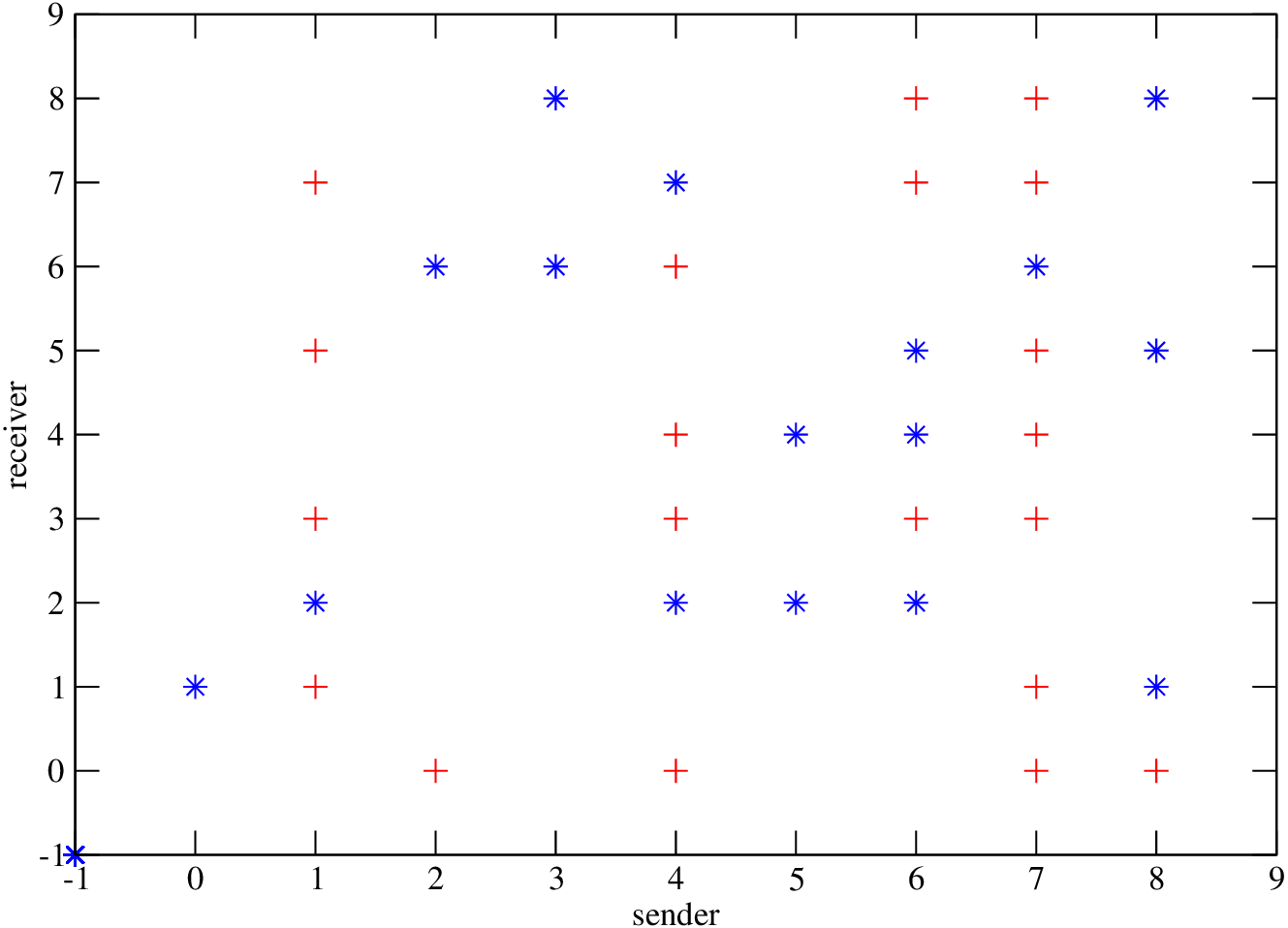}
\hspace{1cm}
\includegraphics[height=6cm,width=6cm,clip=false]{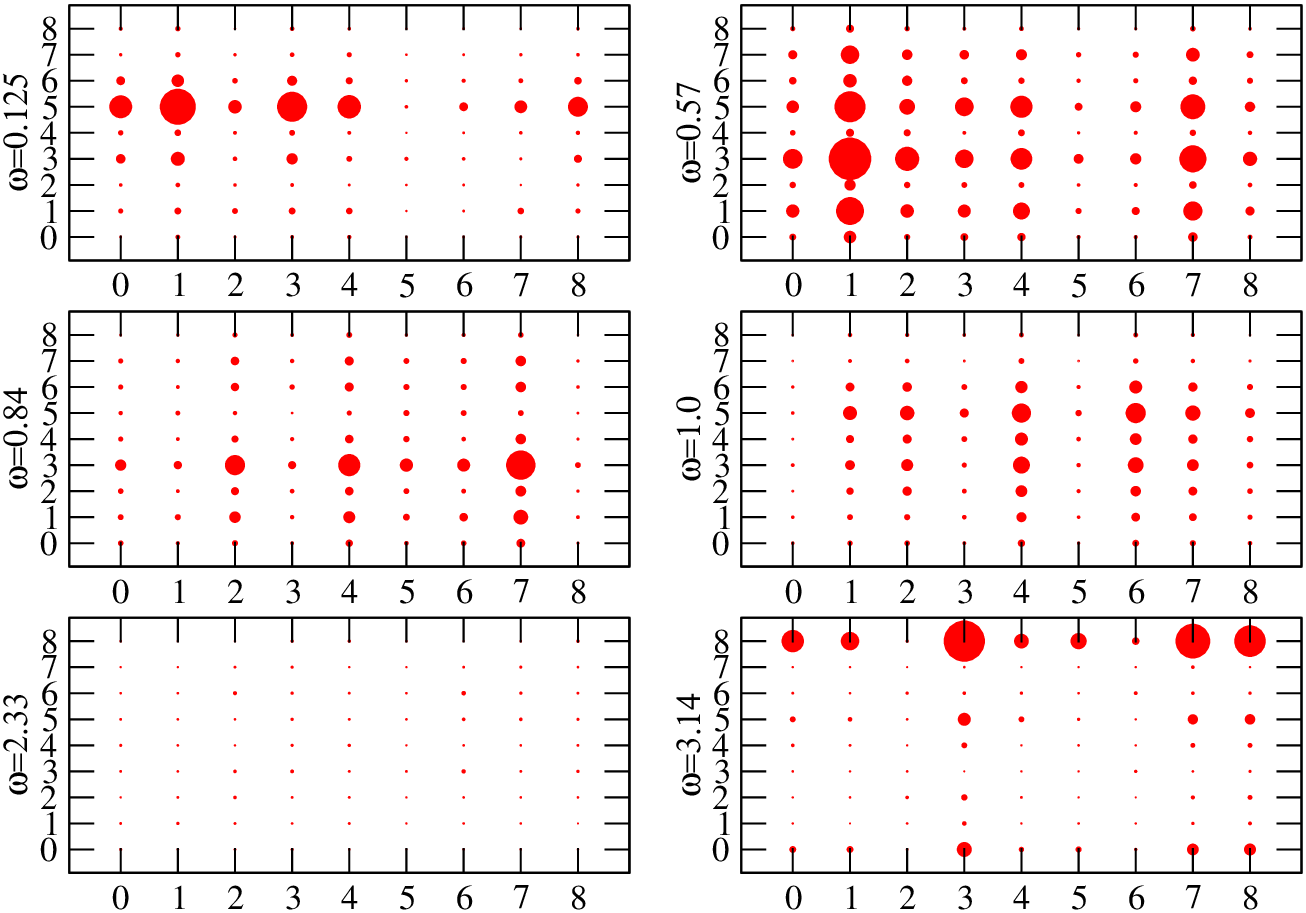}\\

\includegraphics[height=6cm,width=6cm,clip=false]{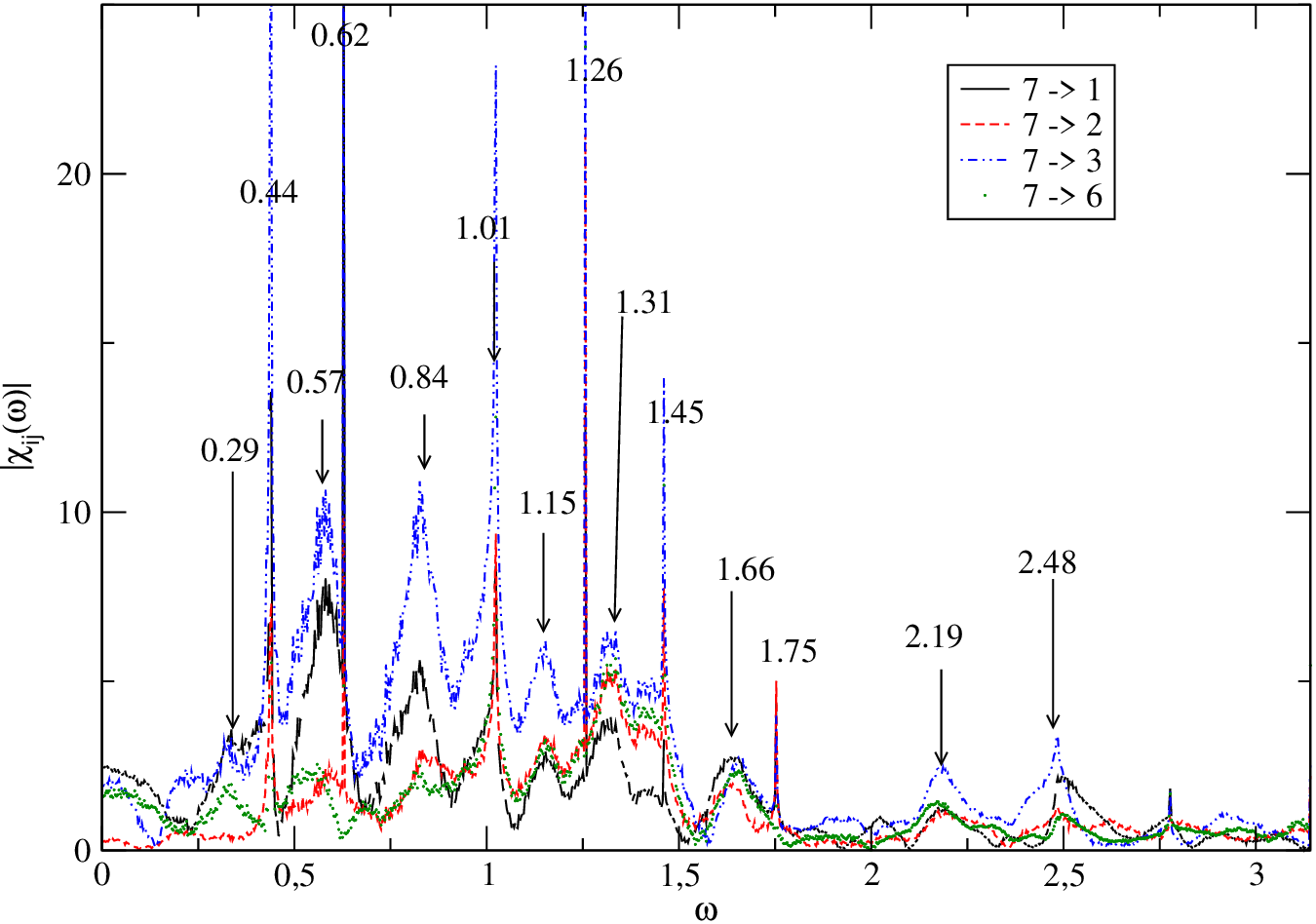}
\hspace{1cm}
\includegraphics[height=6cm,width=6cm,clip=false]{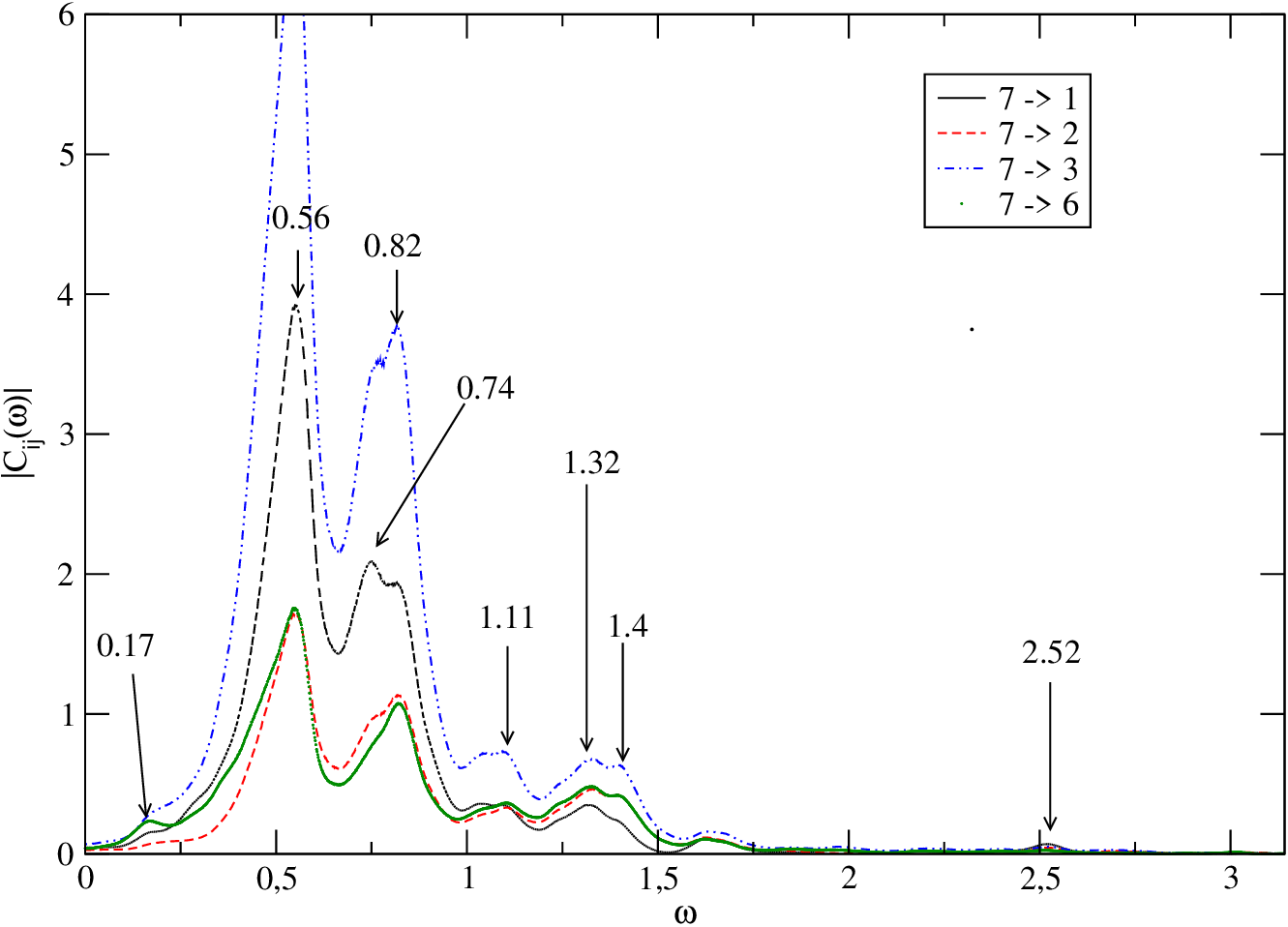}\\

\caption{\label{Fexres} 
\footnotesize{Fig. \ref{Fexres}a (left top) Connectivity matrix. 
Fig. \ref{Fexres}b (right top) Modulus of the susceptibilities for all pairs $(i,j)$ and several  frequencies $\omega$.
The area of the red circle is proportional to the modulus of the susceptibility.
Fig. \ref{Fexres}c  (left bottom) Modulus of the susceptibility for neuron $7$. 
Fig. \ref{Fexres}d.  (right bottom)  Modulus of the corresponding  Fourier transform of the correlations function. 
}} 
\end{center}
\end{figure}

The following case has been analysed in  \cite{cessac-sepulchre:06} for details. 
This is a sparse network  where each unit receives connection
from exactly $K=4$ other units. 
The corresponding network is drawn in Fig. \ref{Fexres}a. 
Blue stars correspond to inhibitory links
and red crosses to excitatory links. In this example the unit $7$ is a ``hub'' in
the sense that it sends links to most  units, while $0$, $2$, $3$ or $5$ send at most two links.
   
In figure \ref{Fexres}b, we have represented the modulus of the susceptibilities for all
pairs $(i,j)$ and different  frequencies $\omega$. This provides a notion of ``causal connectivity'',
related to linear response, which departs strongly from the connectivity provided by weights matrix.

Computing the susceptibility one obtains the curves  shown in Fig.\ref{Fexres}c.
Some resonance peaks are rather high ($\sim 20$) corresponding to 
an efficient
amplification of a signal with suitable frequency. It is also clear
 that the intensity of the resonance has no direct connection with the intensity or the sign
of the coupling and is mainly due to nonlinear effects. For example, there is no direct connection from $0$ to $3$ or $5$
but nevertheless these units react  strongly to a suitable signal injected at  unit $0$.
Let us now compare the Fourier transform of the correlations function $C_{ij}(t)$ for the same pairs
(Fig. \ref{Fexres} d).  One remarks that  these  functions exhibit  less resonance peaks. 
This is explained in the context of Ruelle's linear response theory and is related
to the decomposition of the linear response into stable and unstable contributions,
related to the local splitting of the tangent space (see \cite{cessac-sepulchre:04,cessac-sepulchre:06} for more details).  

%
%
%

\ssu{Conclusion.} The previous analysis leads then us to propose a notion of ``effective'', frequency dependent, connectivity
based on susceptibility curves. For a  given  frequency $\omega$, we plot the modulus of the susceptibility
$|\chi_{ij}(\omega)|$ with a representation assigning to each pair $i,j$ a circle whose size is proportional
to the modulus.  We clearly  see in figure \ref{Fexres} that changing the
 frequency changes the effective
network.

All these effects are due to a combination of topology and dynamics  and they
cannot be read in the connectivity matrix $\cW$. Therefore, the example
of neural networks treated in this section shows convincingly that the analysis
of neural circuits requires a careful investigation of the \textit{combined}
effects induced by non-linear dynamics and  topology of the synaptic
graph. It also shows that the analysis of correlations 
provides less information than a linear response analysis. This is particularly
clear when looking at the resonances curves displayed by linear response and correlations functions.
As we saw, this difference is well understood on theoretical grounds and has deep
relations with salient characteristics of the nonlinear dynamics (saturation in
the transfer function closely related to the refractory period). Using linear
response in neural networks is not new (see for
example \cite{rieke-etal:96} and references therein), but the point of view
 adopted in the present section, is, we believe, less known and
raises new interesting questions.

For example, one may wander what would bring this approach
in spiking neural networks, where the causal action from a
neuron to another can be somewhat ``directly'' read in the timing of pre- and post-synaptic 
neurons spikes.  Another
remaining question is what would the use of linear response analysis
tell us in neural networks having synaptic plasticity.
This issue is briefly addressed in the next section.  

\su{Dynamical effects of synaptic plasticity} \label{adapt}

\ssu{General context}

The notions of neural code and information cannot be separated from
the capacity that neuronal networks have to  evolve and adapt by \textit{plasticity}
mechanisms, and especially \textit{synaptic plasticity}.
 Therefore, understanding the effects of synaptic plasticity on neurons dynamics
is a crucial challenge. 
Especially, addressing the effect of synaptic plasticity in neural networks where dynamics is \textit{emerging} from collective effects and where spikes statistics are \textit{constrained} by this dynamics
seems  to be of central importance.  This issue is subject to two main difficulties.
On  one hand, one must identity the generic dynamical regimes displayed by a neural network
model for different choices of parameters (including synaptic weights). Some
examples of such analysis have been given in section \ref{genedyn}. On the other hand,
one must analyse the effects of varying synaptic weights when applying plasticity rules.
This requires to handle a complex interwoven evolution where neurons dynamics depends on synapses
and synapses evolution depends on neuron dynamics. 

\paragraph{Effects of synaptic adaptation} Three main classes of effects can be anticipated \cite{dauce-etal:98,siri-etal:07,siri-etal:08}.

\ben

\item \textbf{Structural effects.} There is a first, evident, effect
of synaptic plasticity: a rewiring of the neural network. However, this
rewiring is not some random process where edges would be selected or removed
independently of the history. Instead, it results from a complex
process where edges are potentiated or depressed according to the
neuron dynamics, which is itself depending on synaptic weights and external stimuli.
The question is therefore whether one can nevertheless extract some
general characteristics of the network structure evolution  and what is the impact
of this structure evolution on neural network behaviour.

\begin{figure}
\begin{center}
\includegraphics[height=4cm,width=6cm]{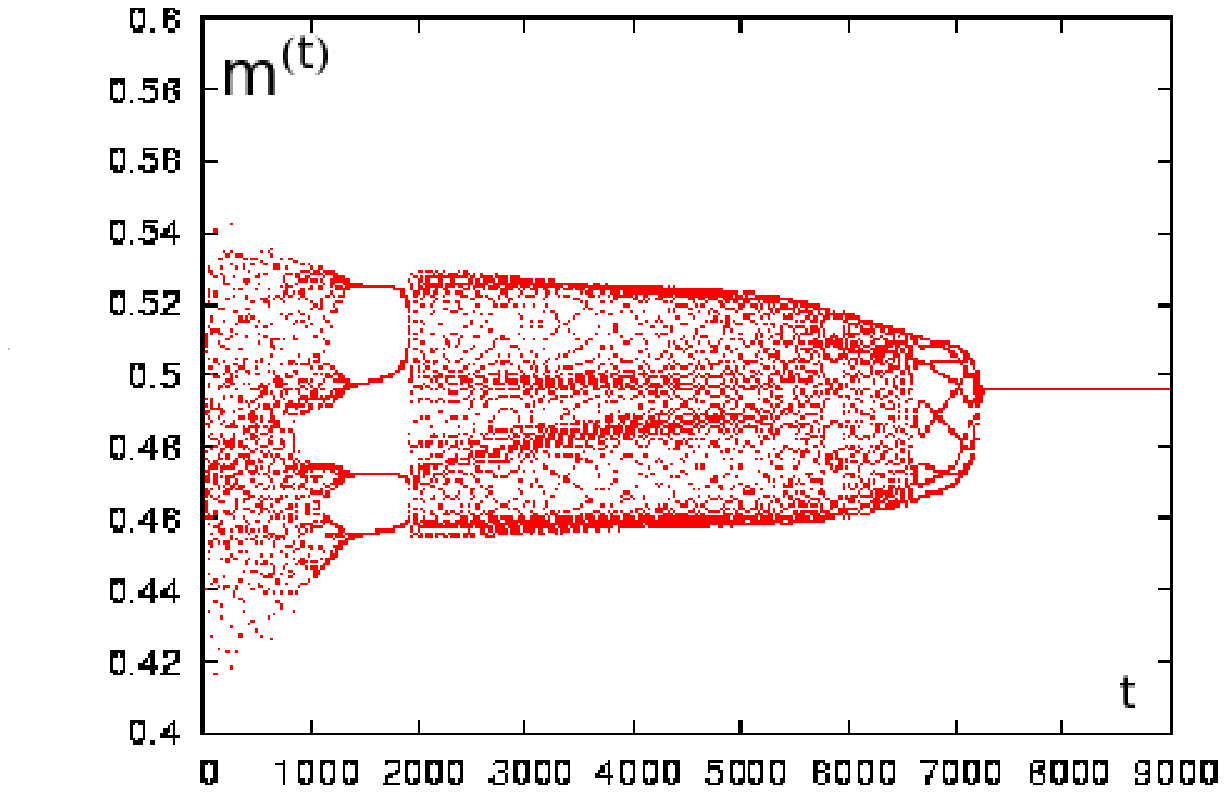}\\
\includegraphics[height=4cm,width=6cm]{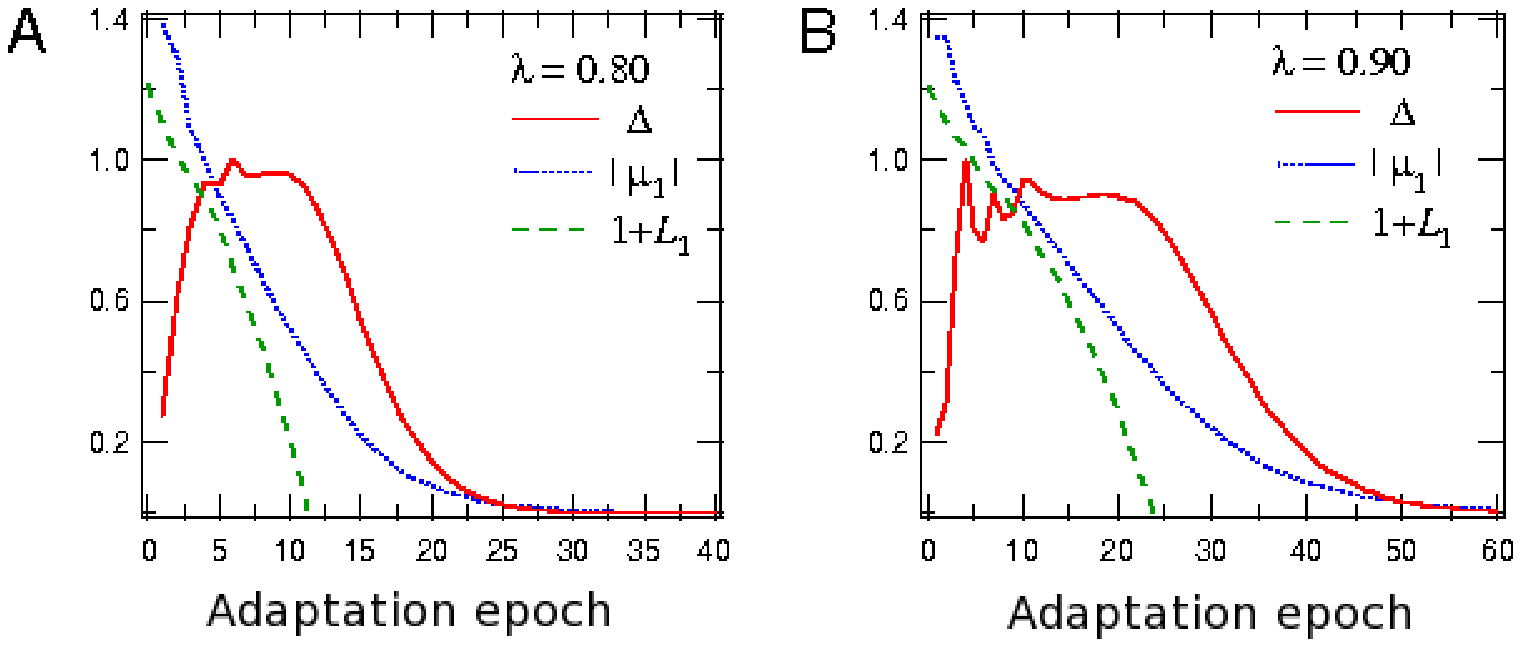}\\
\includegraphics[height=4cm,width=6cm]{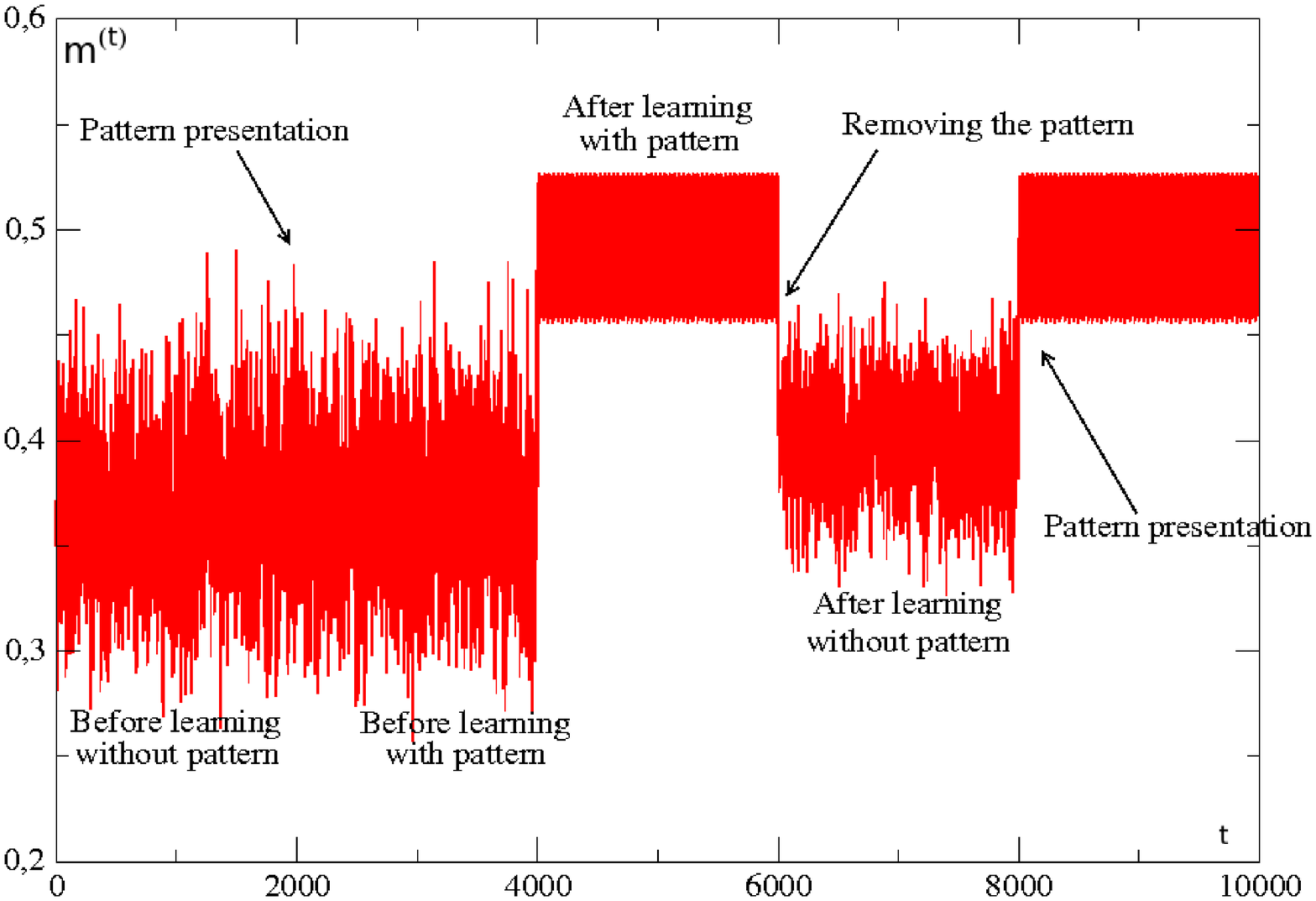}
\vspace{1cm}
\caption{\footnotesize \label{FHebb}
Fig. \ref{FHebb}a (left) Inverse quasi periodicity route induced by learning. 
The plotted quantity is  $m^{(\tau)}(t_0) \deq \frac{1}{N}\sum_{i=1}^N S(V_i(t_0))$,
where $t_0 \in [0,T]$ is fixed and where the adaptation epoch index $\tau$ varies.  Fig. \ref{FHebb}b (middle) Sensitivity to the learned input.
$\mu_1$ is the largest eigenvalue of $ \lb D\F\rb^{(\tau)}$ and $L_1$ is the largest Lyapunov exponent.
 Fig. \ref{FHebb}c (right) Dynamical effect of input presentation before and after adaptation. 
}
\end{center}
\end{figure}

\item \textbf{Dynamical effects.} Changing the synaptic weights, which
are parameters of the dynamical systems (\ref{CDNN}) and (\ref{DNN}), will obviously
have an incidence on dynamics. These effects can be smooth or sharp (bifurcations).
More generally, one expects period of smooth changes interrupted by sharp 
transitions (see fig. \ref{FHebb} for an example of this). Thus, adaptation drives the dynamical system along a \textit{definite}
path in the space parameters, which integrates the whole past, via synaptic
changes. In this respect,
we address a very untypical and complex type of dynamical system. This induces rich
properties such as a wide synaptic adaptation-induced \textit{variability} in the
network response to a given stimulus, with the same set of initial
synaptic weights, simply by changing the initial conditions. 

\item \textbf{Functional effects.} This evolution typically arises
when the system is submitted to inputs or stimuli which constrain 
neuron dynamics and thus synaptic evolution.  For simplicity,  let us think of synaptic
plasticity in the restricted context of ``learning'' some input. Learning should result in the acquisition
of a new ability. The network after learning should be able to ``recognise''
a learnt input, while this was not necessarily the case before
learning. In this context, recognition can be manifested by
a drastic change in the dynamics whenever a learnt input is presented.
Moreover, this effect must be robust and selective. 
Examples of this are presented now.
\een

\paragraph{Coupled dynamics}

 As an illustration we consider now the following coupled dynamics.
Neurons are evolving according to (\ref{DNN}) (we focus here on discrete time dynamics). 
We consider \textit{slow} synapses dynamics.  
Namely, synaptic weights are  constant for $T$  consecutive dynamics steps, where $T$ is large. This defines an
``adaptation epoch''. At the end of the adaptation epoch,  synaptic weights are updated according to (\ref{DSyn}). 
 This  has the consequence
of modifying neurons dynamics and possibly spike trains. The weights are then updated and a new adaptation
epoch begins. We denote by $t$ the update index of neuron states (neuron
dynamics) inside an adaptation epoch, while $\tau$ indicates the update index
of synaptic weights (synaptic plasticity). Call $\X^{(\tau)}(t)$
the state of the neurons  at time $t$ within the adaptation
epoch $\tau$ (we use here the notation $\X$ instead of $\V$ since eq. (\ref{Dcoupled})
holds for model having possibly more variables than the membrane potential for the definition of the neuron state).
  Let $W_{ij}^{(\tau)}$ be the synaptic weights 
from neuron $j$ to neuron at $i$ in the $\tau$-th adaptation epoch. 
At the end of each adaptation epoch, the neuron dynamics indexes are reset, and
$X_i^{(\tau+1)}(0)=X_i^{(\tau)}(T), i=1 \dots N$.
The coupled dynamics writes:

\beq\label{Dcoupled}
\left\{
\baR{ccc}
\XTtp&=&\FgT(\XTt) \\
\dWijT &\deq& \WijTp-\WijT=g\left(\WijT,\omeit,\omejt)\right)
\eaR
\right.
\eeq

Recall that $\bg=(\cW,\Ie)$ (see section \ref{neurasdyn}) and $\gT$ is the set of parameters at adaptation epoch
$\tau$. In the present setting the external current
 is used as a time constant \textit{external stimulus}.
We write it $\bxi$ (see footnote \ref{fxi}).

Let us discuss a few examples.

\ssu{Hebbian learning}\label{SHebb}

This example has been considered in \cite{siri-etal:07,siri-etal:08}.
The goal is to study the role of synaptic plasticity mechanisms inspired by Hebb's
work \cite{hebb:49} and its generalisation, in a situation where the neural network
is submitted to some specific stimulus over a long time. The main is to study
the conjugated effects of stimulus action and synaptic plasticity mechanisms on the neural network
dynamics. Specifically, we want to investigate whether these conjugated effects can lead 
the system to a state where it acquires some ability to ``recognise'' this stimulus.
In the example developed below, this corresponds to drive the system, via an Hebb's-inspired
modification of the synaptic weights, in a region in the parameters space where presenting
the stimulus induces a bifurcation in the dynamics whereas this bifurcation didn't occur
before synaptic adaptation.

Let us before briefly explain what we mean by ``Hebbian learning''.  
D. Hebb has proposed in \cite{hebb:49} a theory of behaviour based on the physiology of the nervous system.
The most important concept to emerge from Hebb's work was his formal statement (known as Hebb's rule) 
of how learning could occur. \\

\textit{When an axon of cell A is near enough to excite a cell B and repeatedly or persistently takes part in firing it, some growth 
process or metabolic change takes place in one or both cells such that A's efficiency, as one of the cells firing B, is increased. 
}\\

Many ``learning rules'' in neural networks are based on Hebb's observations plus a few well established facts.
They rely upon a few recipes that can summarised as \cite{hoppensteadt-izhikevich:97}:

\bit

\item Learning results from modifying synaptic connections between neurons.

\item Learning is local i.e. the synaptic modification depends only upon the pre- and post- synaptic
neurons activity and does not depend upon the activity of the other neurons.

\item The modification of synapses is slow compared with characteristic times of neuron dynamics.

\item If either pre- or post- synaptic neurons or both are silent then no synaptic change takes place
except for (exponential) decay which corresponds to forgetting.  
\eit

The first item
implies that learning results in a modification of the $W_{ij}$'s. The
second one basically says that the synaptic modification of $W_{ij}$ writes
$W_{ij}'=\epsilon h(W_{ij}^T,m_j,m_i)$ where $W_{ij}'$ is the value of the 
synapses $j \to i$ after the learning rule has been applied. 
The numbers $m_i$ ($m_j$) 
denotes the ``state'' or ``activity''
of the neuron $i$ ($j$). How this ``state'' is defined vary according
to the model. 
 The third item  implies then that 
$\epsilon$ is a  small parameter, whose inverse corresponds to the characteristic time for a significant
change of $W_{ij}$.  
If one assumes that $h$ is a smooth function then one may simply 
consider a Taylor expansion of a generic regular function $h$. This gives, up
to the second order in $m_i,mj$.

$$W_{ij}'=\epsilon \left(a_{000}+ a_{100}W_{ij}+ a_{010}m_j + a_{001}m_i + 
a_{011}m_ i m_j + h.o.t. \right)$$

\nid where h.o.t. means ``higher order terms'' such as $W_{ij}m_i m_j$, etc....
Then, the fourth item implies $a_{000}=0$ and leads to introduce a parameter
$\lambda=\epsilon a_{100} \in [0,1]$ that models passive ``forgetting'': if a synapse is not solicited 
its intensity decreases with a decay rate $\frac{1}{\lambda}$.

\sssu{Coupled dynamics.} 

On these bases, consider  the model (\ref{SDNN}) where synaptic weights evolve according
to\footnote{For another implementation of Hebbian rule see eq. (\ref{HebbCorr}).}:

\beq\label{DW}
\WijTp=\lambda
\WijT+\frac{\alpha}{N} \sum_{j=1}^N m_i^{(\tau)} m_j^{(\tau)}H\left[m_j^{(\tau)}\right], 
\eeq 

\nid where $\alpha$ is a small number, controlling the rate of synaptic plasticity.
Here, one associates to the history of neuron $i$'s rate an activity index
$m_i^{(\tau)}$: 

$$m_i^{(\tau)}=\frac{1}{T}\sum_{t=1}^T (\nu_i^{(\tau)}(t) - d_i)$$
 where $\nu_i^{(\tau)}(t)$ is the firing rate of neuron $i$ at time $t$ in the
adaptation epoch $\tau$,  $d_i \in [0,1]$ is a threshold.
The neuron is considered
active during synaptic adaptation epoch $\tau$ whenever $m_i^{(\tau)}>0$, and silent
otherwise. Finally, $H(x)$ is the Heaviside function.

Let us now interpret this equation. The first term is conform to the recipes introduced in the previous
section. The second term is positive whenever neuron $j$ and $i$ are active, increasing the synapse efficacy
$W_{ij}$ (i.e. render it more positive if is excitatory and less negative if it is inhibitory).
If $j$ is active and $i$ inactive the synapse efficacy decreases. Finally, if $j$ is inactive
the second term is zero. We emphasise that this  one possible implementation among many others. \\

Equations (\ref{SDNN}) \& (\ref{DW}) define a dynamical system
where two distinct processes (neuron dynamics and synaptic network
evolution) interact with distinct time scales. This results in a complex
interwoven evolution where neuronal dynamics depends on the synaptic
structure and synapses evolve according to neuron activity. On general
grounds, this process has a memory that is \textit{a priori} infinite
and the state of the neural network depends on the past history.

\sssu{Observed effects of Hebbian synaptic plasticity}

The effect of this Hebbian synaptic adaptation has been explored numerically
in \cite{dauce-etal:98} and mathematically in \cite{siri-etal:08}. 
Assume that the initial synaptic weights are chosen \textit{independently}, at random, with
a Gaussian distribution of mean $\frac{\Wb}{N}$ and variance $\frac{\sigma^2}{N}$.
This choice mimics a situation where no
structure is imposed a priori in the correlations between  synaptic weights.
The idea is to see how synaptic adaptation changes this situation. 
Then, according to section (\ref{bifmeanfield}) dynamics is chaotic provided the gain $g$
of the sigmoid $S$ is large enough. Starting from such a chaotic dynamics,
the following effects have been observed, in correspondence with the three effects
anticipated in the beginning of this section.

\paragraph{Structural effects.} The rewiring of the network by the synaptic adaptation rule (\ref{DW})
reveals a variation in the synaptic weights distribution, and an increase in weights correlations. This
effect is evident but does not give significant hints to interpret the dynamical and functional effects
described below \cite{siri-etal:08}. Also, a computation of standard indicators in complex
graphs analysis, when applied to the synaptic weights matrix, does not show any salient effect.
The only important hint provided by synaptic weights matrix analysis is an increase in the number
and weights of positive feedback loops\footnote{If $e$ is an edge, denote by $o(e)$ the origin of the
edge and $t(e)$ its end. Then a feedback loop (or circuit) is a sequence of edges $e_1, . .
. ,e_k$ such that $o(e_{i+1}) = t(e_i)$, $\forall i = 1 . . . k -1$, and
$t(e_k) = o(e_1)$. Such a circuit is positive (negative) if the product
of its edge's weight is positive (negative).}, which renders the system more cooperative, with 
a strong impact on dynamics\footnote{In short, Hirsch \cite{hirsch:89} showed that cooperative systems, characterised
by the property $DF_{ij}(\X) \geq 0, \, \forall i,j, \, \forall \X \in \cM$, where $DF$ is the Jacobian
matrix, are
convergent. As a matter of fact, this result
holds for Jacobian matrices instead of synaptic weights matrix. But, in the particular
example (\ref{SDNN}) the entry $DF_{ij}$ of the Jacobian matrix
is given by $W_{ij}S'(V_j)$. Thus, it is proportional to $W_{ij}$ with a positive factor $S'(V_j)$.} \cite{hirsch:89}. As suggested in the section \ref{network}
the analysis of synaptic weights matrix is indeed not expected to provide deep insights 
on dynamical effects. On the opposite the analysis of Jacobian matrices reveals important
properties. This is not surprising. A standard procedure for the analysis of
nonlinear dynamical systems starts with a linear analysis. This
holds e.g. for stability and bifurcation analysis but also for the
computation of indicators such as Lyapunov exponents.  The key object for
this analysis are  Jacobian matrices. Moreover, as we saw in the previous section,
Jacobian matrices and their generalisation, the linear response,
 generate a  graph structure that can be interpreted in
causal terms. 
 
\paragraph{Dynamical effects.} As a corollary in the increase of positive feedback loops
it is observed that Hebbian synaptic adaptation leads to a \textit{systematic reduction of the dynamics
complexity} (transition from chaos to fixed point by an inverse
quasi-periodicity route, see fig. \ref{FHebb}a). As a corollary the largest Lyapunov exponent  $\lmT$, which
depends on the synaptic adaptation epoch $\tau$,
decreases from positive to negative values.  Two main effects contribute to this
decay. The first effect is due to passive LTD term in (\ref{DW}).
The second one is related to an increase in the level of neurons' saturation.
Basically, cooperativity between neurons has a tendency to either render
them more silent, or more active. In both case, this ``pushes'' them 
to the saturated part of the sigmoidal transfer function, reducing the average value of
$S'(V_i)$. Since the entry $ij$ of the Jacobian matrix, $DF_{ij}(\V)=W_{ij}S'(V_i)$
an increase in the saturation of neurons has the effect of decreasing the spectral
radius of $D\F(V)$ with a computable impact on the maximal Lyapunov exponent.

\paragraph{Functional effects.}

This property has been exploited for pattern
retrieval. Label by $\V$  the neuron state  when the (time constant) input (external current) $\XI$ is applied to the
network (see eq. (\ref{SDNN})) and by $\V'$ the neuron state without $\XI$.
  The removal of $\XI$ modifies the attractor structure and the average
value of observables. More precisely, let $\phi$ be some suitable function and
 call:

\beq\label{Deltatauphi}
\Delta^{(\tau)}\left[\phi\right]=\lb\phi(\V')\rb^{(\tau)}
-\lb\phi(\V)\rbT
\eeq
\nid where $\lb\phi(\V')\rb^{(\tau)}$ is the (time or SRB) average value of $\phi$
without $\XI$ and $\lb\phi(\V)\rbT$ the average value in the presence of
$\XI$. Two cases can arise.\\

In the first case, the system is away from a bifurcation point and
removal results in a variation of $\Delta^{(\tau)}\left[\phi\right]$ that
remains proportional to $\XI$, provided $\XI$ is sufficiently small.  In the
present context, the linear response theory (see section \ref{network}) predicts that the variation
of the average value of $\V$ is given by\footnote{We consider here the case $\phi(\V)=\V$
for simplicity. For a general (differentiable) function $\phi$, the corresponding formula
is \cite{ruelle:99}:

$$\Delta^{(\tau)}\left[\phi\right]=-\sum_{n=0}^{+\infty} \lb D\F^n\nabla \phi\rb^{(\tau)} \XI $$

 }
 \cite{cessac-sepulchre:06,cessac-sepulchre:07}:
\beq \label{RepLin}
\Delta^{(\tau)}\left[\V\right]= -\chi^{(\tau)} \XI
\eeq
\nid where
\beq\label{chiTmat}
\chi^{(\tau)}=\sum_{n=0}^\infty \lb D\F^n\rb^{(\tau)}
\eeq
is a matrix whose entries are given by eq. (\ref{chiij})
in section \ref{network}.
Note therefore that $\Delta^{(\tau)}\left[\V\right]=-\XI - M^{(\tau)}\XI$
where the matrix $M^{(\tau)}=\sum_{n=1}^\infty \lb D\F^n\rb^{(\tau)}$
integrates dynamical effects. The application of $\XI$
implies a reorganisation of the dynamics which results in a complex
formula for the variation of $\lb \V \rbT$, even if the dominant term is
$\XI$, as expected. More precisely, as emphasised several times above,
one remarks that each path in the sum $\sum_{\gamma_{ij}(n)}$ is
weighted by the product of a topological contribution depending
only on the weights $W_{ij}$ and on a \textit{dynamical} contribution.
The weight of a path $\gamma_{ij}$ depends on the average value of $\lb
\prod_{l=1}^{n}f'(u_{k_{l-1}}(l-1))\rb^{(\tau)}$ thus on
\textit{correlations} between the state of saturation of the units $k_0,
\dots, k_{n-1}$ at times $0, \dots, n-1$.

In the second case the system is close to a bifurcation point and the presentation/removal
of the input induces a sharp variation (bifurcation) in dynamics. 
In this case, eq. (\ref{RepLin}) does not apply anymore.
We expect therefore input presentation/removal
to have a maximal effect close to bifurcations points.
This can be revealed by studying the quantity $\Delta^{(\tau)}\left[\phi\right]$ in eq. (\ref{Deltatauphi})
for the observable $\phi(V_i)=S'(V_i)$, which measures the level of saturation of neuron 
$i$ in the state $V_i$. 
Indeed, this quantity  is maximal when $V_i$
is in the central  part of the
sigmoidal transfer function,  where neuron $i$ is the most sensitive to 
small variations. Hence this quantity, called $\dtl$, measures how neuron excitability is modified when
the input is removed. The evolution of $\dtl$ during learning
following rule eq. (\ref{DW}) is shown on fig.~\ref{FHebb}b
(full lines) for two values of the passive forgetting rate $\lambda$.
$\dtl$ is found to increase to a maximum at early learning epochs, while
it vanishes afterwards. Interestingly, comparison with the decay of the
leading eigenvalue $\mu_1$ (dotted lines) of the average Jacobian matrix $\lb D\F\rb^{(\tau)}$  shows
that the maximal values of $\dtl$ are obtained when
$|\mu_1|$ is close to $1$. This also corresponds to a vanishing
of the maximal Lyapunov exponent.
Hence, these numerical
simulations confirm that sensitivity to input removal is maximal when
the leading eigenvalue is close to $1$. Therefore, \emph{``Hebb-like''
learning drives the global dynamics in a region of the parameters space
were sensitivity to the stimulus, manifested by a drastic change in the average of functions hence
by a drastic change in the underlying dynamics,   is maximal.} This
property may be crucial regarding memory properties of recurrent neural networks, which must
be able to detect, through their collective response, whether a learnt
input is present or absent. This property is obtained at the frontier
where the strange attractor begins to destabilise ($|\mu_1|=1$), hence
at the so-called ``edge of chaos''.

Note that continuing adaptation after this phase of highest sensitivity, leads
to a decay of sensitivity and to a stabilisation of the system to a fixed point
(fig. \ref{FHebb}a). This is not surprising. Insisting too much on adaptation
to this stimulus drive the system to a state where activity pattern is essentially identical
to the input, with no room any more for spontaneous activity.

%
%
%
%
%
%
%

\paragraph{Conclusion.} In fig. \ref{FHebb}c we have represented the effect of presenting/removing the input
before adaptation, and after adaptation, in the region where the reactivity is maximal.
Clearly, the presentation of $\XI$ after adaptation induces a sharp transition in dynamics, which
is not occurring before adaptation. Moreover, this effect, inherited via learning,
 is robust to a small amount of noise, and selective (it does not occur for drastically different patterns) \cite{dauce-etal:98}.
Though established in the context of a rather simple ``neural'' model, these results
raise interesting questions and comments. They suggest that adaptation corresponds to some
path in the space of parameters of (\ref{SDNN}) leading the system in a region were it
\textit{acquires} sensitivity to the input it has adapted to, this sensitivity being
manifested by sharp and rapid variations in neurons dynamics. 
An obvious question is does this effect generalise to more complex (and realistic)
architecture ? This is an ongoing research field. 

\ssu{Effects of synaptic plasticity on spike trains statistics.}\label{plaststat}

Synaptic plasticity  acts also on the statistics of spike trains.
Let us now briefly mention recent works where the effects of plasticity
on spike train statistics is analysed. 

\paragraph{Synapses update as an integration over spikes trains.} Let us reconsider the equation (\ref{DSyn})
for synaptic weights dynamics. The synaptic variation $\dWij$ is the integrated response of the synapse from neuron $j$ to neuron $i$
when neuron $j$ sends a spike sequence $\omejt$ and neuron $i$ fires according to $\omeit$. This
response is not a deterministic function, while (\ref{DSyn}) is deterministic.
As a matter of fact, the explicit form of $g$ 
is usually derived from phenomenological considerations
as well as experimental results where synaptic changes can be induced by \textit{specific} simulations conditions,
defined through  the firing frequency of pre- and post-synaptic
neurons \cite{bliss-gardner:73,dudek-bear:93}, the membrane potential of the post-synaptic
neuron \cite{artola-etal:90}, spike timing \cite{levy-stewart:83,markram-etal:97,bi-poo:01}
(see \cite{malenka-nicoll:99} for a review). 
 Thus, these results are usually based
on a repetition of experiments involving the excitation
of pre- and post-synaptic neurons by specific spike trains. The phenomenological
plasticity rules derived from these experiments are therefore 
of \textit{statistical} nature. Namely, they do not tell us what will be the exact changes
induced on synapses when this or this spike train is applied
to pre- and post-synaptic neuron. Instead, they provide us the average synaptic change. 
 Thus, the function
$g(W_{ij},\omeit,\omejt)$ in (\ref{DSyn}) is typically a statistical average of the synaptic response
when the spike train of neuron $j$ (resp. $i$)
is $\omejt$ (resp. $\omeit$), and the actual synaptic weight value is $W_{ij}$.

\paragraph{Slow synaptic update.} On this basis let us assume that $g$
is obtained via a \textit{time average} $\pTo$ (see eq. (\ref{pTo}) in section \ref{SpikesStat}) 
of some function $\phi$ having a form depending on the type of ``rule''considered (e.g. Hebbian or STDP).
Namely, $g$ has the form:

\beq\label{gcondensed}
g\left(W_{ij},\omeit,\omejt \right) \equiv\epsilon\pTo\left[ \phi_{ij}(\Wij,.) \right].
\eeq

\nid where $\epsilon$ is a  parameter that will be typically small.
In general $\phi_{ij}$ can be expanded in terms of singlets, pairs, triplets, etc
of spikes \cite{gerstner-kistler:02,cessac-rostro-etal:09}.

\paragraph{Statistical effects of synaptic plasticity}
The coupled dynamics (\ref{Dcoupled}) 
 has an impact on the set of admissible raster plots  and
 the spikes train statistics. Typically, the
 empirical average constructed via the  raster plot $\omega^{(\tau)}$
in the adaptation epoch $\tau$, changes from $\pToT \to \pTpoT$.
Thus, the adaptation dynamics results in
a sequence of  empirical measures $\left\{\pToT \right\}_{\tau=1}^\infty$ and
 the corresponding statistical model also evolves.
 Let us characterise this evolution using the tools introduced
in section \ref{SpikesStat}. 
The main idea is to  make the assumption that each $\pToT$ can be approximated by a Gibbs
measure $\npT$ with potential $\psiT$. Then synaptic adaptation writes:
\beq \label{dWij}
\dWijT  = \epsilon\npT\left[\phi_{ij}(\WijT,.) \right].
\eeq

The synaptic update results in a change of parameters $\bg$, $\gTp=\gT+\delta \gT$
where $\delta \gT$ is assumed to be small (this is the role of the constant
$\epsilon$ in (\ref{DSyn})). This induces  variation in statistical properties
of raster plots (e.g. the topological pressure (\ref{pres}) -see \cite{cessac-rostro-etal:09} for details).
These variations can be smooth or not. 

\paragraph{Smooth variations} If they are smooth one can
show that there exist a function $\FpT(\cW)$ such that the adaptation rule (\ref{dWij}) can be written in the form:

\beq \label{dWgrad}
\dWT= \epsilon \nabla_{\tiny{\cW=\cWT}} \FpT(\cW).
\eeq

 Moreover, this quantity decay under synaptic adaptation. Thus,
the adaptation rule (\ref{dWgrad}) is a \textit{gradient} system where the function $\FpT$ decreases
when iterating synaptic adaptation rules.
Were the transition $\tau \to \tau+1$ to be smooth for all $\tau$, would $\FpT$ reach a minimum\footnote{In implementing
synaptic update rule, one adds conditions ensuring that weights do not diverge. This condition
ensures that $\FpT$ is bounded from below, $\forall \tau$} at some $\cW^\ast$ as $\tau \to \infty$.
Such a minimum corresponds to $\nabla_{\tiny{\cW^\ast}} \FpT=0$, thus to 

$$\dWij^\ast=\mpgs\left[\phi_{ij}\right]=0, \quad, \forall i,j=1 \dots N,$$

\nid according to eq. (\ref{dWgrad}). Hence, this minimum
corresponds to a \textit{static distribution} for the synaptic weights. 

\paragraph{Static synaptic weights.}\label{static}

Since this imposes  a condition on the average value of the $\phi_{ij}$'s,
 this imposes as well the statistical model as a \textit{Gibbs measure $\mpg$ with a  potential}:

\beq\label{psis}
\bpsis=\Phi+\bl^\ast.\bphi,
\eeq 

\nid where $\bpsis=\left(\psi^\ast_{ij} \right)_{i,j=1}^N$. The potential $\Phi$ in (\ref{psis}) is such that  $\Phi(\tom)=0$ is $\tom$ is admissible and
$\Phi(\tom)=-\infty$ if it is forbidden, so that forbidden raster plots have zero probability.
We use the notation $\bphi=\left(\phi_{ij} \right)_{i,j=1}^N$, $\bl^\ast=\left(\lambda^\ast_{ij} \right)_{i,j=1}^N$ 
and $\bl^\ast.\bphi=\sum_{i,j=1}^N \lambda^\ast_{ij}\phi_{ij} $. The statistical parameters
$\lambda_{ij}^\ast$, are given by eq. (\ref{Gener}) in section \ref{Statmod}.

As a conclusion, the statistical model, in the sense of section \ref{Statmod},
 is a Gibbs distribution such that the probability
of a spin block $R$ of depth $n$ obeys eq. (\ref{PRS}) with a potential $\bpsi^\ast$
(see eq. (\ref{PRSSTDP}) for an explicit form.))
When this corresponds to the asymptotic state for a synaptic adaptation process,
this potential provides us the form of the statistical model \textit{after adaptation}, and \textit{integrates all past 
changes in the synaptic weights}.

Moreover, it has a deep implication. Since the Gibbs distribution obeys the variational
principle (\ref{pres}) with $\mpgs[\bphi]=0$ , the probability distribution $\mpgs$ \textit{has maximal entropy}.  
In other words, synaptic adaptation rules of type (\ref{dWij}), when they converge,
 drive the system toward a dynamics where the statistical entropy of spike train
is maximal, taking into account the constraints\footnote{We mean that, as emphasised several
times in the paper, dynamics is not able to produce all possible spikes trains. Henceforth,
statistical entropy must be maximised on a subset of spike trains which can be relatively small
compared to the whole set of possible spike trains ($2^{NT}$ possible spike trains
of length $T$ for a system of $N$ neurons). Note also that ``maximal entropy'' does not
mean ``equi-probability'' here.} imposed by dynamics.

\paragraph{Singular variations.}\
The synaptic weights variations can induce a change in the set of admissible raster plots
that the system is able to display.
When this happens the set of admissible raster plots is suddenly
modified by the synaptic adaptation mechanism. Formerly forbidden sequences become
allowed, formerly allowed sequences become forbidden, but also a large core of legal 
sequences may remain legal. These changes depend obviously on the detailed form 
of neuron dynamics (\ref{DNN}) and of the synaptic update mechanism (\ref{DSyn}).
An interesting situation occurs when the set of admissible
raster plots obtained after adaptation belongs to $\SgT \cap \SgTp$.
In this case, adaptation plays the role of a \textit{selective mechanism}
where the set of admissible raster plots, viewed as a neural
code, is gradually reducing, producing
after $n$ steps of adaptation a 
set $\cap_{m=1}^n \Sgm$ which can be 
rather small. 
If we consider the situation where (\ref{DNN}) is
a neural network submitted to some stimulus, where
a raster plot $\tom$ encodes the spike response to the stimulus,
 then $\Spg$ is the set of all
possible raster plots encoding this stimulus.
Adaptation results in a reduction of the possible
coding, thus reducing the variability in the possible
responses.

\paragraph{Example: Spike Time Dependent Plasticity.}
As an example we consider a neural network of type (\ref{DgIF})  with an adaptation rule  inspired from (\ref{STDPFC})
with an additional term $\ld \WijT$, $-1 <\ld <0$, corresponding to passive
LTD.

\beq\label{Rexample}
\dWijT=\epsilon
\left[\ld \WijT+  \frac{1}{T} \sum_{t=T_s}^{T+T_s} \omjt(t) \sum_{u=-T_s}^{T_s} f(u) \, \omit(t+u)\right],
\eeq

\nid where $f(x)$ is given by (\ref{fSTDP}) and with:

$$T_s \deq 2 \max(\tau_+,\tau_-).$$

Set :

\beq\label{phiij_ex}
\phi_{ij}(\Wij,\tom)=\ld \Wij+\omega_j(0)\sum_{u=-T_s}^{T_s} f(u) \, \omei(u),
\eeq 
then (\ref{Rexample}) has the form (\ref{gcondensed}),
$\dWijT=\epsilon \pToT\left[\phi_{ij}(\Wij,.) \right]$.

\paragraph{Static weights.}

Thanks to the soft bound term $\ld\Wij$ the synaptic adaptation rule admits a static solution given by:

\beq\label{Wstatex}
\Wij=-\frac{\sum_{u=-T_s}^{T_s} f(u) \pToT\left[\omega_j(0)\omei(u)\right]}{\ld}.
\eeq

The sign of $\Wij$ depend on the parameters $A_-,A_+,T_s$, but also on the relative
strength of the terms $\pToT\left[\omega_j(0) \, \omei(u)\right]$. 
Note that  this equation may have several solutions.

\paragraph{Spike train statistics in a static weights regime.}

 As emphasised  in section \ref{Statmod} and \ref{static}, when the synaptic
adaptation rule converges to a fixed point, the corresponding 
statistical model is a Gibbs
measure with a potential

$$\bpsis=\Phi +\sum_{i=1}^N\sum_{j=1}^N \lambda^\ast_{ij}\phi_{ij},$$ 

\nid where the value $\lambda_{ij}^\ast$ of the Lagrange multipliers
 is constrained by the
relation (\ref{Wstatex}). Henceforth,
the probability of an \textit{admissible} spike bloc $R$ is given by:

\beq\label{PRSSTDP}
P\left[R|S \right]= 
\frac{1}{Z_n\left[\lambda^\ast_{1,1}(S),\dots,\lambda^\ast_{N,N}(S)\right]}
\exp\left[\sum_{i=1}^{N}\sum_{j=1}^N \lambda^\ast_{ij}(S) 
\sum_{t=0}^{n-1} \sum_{u=-T_s}^{T_s}  f(u) \omej(t)\omei(u+t)\right].
\eeq

\nid Note that the term $\ld \Wij$ arising in the definition of the potential
can be removed thanks to the normalisation constant $Z_n\left[\lambda_1(S),\dots,\lambda_l(S)\right]$.

\paragraph{Conclusion}
At the end of section \ref{SpikesStat} we were asking
whether it is possible to propose a canonical form 
for spike trains statistical models relying on some generic principle ?
Here, we have exhibited an example where such a construction can be made. 
Our argumentation suggests
that  Gibbs measures may be good statistical models for
a neuron dynamics resulting from slow adaptation rules where
synaptic weights converge to a fix value. In this case,
synaptic weights contains the whole history of the neural network,
expressed in the structure of the generating function of cumulants
(the topological pressure) and in the structure of allowed/forbidden
raster plots (potential $\Phi$ in (\ref{psis})). This theoretical results, confirmed by numerical
experiments \cite{cessac-rostro-etal:09}, can be compared with
the recent paper of Schneidman and collaborators, already quoted
in section \ref{SpikesStat}, proposing a Gibbs distribution with
Ising like potential to match empirical data on the salamander retina \cite{schneidman-etal:06}.
Our approach suggests that Gibbs distribution could be ubiquitous and that
the corresponding potential can be inferred according to the plasticity
mechanisms at work, defining the ``rule''.

\su{Conclusion} \label{conclusion}

In this paper we have presented a series of results
related to questions, that, we believe, are central
in the computational neuroscience community.
These questions, when addressed from the point of
dynamical systems theory, shed new light on neural
network dynamics with possible outcomes
towards experimentation. Interestingly enough, the tools and concepts
used here however are not restricted to the field of neural networks
but could be applied to other type of ``complex'' systems.

We now would like now to point out some ``challenging'' points raised in this
paper, which, to our opinion, are central at the current state of the art,
in the neuroscience community, especially for those people who want to
use models and their analysis to
``understand some fundamental keys ruling the behaviour of neural networks''.
We also believe that some of these questions address also to experimentalists community.

\paragraph{Finite size corrections.} When dealing with large populations of networks,
theoretical methods such as mean-field approaches, neural mass models, large deviations, 
use a limit $N \to \infty$ where $N$ is the number of neurons. When dealing with ``concrete''
neural systems the number is finite and finite size systems can have a behaviour that
departs strongly from the limiting system. Computing these finite size corrections
is an open problem in the whole scientific community (not only neuroscience).

\paragraph{Finite time corrections.} A similar question holds for finite time effects.
While many theoretical methods assume stationarity in the data, concrete experiments
handle  non stationary or transient dynamics. There exists currently empirical methods
to tackle this problem such as sliding time windows. But, to the best of our knowledge
there are rather few methods (i)  to estimate the width of this window which must be large
enough to ensure reliable statistical estimates and smaller than relevant characteristic
time scales of the non-stationary dynamics; (ii) to define a priori the statistical models (resp. the set
of observable) used to characterize dynamics; (iii) to propose and compute reliable indicators that guarantee the
  liability of the result.  

\paragraph{Correlated weights.}  Synaptic weights are non independent in  real
neuronal networks since e.g. plasticity build correlations, that can have long range in space and time.
 How to characterize these correlations ? How to measure them ?
How to handle them in a model ? For example, the mean-field approaches
developed in section \ref{meanfield} cannot be extended in a straightforward
way to this case. Here again there is a need to invent new theoretical methods.
Promising results using spectral properties of graphs  \cite{jost-joy:02,atay-biyikoglu-etal:06,jost-jalan-etal:06,afraimovich-bunimovich:07,blanchard-volchenkov:09} combined to such linear responses methods as developed
in section \ref{network}
could provide a breakthrough.

\paragraph{Continuous time.} Most phenomenological models of neurons, such as Hodgkin-Huxley's,
use a description with ordinary differential equations assuming a continuous time. 
This comes from the description of neurons in terms of membrane potential, a notion
which corresponds to an integration over space $dx$ and time scale $dt$ where
classical equations of electrodynamics holds (Kirchhoff and Ohm's law). Also,
as we saw, Hodgkin-Huxley's equations for example, corresponds to integrating
the activity of ionic channel at space and time scales so that the notion of probability  
of open/closed channel have a meaning and, moreover, so that the Markovian approach used
for the master equations (\ref{HHn},\ref{HHm},\ref{HHh},) in section \ref{Models} holds. 
In other words, writing neurons
dynamics at the scale of ionic channel would requires the use of a different physics. 
Though this remarks looks obvious it prevents one to take the continuum limits
$dx \to 0$ or $dt \to 0$ without caution. On the opposite, the ``spike'' description of neurons
is discrete and assume a time discretisation $\delta$ (see section \ref{Models}) such that
a given neuron can produce, at most, one spike within this time delay. This time discretisation
is essential for the definition of raster plots, as we saw. The existence of a minimal time
scale $\delta$ can be defended using arguments from physics and neuronal characterisation (see
\cite{cessac-vieville:08}). However, time discretisation as done in section \ref{Models}
for the definition of raster plots, imposes a time grid which can be contested as well
since one may argue that a neuron spike can occur at ``any'' time (with the restriction
discussed above on ``continuous'' time)  while
the time grid set it to a time multiple of $\delta$ \cite{kirst-timme:09}.
So, a central question is: does the effects induced by this slight error on a spike time
matters ? This question can be addressed in the realm of models (like gIF models) and the answer is:
``it depends''. Indeed, according to the control parameters values such an error can be damped
(then it is harmless) or amplified by dynamics. This is precisely the discussion on contraction and initial
conditions sensitivity presented in section \ref{genedynIF}. Here, the mathematical
analysis relies strongly on the simple structure of IF models. But the question
of time discretisation and ``How precise is the timing of action potentials'' \cite{kirst-timme:09}
must be certainly addressed in a more general context and for more general models.

\paragraph{Neural code and predictability} Another interesting issue, raised by the dynamical
system point of view is the following. When building models which reproduce the dynamics
of neural assemblies, one is faced to the question ``how well does this model approximate
real neural systems''. 
 Let us state in a different way. Assume that we have built some
artificial neural network that mimics some part of the brain and assume (Gedanken experiment)
 that we are able to remove this part of the brain and to replace it by our artificial system,
what do we need to ensure that this ''cyborg brain'' works ``as'' the original one ? This is a (too) wide
open question but let us focus the discussion
on ``spikes'' aspects  for simplicity. Our artificial system receives spikes from other parts of the
brain and respond with spikes trains that it sends to various brain areas. Must this device
be able to reproduce \textit{exactly}, spikes par spikes, the response of the piece of brain that it replaces ?
It is possible to reproduce exactly the response of one neuron to Poisson stimuli \cite{jolivet-et-al:06}, and also
at the level of a network, to reproduce exactly finite spike trains coming from biological neurons 
\textit{over a finite time}
\cite{rostro-cessac-etal:09a}. But, 
from the dynamical system point of view this appears to be too restrictive since, as we saw,
in many cases dynamics is chaotic.  If one jitters
the time of a single spike, the output of the network can change
dramatically. Thus this property is not robust and reliable. 
 So which characteristics of the spike train must be reproduced ?
This question seems a clear challenge for a not so near future. This also
raise questions such as: Are all details important in modelling neural networks ?
Which details can be neglected and what is imposed by the biological structure ? \\

Though the analysis presented here is rather simple compared to
the overwhelming richness of brain dynamics, we hope that this work will be
useful for readers interested in this beautiful ongoing research field,
computational neuroscience.

 \pagebreak
\bibliographystyle{abbrv}

\bibliography{string,odyssee,biblio}

\end{document}